\documentclass[english,aps,prb,superscriptaddress,twocolumn]{revtex4-1}
\usepackage[T1]{fontenc}
\usepackage[latin9]{inputenc}
\usepackage{color}
\usepackage{float}
\usepackage{amsmath}
\usepackage{graphicx}
\usepackage{setspace}
\usepackage{esint}

\makeatletter

\newcommand{\lyxdot}{.}

\usepackage{braket}
\PassOptionsToPackage{caption=false}{subfig}
\renewcommand\[{\begin{equation}}
\renewcommand\]{\end{equation}}

\makeatother

\usepackage{babel}
\begin{document}

\title{The Anderson--Holstein Model in Two Flavors of the Non--Crossing
Approximation}

\author{Hsing-Ta Chen}

\affiliation{Department of Chemistry, Columbia University, New York, New York
10027, U.S.A.}

\author{Guy Cohen}

\affiliation{Department of Chemistry, Columbia University, New York, New York
10027, U.S.A.}

\affiliation{Department of Physics, Columbia University, New York, New York 10027,
U.S.A.}

\author{Andrew J. Millis}

\affiliation{Department of Physics, Columbia University, New York, New York 10027,
U.S.A.}

\author{David R. Reichman}

\affiliation{Department of Chemistry, Columbia University, New York, New York
10027, U.S.A.}
\begin{abstract}
The dynamical interplay between electron--electron interactions and
electron--phonon coupling is investigated within the Anderson--Holstein
model, a minimal model for open quantum systems that embody these
effects. The influence of phonons on spectral and transport properties
is explored in equilibrium, for non-equilibrium steady state and for
transient dynamics after a quench. Both the particle--hole symmetric
and the more generic particle--hole asymmetric cases are studied.
The treatment is based on two complementary non-crossing approximations,
the first of which is constructed around the weak-coupling limit and
the second around the polaron limit. In general, the two methods disagree
in nontrivial ways, indicating that more reliable approaches to the
problem are needed. The frameworks used here can form the starting
point for numerically exact methods based on bold-line continuous-time
quantum Monte Carlo algorithms capable of treating open systems simultaneously
coupled to multiple fermionic and bosonic baths.
\end{abstract}
\maketitle

\section{Introduction\label{sec:Introduction}}

The interaction between electrons and phonons plays an essential role
in condensed matter physics: it is for example the fundamental factor
responsible for the resistivity of conduction electrons in crystals
at relatively high temperatures and the onset of superconductivity
at low temperatures.\cite{AshcroftandMermin1976} In non-equilibrium
molecular electronics experiments,\cite{Aradhya2013,Nitzan2003,Qiu2003}
electron--phonon interactions are ever present and have major implications\cite{Joachim2005,Chen2003}
which can be exploited in the design of phononic devices.\cite{Li2012,Dubi2011}
In addition, the interplay between electron--electron interactions
(responsible for Coulomb blockade and the Kondo effect) and electron--phonon
scattering leads to novel and subtle behaviors.\cite{Scott2010,Zimbovskaya2011}
For example, conductance side peaks replicating the Kondo resonance\cite{Yu2004,Scott2010,Rakhmilevitch2014}
and negative differential resistance at voltages corresponding to
the vibrational energy of the molecule\cite{Gaudioso2000} have been
observed. In a broader sense, explicating the role played by electron--phonon
interactions in strongly correlated materials remains a fertile area
of research, where recent interest has focused, for example, on the
role played by phonons in fulleride,\cite{Capone2009} cuprate\cite{DalConte2012}
and pnictide superconductors\cite{Gadermaier2010,Gadermaier2014}
and the control of superconductivity and metal--insulator transitions
in correlated materials via strong laser fields.\cite{Perfetti2006,Perfetti2008,Capone2010,Fausti2011,Kaiser2014}

\textbf{}

A standard model that simultaneously describes both electronic interactions
and electron--phonon coupling in nanoscale devices is the Anderson--Holstein
model.\cite{Anderson1961,Holstein1959,Hewson2002} This model consists
of a single interacting site (sometimes called the dot or impurity)
coupled to a non-interacting electron reservoir (or reservoirs) and
to a set of localized phonon modes. The Anderson--Holstein model can
be considered a minimal description of the essential aspects of a
correlated electron system interacting with phonon excitations, and
has been used to describe vibrational effects in molecular electronics.\cite{Hewson2002,Cornaglia2004,Werner2007,Han2010}
Furthermore, within the framework of dynamical mean-field theory
(DMFT),\cite{Georges1996} the characterization of a strongly correlated
material with active phonon degrees of freedom may be effectively
reduced to the Anderson--Holstein model and its variants.\cite{Werner2007,Werner2013,Golez2015}

Despite the importance of the Anderson--Holstein model, there is surprisingly
little known about its real--time dynamical properties outside of
simple limits where perturbation arguments can be made. The case
of zero on-site electron--electron interactions can describe some
phenomena associated with the electron--phonon interaction,including
non--equilibrium transient dynamics, inelastic transport, and phonon-induced
side peaks.\cite{Paaske2005b,Jovchev2013,SeoaneSouto2014,Albrecht2013a,Albrecht2015}
This limit has been widely considered in the literature; despite its
simplicity, it is non-trivial to solve, especially out of equilibrium.
A variety of techniques have been used to analyze this model, including
perturbation theory in the electron--phonon coupling,\cite{Mitra2004}
a semi-classical treatment,\cite{Mitra2005} and master--equation
approaches.\cite{Mitra2004,Hartle2009,Schultz2009,Esposito2009,Esposito2010,Dou2015}
Semi-analytical approximations within nonequilibrium Keldysh Green's
functions (NEGF),\cite{Tikhodeev2001,Mii2002,Galperin2004,Ueda2006,Dash2010,Dash2011,Dong2013}
the equation-of-motion (EOM) approach,\cite{Galperin2006a,Galperin2007,Monreal2010a,Sayyad2015}
an interpolative ansatz,\cite{Martin-Rodero2008} and a recent dressed
tunneling approximation\cite{SeoaneSouto2014} have been applied to
the model in various limits. Numerically exact methods have also been
applied, including real-time Quantum Monte Carlo (QMC),\cite{Muhlbacher2008,Albrecht2013a,Albrecht2015,Klatt2015}
iterative path integral schemes\cite{Hutzen2012,Simine2013a,Simine2014}
and the multi-layer multi-configuration time-dependent Hartree (ML-MCTDH)
method.\cite{Albrecht2012,Wang2013}

Treatment of the combined effect of electron--electron and electron--phonon
interactions is simplest when the on-site Coulomb repulsion is effectively
infinite ($U\rightarrow\infty$). In this limit, some methods used
to treat the non-interacting case can be adopted and generalized,
including certain Monte Carlo approaches,\cite{Han2010,Albrecht2013}
the equation-of-motion technique,\cite{Entin-Wohlman2005,Koch2005,Koch2006,Monreal2009}
a decoupling scheme for NEGF,\cite{Swirkowicz2008} and the slave-boson
technique \cite{Goker2011,Roura-Bas2013}. Studies of the infinite--$U$
Anderson--Holstein model predict non-trivial effects, such as the
appearance of Kondo replicas above and below the chemical potential
and negative differential resistance associated with the destruction
of the Kondo resonance.\cite{Gaudioso2000,Han2010} However it remains
unclear if these predictions are valid outside of linear response
from equilibrium, and in general neither the $U=0$ nor $U\rightarrow\infty$
limits describe the bulk of interesting cases of experimental relevance.

Only a handful of approaches are capable of calculating properties
of a generic Anderson--Holstein model outside of the idealized limits
discussed above. Approximate methods, such as the master equation
approach, can accurately describe transport phenomena at high temperatures
and large voltages.\cite{Hartle2011a} The ML-MCTDH method is numerically
exact, but has difficulty converging for strong electron--phonon coupling
or far from equilibrium.\cite{Wilner2013,Wilner2014} The numerical
renormalization group (NRG) can also be extended to include electron--phonon
interactions, but remains difficult to apply out of equilibrium and
is generally reliable only for the low energy properties of the system.\cite{Hewson2002,Cornaglia2004,Cornaglia2005,Paaske2005b,Cornaglia2007,Eidelstein2013,Laakso2014}
 The auxiliary-field QMC method has been used to calculate the density
of states under the influence of the phonons in imaginary time,\cite{Arrachea2005}
but application to dynamics involves an uncontrolled analytical continuation
which is problematic at certain parameters,\cite{Cohen2014} and the
Matsubara formulation is only valid for equilibrium and linear response
properties.  Real time QMC provides an alternative numerically exact
approach which has the ability to describe transient dynamics and
non-equilibrium transport properties over a wide range of parameters.\cite{Konig1996,Muhlbacher2008,Schiro2009,Werner2013,Gull2011}
In combination with reduced dynamics techniques\cite{Cohen2011,Cohen2013}
it can sometimes be used to obtain results over very long timescales.\cite{Cohen2013a}
However, real time QMC is generically plagued by a dynamical sign
problem which limits the accessible timescales. Although not the direct
focus of this manuscript, we note that the approaches described here
can provide a foundation to allow for an amelioration of the sign
problem in QMC simulations.\cite{Gull2010,Cohen2014}

The self-consistent resummation of particular classes of interaction
terms may allow for an extension of the domain of validity provided
by bare perturbation theory. A prominent example is provided by the
non-crossing approximation (NCA).\cite{Bickers1987b,Pruschke1989}
The NCA is a semi-analytical method based on the resummation to all
orders of a specific subset of diagrams (those that do not cross temporally
on the Keldysh contour) associated with the hybridization between
the impurity and the non-interacting leads. It provides a computationally
inexpensive approach for solving generic impurity models out of equilibrium.\cite{Wingreen1994}
NCA is exact in the atomic limit, and works best in the limit of infinite
$U$ and finite $\epsilon$. The approximation does not fully capture
low energy properties and does not correctly reproduce the noninteracting
limit. But despite the quantitative inaccuracies, the NCA qualitatively
predicts the emergence and some properties of the Kondo resonance,
and is generally accurate for high-energy features. While the NCA
as a \textquotedbl{}stand alone\textquotedbl{} approximation may quantitatively
fail, higher order approximations (e.g. one--crossing approximation)
based on the same principles have been used,\cite{Haule2001,Eckstein2010}
and recent numerically exact QMC approaches have been formulated that
sample corrections to the NCA in a numerically exact way.\cite{Gull2010,Gull2011a,Cohen2014,Cohen2014a,Cohen2015}

The NCA has been extended to include the electron--phonon coupling,
via the slave-boson technique,\cite{Goker2011,Roura-Bas2013} in
nonequilibrium DMFT studies,\cite{Werner2013,Golez2015} and within
a pseudoparticle picture.\cite{White2012} A first goal of our work
is to clearly formulate two complementary NCA-like approximations
in the full many-body basis of the impurity, in a form suitable for
studying the non-equilibrium behavior of the Anderson--Holstein model,
and to compare and contrast the predictions of these distinct self-consistent
procedures. A second goal is to clearly delineate the diagrammatic
rules associated with each self-consistent resummation on the Keldysh
contour so that future exact QMC schemes which sample remaining diagrams
may be explicitly formulated. The outline of this paper is as follows.
In Sec.~\ref{sec:Coupling-Expansion} we introduce the Anderson--Holstein
model and provide the needed formalism. In Sec.~\ref{sec:Two-NCAs},
two distinct types of NCA-like approximation are described. In Sec.~\ref{sec:Results},
we present and compare results for transient dynamics, steady state
spectral function and differential conductance for a generic Anderson--Holstein
model in the Kondo regime. A summary and conclusion are presented
in Sec.~\ref{sec:Conclusions}.

\section{Coupling Expansion for Anderson--Holstein Model\label{sec:Coupling-Expansion}}

\subsection{Model and definitions}

We consider a single spin-degenerate impurity or quantum dot level
with a linear coupling to a phonon bath and to a pair of metallic
leads which will be referred to as ``left'' ($L$) and ``right''
($R$). This model is described by the nonequilibrium Anderson--Holstein
Hamiltonian\cite{Hewson2002,Cornaglia2004,Werner2007} 
\begin{equation}
H=H_{d}+H_{b}+V_{b}+\sum_{\ell\in L,R}\left(H_{\ell}+V_{\ell}\right).
\end{equation}
The electronic part of the dot Hamiltonian, $H_{d}$, is 
\begin{equation}
H_{d}=\sum_{\sigma=\uparrow,\downarrow}\epsilon_{\sigma}n_{\sigma}+Un_{\uparrow}n_{\downarrow},
\end{equation}
where $\epsilon_{\sigma}$ denotes the energy of singly-occupied states
and $U$ is the Coulomb interaction. The operators $d_{\sigma}^{\dagger}$
creates an electron of spin $\sigma$ on the dot and the occupation
$n_{\sigma}=d_{\sigma}^{\dagger}d_{\sigma}$.

The local phonon bath Hamiltonian is 
\begin{equation}
H_{b}=\sum_{q}\omega_{q}b_{q}^{\dagger}b_{q}.
\end{equation}
Here the $b_{q}^{\dagger}$ are phonon creation operators, and $\omega_{q}$
is the frequency associated with a phonon mode $q$. We will typically
assume that the phonons are initially in equilibrium, such that the
occupation of the phonon modes is given by the Bose--Einstein distribution
$\langle b_{q}^{\dagger}b_{q}\rangle=\frac{1}{e^{\beta_{d}\omega_{q}}-1}$,
$\beta_{d}$ being the inverse temperature of the phonon bath. The
electron--phonon coupling Hamiltonian $V_{b}$ is 
\begin{equation}
V_{b}=\sum_{q}\lambda_{q}(b_{q}^{\dagger}+b_{q})\left(n_{d}-\delta\right),\label{eq:electron_phonon_coupling_hamiltonian}
\end{equation}
where $n_{d}=\sum_{\sigma}n_{\sigma}$ is the total electronic occupation
of the dot and $\lambda_{q}$ the coupling strength between the dot
and phonon mode $q$. The parameter $\delta$ is of no physical significance,
in the sense that it may be absorbed into a redefinition of the zero
point of the oscillator coordinate. However, it is convenient to set
$\delta=1$, so that $\epsilon=0$ describes the particle--hole symmetric
dot, and we will primarily consider this case. We will also investigate
the case $\delta=0$, which provide a more convenient description
of a molecular junction in which polaron formation is linked to the
presence of extra electrons on the dot. In either case, the electron--phonon
coupling is characterized by a spectral density $J(\omega)\equiv\frac{\pi}{2}\sum_{q}\frac{\lambda_{q}^{2}}{\omega_{q}}\delta(\omega-\omega_{q})$.

The left and right lead Hamiltonians are 
\begin{equation}
H_{\ell}=\sum_{k\in\ell}\sum_{\sigma}\epsilon_{k}c_{k\sigma}^{\dagger}c_{k\sigma},
\end{equation}
with $\ell\in\left\{ L,R\right\} $ and the index $k$ denoting a
level within a lead. We assume the leads to be non-interacting, such
that they are fully described by the dispersion relation $\epsilon_{k}$
and the creation operators $c_{k\sigma}^{\dagger}$. The leads are
taken to each be initially isolated and at an equilibrium state with
density matrix $\rho_{\ell}$, and their thermodynamic properties
characterized by an inverse temperature $\beta_{\ell}$ and a chemical
potential $\mu_{\ell}$. The initial density of states is then described
by a Fermi--Dirac distribution, $\langle c_{k\sigma}^{\dagger}c_{k\sigma}\rangle=f_{\ell}(\epsilon_{k})=\frac{1}{e^{\beta_{\ell}(\epsilon_{k}-\mu_{\ell})}+1}$.The
hybridization $V_{\ell}$ between the dot and lead electrons is described
by the dot-lead coupling Hamiltonian 
\begin{equation}
V_{\ell}=\sum_{k\in\ell}\sum_{\sigma}\left[t_{k}d_{\sigma}c_{k\sigma}^{\dagger}+t_{k}^{*}d_{\sigma}^{\dagger}c_{k\sigma}\right],
\end{equation}
where $t_{k}$ enumerates the coupling strength between the dot and
level $k$ of lead $\ell$. We define a coupling density $\Gamma_{\ell}(\omega)=2\pi\sum_{k\in\ell}|t_{k}|^{2}\delta(\omega-\epsilon_{k})$,
which fully characterizes the $t_{k}$ within this model.

In steady state the dynamical response of a system is characterized
by its spectral function
\begin{equation}
A(\omega)=\frac{i}{2\pi}\mathrm{Tr}\left\{ G^{r}(\omega)-G^{a}(\omega)\right\} ,\label{eq:steady_state_spectral_function}
\end{equation}
which may be considered a probe of the density of electron and hole
excitations as a function of energy. To calculate the spectral function
at frequency $\omega^{\prime}$, we use the auxiliary current method\cite{Cohen2014,Cohen2014a}
by appending two auxiliary leads to the model, $H\rightarrow H+H_{A}+V_{A}$,
where $H_{A}=\sum_{k\in A}\epsilon_{k}a_{k}^{\dagger}a_{k}$ and $V_{A}=\sum_{k\in A}\sum_{\sigma}\left[t_{k}d_{\sigma}a_{k}^{\dagger}+t_{k}^{*}d_{\sigma}^{\dagger}a_{k}\right]$.
These auxiliary leads are coupled to the dot at the single frequency
$\omega'$ with a spectral density $\Gamma_{A}^{\omega'}(\omega)=\eta\delta(\omega-\omega')$.
One lead is kept fully occupied, such that $f_{A1}(\omega)=1$; the
other lead is kept empty, such that $f_{A0}(\omega)=0$. We can calculate
the \emph{auxiliary} spectral function $A\left(\omega;t\right)$ at
any finite time by the following relation: 
\begin{equation}
A\left(\omega;t\right)=\lim_{\eta\rightarrow0}-\frac{2h}{e\pi\eta}\left[I_{A1}^{\omega}\left(t\right)-I_{A0}^{\omega}\left(t\right)\right].
\end{equation}
Here, $I_{A0}^{\omega}\left(t\right)$ and $I_{A1}^{\omega}\left(t\right)$
are the currents flowing out of lead $A0$ and $A1$, respectively,
at time $t$. At long times, the auxiliary spectral function approaches
the steady state spectral function, Eq.~\ref{eq:steady_state_spectral_function}.
While at finite times the auxiliary spectral function does not conform
to the standard definition of a spectral function in terms of a Fourier
transform of a correlation function, it retains the appealing physical
interpretation as a measure of the single-particle excitation density
in energies, and could in principle be accessed experimentally by
way of three-lead experiments.\cite{Lebanon2001,Sun2001,Cohen2014,Cohen2014a}

We shall also be interested in the differential conductance, 
\begin{equation}
G(V)=\frac{\mathrm{d}}{\mathrm{d}V}(I_{L}-I_{R}).
\end{equation}
which is directly accessible in transport experiments. Here, $V=\mu_{L}-\mu_{R}$
is the bias voltage between the two leads. The current $I_{\ell}(t)$
out of lead $\ell$ is given by $I_{\ell}(t)=\langle\mathcal{I}_{\ell}(t)\rangle$,
where the current operator for a given lead, 
\begin{equation}
\mathcal{I}_{\ell}=\dot{N}_{\ell}=i\sum_{k\in\ell}\left(t_{k}c_{k\sigma}^{\dagger}d_{\sigma}-t_{k}^{*}c_{k\sigma}d_{\sigma}^{\dagger}\right),
\end{equation}
describes the rate at which carriers flow out of that lead. The differential
conductance is often interpreted as an estimator for the equilibrium
spectral function of the model. However, this interpretation is only
valid if the spectral function is independent of the bias voltage.
In practice, the two quantities may be qualitatively different.\cite{Cohen2014a}

\subsection{Coupling expansion: general formalism}

We now formulate a double expansion in the electron--phonon and dot--lead
couplings. A brief review will be provided here for completeness;
we refer readers interested in a more detailed technical outline of
the formalism and algorithm elsewhere.\cite{Cohen2014} We begin by
recasting the Hamiltonian as $H=H_{0}+V$. $H_{0}$ describes the
isolated dot and bath subsystems, while $V=V_{b}+\sum_{\ell}V_{\ell}$
describes the coupling Hamiltonian.

The expectation value of an operator ${\cal O}$ at time $t$ can
be written in the form $\langle\mathcal{O}(t)\rangle=\langle e^{iHt}\mathcal{O}e^{-iHt}\rangle=\langle U^{\dagger}(t)\mathcal{O}_{I}(t)U(t)\rangle$,
where $U(t)=e^{iH_{0}t}e^{-iHt}$ and $\mathcal{O}_{I}(t)=e^{iH_{0}t}\mathcal{O}e^{-iH_{0}t}$.
The subscript $I$ denotes an operator in the interaction picture.
We also define thermal averaging by way of the notation $\langle O\rangle\equiv\mathrm{Tr}\left\{ \rho O\right\} $,
with the averaging performed with respect to the uncorrelated initial
density matrix formed by the product of subsystem density matrices:
$\rho=\rho_{d}\otimes\prod_{\ell}\rho_{\ell}\otimes\rho_{b}$. Thus
the dynamics that appear in the following are not in equilibrium and
illustrate the approach to equilibrium in the appropriate limits.
Other than in some very special cases, a \emph{finite} system coupled
to an infinite thermal bath which is allowed to evolve in time is
generally found to reproduce the steady state results at long times.
Moreover, this is often the only rigorous way to construct the correct
nonequilibrium steady state in open quantum systems. Initial correlations
allow the system to be thermalized at time zero. Within DMFT,\cite{Eckstein2010,Werner2007,Werner2010a,Werner2013,Golez2015}
one deals with an \emph{infinite} interacting system which is not
coupled to a bath, and the role of the initial correlations therefore
becomes more important. They are needed to model an initially thermalized
system, which might be thought of as a system that had been weakly
coupled to a bath and allowed to relax before the beginning of the
calculation. 

We now describe the details of a Dyson expansion for the reduced propagator
on the Keldysh contour. We can expand $U(t)$ in a Dyson series
\begin{equation}
\begin{split}U(t)= & \sum_{n=0}^{\infty}\left(-i\right)^{n}\int_{0}^{t}dt_{1}\int_{0}^{t_{1}}dt_{2}\cdots\int_{0}^{t_{n-1}}dt_{n}\\
 & ~\times V_{I}\left(t_{1}\right)V_{I}\left(t_{2}\right)\cdots V_{I}\left(t_{n}\right),
\end{split}
\end{equation}
such that the propagator can be expressed as $e^{-iHt}=e^{-iH_{0}t}U(t)$.
We adopt the many-body atomic states of the isolated dot, $\left\{ |\alpha\rangle\right\} =\left\{ |00\rangle\equiv|0\rangle,|\uparrow\rangle\equiv|1\rangle,|\downarrow\rangle\equiv|2\rangle,|\uparrow\downarrow\rangle\equiv|3\rangle\right\} $,
as a basis, and define the reduced propagator matrix element $G_{\alpha\beta}(t)\equiv\left\langle \alpha\left|\mathrm{Tr}_{B}\left\{ \rho e^{-iHt}\right\} \right|\beta\right\rangle $.
The trace is taken over the lead and phonon degrees of freedom: $\mathrm{Tr}_{B}\equiv\mathrm{Tr}_{\ell}\mathrm{Tr}_{b}$.
The remaining quantity is reduced to the dimensionality of the (many-body)
dot subspace. We also define the unperturbed reduced propagator $G_{\alpha\beta}^{(0)}(t)\equiv\left\langle \alpha\left|\mathrm{Tr}_{B}\left\{ \rho e^{-iH_{0}t}\right\} \right|\beta\right\rangle $.
$G_{\alpha\beta}^{\left(0\right)}$ is diagonal for the model treated
here, and takes the form $G_{\alpha\beta}^{(0)}(t)=\Phi(t)\delta_{\alpha\beta}e^{-iE_{\alpha}t}$.
The state energy $E_{\alpha}$ is evaluated from the isolated dot
Hamiltonian. The factor $\Phi(t)=\mathrm{Tr}_{B}\left\{ \rho e^{-i(H_{0}-H_{d})t}\right\} $
is related to fluctuations in the noninteracting baths, and is independent
of the dot state. It is exactly canceled when considering any quantity
defined on the two branch Keldysh contour, and can therefore be safely
ignored.

\begin{figure}
\begin{centering}
\includegraphics{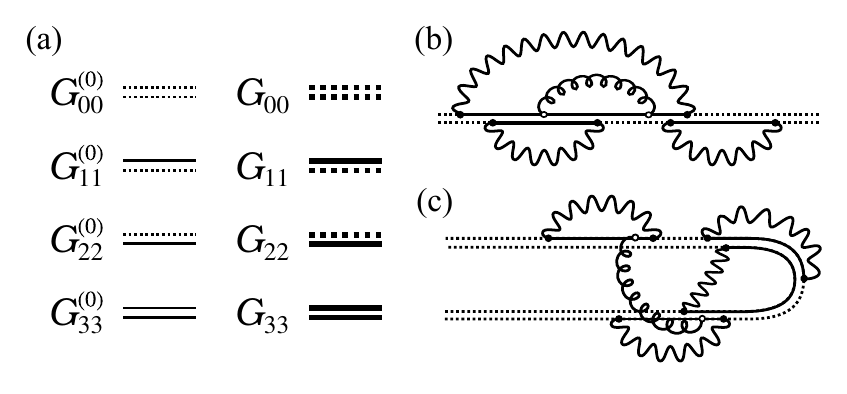}
\par\end{centering}

\caption{(a) The elements of the unperturbed propagator $G_{\alpha\alpha}^{(0)}$
(left column, thin lines) and of the NCA propagator $G_{\alpha\alpha}$
(right column, bold lines). The upper line represents spin up occupation
and the lower line spin down occupation. The dotted line signifies
that the spin level is unoccupied, while a solid line marks it as
occupied. (b) An example of a diagram included in the reduced propagator
$G_{00}$. Electronic hybridization lines are shown as wiggly lines,
and phonon interaction lines as gluon lines. (c) An example of a diagram
on the Keldysh contour with inter-branch lines and a special hybridization
line ending at the final time, corresponding to a contribution to
the current.\label{example-diagram}}
\end{figure}

The full, or perturbed, reduced propagator $G_{\alpha\beta}\left(t\right)$
is also diagonal. Contributions to it from the coupling Hamiltonian
are nonzero only when the creation and annihilation operators occur
in pairs, such that only even orders must be accounted for: 
\begin{equation}
\begin{split}G_{\alpha\alpha}(t)= & G_{\alpha\alpha}^{(0)}(t)-\int_{0}^{t}dt_{1}\int_{0}^{t_{1}}dt_{2}\\
 & \times\langle\alpha|\mathrm{Tr}_{B}\left\{ \rho e^{-iH_{0}t}V_{I}(t_{1})V_{I}(t_{2})\right\} |\alpha\rangle\\
 & +\cdots.
\end{split}
\end{equation}
This series can be represented as a summation of diagrams in which
the coupling Hamiltonian appears an even number of times. An example
diagram is shown Fig.~\ref{example-diagram}: in (a), the representation
of $G_{\alpha\alpha}^{\left(0\right)}$ (thin lines) and $G_{\alpha\alpha}$
(bold lines) in terms of pairs of solid and dashed lines is shown.
In (b) a diagram is shown which contains Fermion hybridizations, denoted
by wiggly lines which change the dot population, and phonon interactions,
denoted by wavy lines with loops which do not change the population
(and may appear only within certain dot states, as detailed below).

The reduced propagator satisfies a causal Dyson equation of the form
\begin{equation}
\begin{split}G_{\alpha\alpha}(t)= & G_{\alpha\alpha}^{(0)}(t)+\int_{0}^{t}dt_{1}\int_{0}^{t_{1}}dt_{2}\\
 & \times G_{\alpha\alpha}^{(0)}(t-t_{1})\Sigma_{\alpha\alpha}(t_{1},t_{2})G_{\alpha\alpha}(t_{2}),
\end{split}
\end{equation}
where all non-trivial aspects of the problem are contained in the
(proper) self energy $\Sigma_{\alpha\alpha}(t_{1},t_{2})$. Solving
the Dyson equation self-consistently is in itself an inexpensive computation
if the self energy is known. Within the hybridization expansion for
the phonon-free version of the model, the simplest approximation to
the self energy includes only a single pair of coupling Hamiltonians:
\begin{eqnarray}
\Sigma_{\alpha\alpha}^{\mathrm{\text{2BA}}}(t_{1}-t_{2}) & = & -\langle\alpha|\mathrm{Tr}_{b}\left\{ \rho Ve^{-iH_{0}(t_{1}-t_{2})}V\right\} |\alpha\rangle\\
 & = & \sum_{\beta}G_{\beta\beta}^{\left(0\right)}(t_{1}-t_{2})\times\Delta_{\alpha\alpha}^{\beta}(t_{1}-t_{2}),\nonumber 
\end{eqnarray}
where the hybridization function is defined as
\begin{equation}
\Delta_{\alpha\alpha}^{\beta}(t_{1}-t_{2})\equiv{\color{red}-}\langle\alpha|\mbox{Tr}_{b}\left\{ \rho V_{I}(t_{1})|\beta\rangle\langle\beta|V_{I}(t_{2})\right\} |\alpha\rangle.
\end{equation}
This is known as the second-order Born approximation (2BA). The non-crossing
approximation (NCA), also known as the self-consistent Born approximation
(SCBA), takes the same form, but inserts the full propagator $G$
into the self energy: 
\begin{equation}
\Sigma_{\alpha\alpha}^{\text{NCA}}(t_{1}-t_{2})=\sum_{\beta}G_{\beta\beta}(t_{1}-t_{2})\times\Delta_{\alpha\alpha}^{\beta}(t_{1}-t_{2}).
\end{equation}
 With this self energy, we can obtain an approximate propagator containing
an infinite, but partial, subset of the diagrams contributing to the
reduced propagator, namely all diagrams in which hybridization lines
do not cross each other. In the following section, two ways of generalizing
this idea to the full Anderson--Holstein model will be described.

So far, in order to simplify the discussion, we have limited our attention
to a reduced propagator living on a single branch of the Keldysh contour.
To calculate a physical observable, we must consider a two-branch
Keldysh contour with the observable operator ${\cal O}$ placed at
the final time $t$, and take into account diagrams with lines crossing
between the two branches. To this end, we define a vertex function
of the observable $\mathcal{O}$, with the two time variables $t$
and $t^{\prime}$ placed on opposite branches of the contour. With
$t'\rightarrow t$, this object yields the physical expectation value
of observable $\mathcal{O}\left(t\right)$. In particular, the current
out of the lead $\ell$ can be obtained from $I_{\ell}(t)=\langle\mathcal{I}_{\ell}(t)\rangle$,
where the current operator 
\begin{equation}
\mathcal{I}_{\ell}=\dot{N}_{\ell}=i\sum_{k\in\ell}\left(t_{k}c_{k\sigma}^{\dagger}d_{\sigma}-t_{k}^{*}c_{k\sigma}d_{\sigma}^{\dagger}\right)
\end{equation}
and the $c$ and $d$ operators are understood to be at the tip of
the Keldysh contour.

Because $\mathcal{I}_{\ell}$ is composed of the same operators appearing
in the dot--bath hybridization Hamiltonian, within the coupling expansion
the current can be obtained by summing over diagrams which have a
special hybridization line placing the current operator at the final
time of the Keldysh contour. An example of such a diagram is given
in Fig.~\ref{example-diagram} (c).

\section{Two types of NCA for electron--phonon coupling\label{sec:Two-NCAs}}

In this section we lay out the construction of NCA-like approximations
in two limits: First, a \emph{bare NCA} based on \textit{\emph{self-consistently
resummed }}\emph{second order} perturbation theory for the electron--phonon
and dot-lead and electron--phonon Hamiltonians. Second, a \emph{dressed
NCA} in which the Hamiltonian is modified by a Lang-Firsov transformation
so that the coupling Hamiltonian becomes a phonon-dressed dot-lead
coupling, and includes non-crossing diagrams composed of phonon-dressed
hybridization lines.\cite{Werner2007,Werner2013} Both approximations
can be extended to higher orders, or used as the preliminary step
withing a numerically exact bold-line QMC algorithm. We initially
formulate these two types of NCA for the symmetric Anderson--Holstein
model in the following two subsections, then discuss the asymmetric
case.

\subsection{Weak coupling perturbation theory}

The \emph{bare NCA} approximation is specified by the following equations
\begin{equation}
\mathbf{G}^{-1}=\mathbf{G}_{0}^{-1}-\boldsymbol{\Sigma}^{\ell}-\boldsymbol{\Sigma}^{b},
\end{equation}
with $\mathbf{G}$, $\mathbf{G}_{0}$ and $\boldsymbol{\Sigma}$ matrices
(diagonal, in the cases of interest here) in the Hilbert space of
the decoupled dot, and the lead ($\ell$) and phonon ($b$) self energies$\boldsymbol{\Sigma}$
given by

\begin{equation}
\Sigma_{\alpha\alpha}^{\ell}(t_{1},t_{2})=\sum_{\beta}G_{\beta\beta}(t_{1},t_{2})\times\Delta_{\alpha\alpha}^{\beta}(t_{1},t_{2})
\end{equation}
\begin{equation}
\Sigma_{\alpha\alpha}^{b}(t_{1},t_{2})=G_{\alpha\alpha}(t_{1},t_{2})\times\Lambda_{\alpha\alpha}(t_{1},t_{2})
\end{equation}
with the lead hybridization function
\begin{align}
\Delta_{\alpha\alpha}^{\beta} & (t_{1},t_{2})=\nonumber \\
 & \sum_{\sigma}\langle\alpha|d_{\sigma}|\beta\rangle\langle\beta|d_{\sigma}^{\dagger}|\alpha\rangle\sum_{k\in\ell}|t_{k}|^{2}\mbox{Tr}_{\ell}\left[\rho_{\ell}c_{k\sigma}^{\dagger}(t_{1})c_{k\sigma}(t_{2})\right]\nonumber \\
+ & \sum_{\sigma}\langle\alpha|d_{\sigma}^{\dagger}|\beta\rangle\langle\beta|d_{\sigma}|\alpha\rangle\sum_{k\in\ell}|t_{k}|^{2}\mbox{Tr}_{\ell}\left[\rho_{\ell}c_{k\sigma}(t_{1})c_{k\sigma}^{\dagger}(t_{2})\right].
\end{align}
We also define the lesser and greater hybridization functions $\Delta_{\ell}^{<,>}(\tau_{1},\tau_{2})=\sum_{k\in\ell}|t_{k}|^{2}\mbox{Tr}_{\ell}\left[\rho c_{k\sigma}^{\dagger}(\tau_{1})c_{k\sigma}(\tau_{2})\right]$
for each lead $\ell$ and times $\tau_{1}$, $\tau_{2}$ on the Keldysh
contour. $\Delta_{\ell}^{>}$ is used when $\tau_{1}$ precedes $\tau_{2}$,
and $\Delta_{\ell}^{<}$ is used otherwise. The dot--lead hybridization
function for each lead can be expressed in terms of the coupling densities
$\Gamma_{\ell}(\omega)$ and the initial occupation of that lead:

\begin{equation}
\Delta_{\ell}^{>}(t_{1},t_{2})=i\int_{-\infty}^{\infty}\frac{d\omega}{\pi}e^{-i\omega(t_{1}-t_{2})}\Gamma_{\ell}(\omega)\left[1-f_{\ell}(\omega-\mu_{\ell})\right],
\end{equation}
\begin{equation}
\Delta_{\ell}^{<}(t_{1},t_{2})=-i\int_{-\infty}^{\infty}\frac{d\omega}{\pi}e^{-i\omega(t_{1}-t_{2})}\Gamma_{\ell}(\omega)f_{\ell}(\omega-\mu_{\ell}).
\end{equation}

We similarly define the phonon hybridization function
\begin{align}
\Lambda_{\alpha\alpha} & (t_{1},t_{2})=\langle\alpha|\left(n_{d}(t_{1})-\delta\right)\left(n_{d}(t_{2})-\delta\right)|\alpha\rangle\times\nonumber \\
 & \sum_{q}\lambda_{q}^{2}\mathrm{Tr}_{b}\left[\rho_{b}\left(b_{q}^{\dagger}(t_{1})+b_{q}(t_{1})\right)\left(b_{q}^{\dagger}(t_{2})+b_{q}(t_{2})\right)\right].
\end{align}

This is analogous (but not identical) to the pseudoparticle NCA approximation
of ref.~\onlinecite{White2012}. Since the the electron--phonon coupling
$V_{b}$ does not modify the electronic state of the dot, one can
write $\langle\alpha|\left(n_{d}(t_{1})-\delta\right)\left(n_{d}(t_{2})-\delta\right)|\alpha\rangle=\left(n_{d}^{(\alpha)}-\delta\right)^{2}$.
We also define the bath correlation function, $B_{q}(t_{1},t_{2})=\mathrm{Tr}_{b}\left[\rho_{b}\left(b_{q}^{\dagger}(t_{1})+b_{q}(t_{1})\right)\left(b_{q}^{\dagger}(t_{2})+b_{q}(t_{2})\right)\right]$.
It can be expressed in terms of the frequency $\omega_{q}$ and the
inverse temperature $\beta$ of the local phonon modes, $B_{q}(t)=\coth(\beta\omega_{q}/2)\cos\left(\omega_{q}t\right)-i\sin\left(\omega_{q}t\right)$,
if we consider a bath initially composed of free harmonic phonon modes.
Thus, it is possible to recast the phonon hybridization function as
$\Lambda_{\alpha\alpha}(t_{1}-t_{2})=\left(n_{d}^{(\alpha)}-\delta\right)^{2}\times\Lambda_{b}(t_{1}-t_{2})$,
where 

\begin{equation}
\Lambda_{b}(t_{1}-t_{2})=\sum_{q}\lambda_{q}^{2}B_{q}(t_{1}-t_{2}).
\end{equation}
Just as the electronic hybridization function is described by $\Gamma_{\ell}\left(\omega\right)$,
the phonon bath is usually characterized by its spectral density,
$J(\omega)=\frac{\pi}{2}\sum_{q}\left(\lambda_{q}^{2}/\omega_{q}\right)\delta(\omega-\omega_{q})$.
In particular, 
\begin{equation}
\Lambda_{b}(t_{1}-t_{2})=\frac{2}{\pi}\int d\omega J(\omega)\omega B_{\omega}(t_{1}-t_{2}).
\end{equation}

\begin{figure}
\begin{centering}
\includegraphics{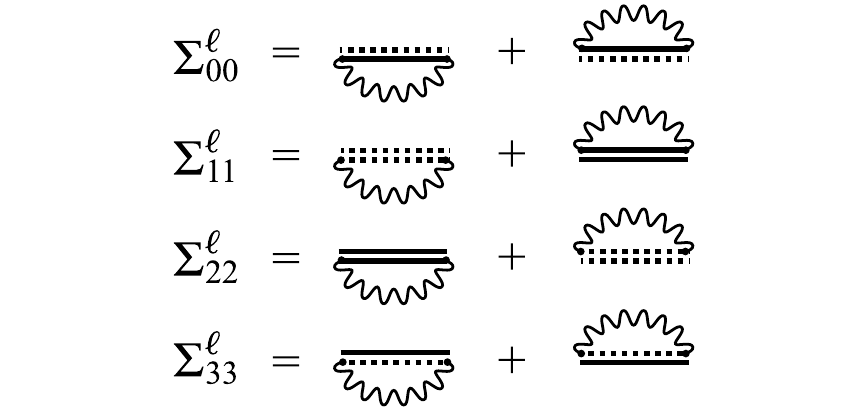}
\par\end{centering}

\caption{The electron hybridization diagrams included in the bare NCA self
energy, where the wiggly lines denote electronic dot--lead hybridization
lines. The pairs of straight lines represent the dot's electronic
state, with the two lines standing for the two possible spins: a solid
line represents an occupied spin level, whereas dashed lines stand
for empty spin levels.\label{self-energy-bare-ele}}
\end{figure}

\begin{figure}
\begin{centering}
\includegraphics{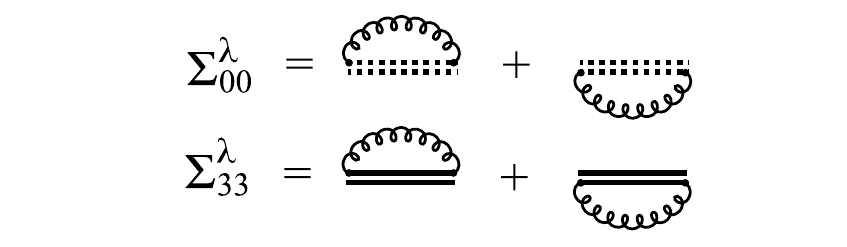}
\par\end{centering}

\caption{The phonon interaction diagrams for the bare NCA self energy in the
symmetric case $\delta=1$. The curly lines denote phonon interaction
lines, and straight lines are as in Fig.~\ref{self-energy-bare-ele}.\label{self-energy-bare-pho}}
\end{figure}

Fig.~\ref{self-energy-bare-ele} and Fig.~\ref{self-energy-bare-pho}
illustrate the diagrams included in the self energy of the \textit{bare
NCA} approach (for the symmetric case $\delta=1$). The wiggly lines
in Fig.~\ref{self-energy-bare-ele} denote the dot--lead hybridization
$\Delta_{\alpha\alpha}^{\beta}(t_{1}-t_{2})$, while the phonon lines
of Fig.~\ref{self-energy-bare-pho} symbolize the phonon coupling
$\Lambda_{\alpha\alpha}(t_{1}-t_{2})$. The computation of the Green's
function from the Dyson equation using this approximate self energy
embodies a self-consistent perturbative expansion including the lowest
order skeleton diagrams in both the dot--lead hybridization and electron--phonon
coupling. We expect this bare NCA approach to be more applicable in
the regime where both $\lambda$ and $\Gamma$ are small. Additionally,
the Green's function resulting from the bare NCA does not contain
certain multi--phonon excitations, related to crossing diagrams, which
might be expected to affect the dot electron if the phonon relaxation
is slow. This implies that the bare NCA is more accurate in the limit
of the fast phonon bath.

\subsection{Strong coupling perturbation theory}

In this section, we present a version of the non-crossing approximation
more suitable to strong coupling between the dot and the phonon bath
to the propagator formulation. This approach, which we will refer
to as the \emph{dressed NCA}, has previously been employed within
a standard Green's function formulation in ref.~\onlinecite{Werner2007,Werner2013}.

We begin by performing the unitary Lang--Firsov transformation $\tilde{H}=SHS^{-1}$
with $S=e^{\frac{\lambda}{\omega_{0}}(b^{\dagger}-b)n_{d}}$, which
eliminates the explicit electron--phonon coupling in the Hamiltonian.
We set the unperturbed Hamiltonian to be $H_{0}=H_{d}+H_{b}+V_{b}$.
After the transformation, this becomes

\begin{equation}
\tilde{H}_{0}=\tilde{\epsilon}_{d}\tilde{n}_{d}+\tilde{U}\tilde{n}_{\uparrow}\tilde{n}_{\downarrow},
\end{equation}
\begin{equation}
\tilde{V}_{\ell}=\sum_{k\in\ell}\sum_{\sigma}\left[t_{k}\tilde{d}_{\sigma}c_{k\sigma}^{\dagger}+t_{k}^{*}\tilde{d}_{\sigma}^{\dagger}c_{k\sigma}\right].
\end{equation}
In the above expressions, the bare dot energy $\epsilon$ and the
Coulomb interaction strength $U$ are replaced by the renormalized
quantities 
\begin{equation}
\tilde{\epsilon}=\epsilon+(2\delta-1)\lambda^{2}/\omega_{0},
\end{equation}
\begin{equation}
\tilde{U}=U-2\lambda^{2}/\omega_{0}.\label{eq:normalized_U}
\end{equation}
Also, the dot electron creation and annihilation operators become
\begin{equation}
\tilde{d}_{\sigma}=e^{-\frac{\lambda}{\omega_{o}}(b^{\dagger}-b)}d_{\sigma},
\end{equation}
\begin{equation}
\tilde{d}_{\sigma}^{\dagger}=e^{\frac{\lambda}{\omega_{o}}(b^{\dagger}-b)}d_{\sigma}^{\dagger}.
\end{equation}
All pairs of hybridization events are therefore connected by an infinite
set of phonon hybridization lines generated by these exponential phonon
displacement operators.

\begin{figure}
\begin{centering}
\includegraphics{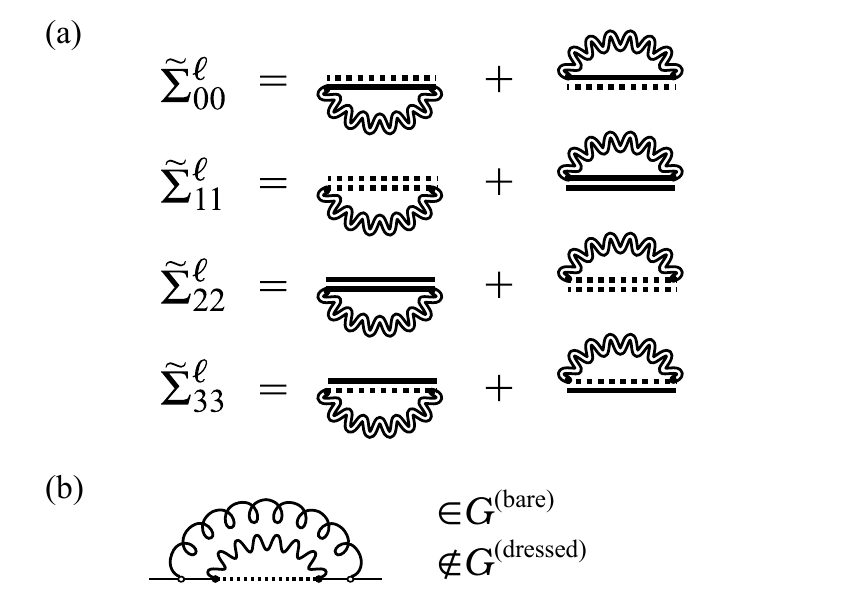}
\par\end{centering}

\caption{(a) The diagrams representing the different matrix elements of the
dressed NCA self energy. The wiggly double lines denote electron hybridization
lines dressed by phonon interactions. (b) An example of a bare NCA
diagram of the lowest order is not included in the dressed NCA diagrams.
\label{fig:self-energy-dressed}}
\end{figure}

Within the dressed NCA approximation for the self energy, we consider
only the dressed phonon lines appearing along the noncrossing fermionic
hybridization lines, as illustrated in Fig.~\ref{fig:self-energy-dressed}.
With this assumption, the effect of the electron--phonon interaction
is simply to reweigh each fermionic hybridization line with a phonon-dependent
factor, such that the NCA self energy takes the form
\begin{eqnarray}
\tilde{\Sigma}_{\alpha\alpha}^{\ell}(t_{1}-t_{2}) & = & w(t_{1}-t_{2})\\
 &  & ~\times\sum_{\beta}\Delta_{\alpha\alpha}^{\beta}(t_{1}-t_{2})G_{\beta\beta}^{(0)}(t_{1}-t_{2}).\nonumber 
\end{eqnarray}
The phonon weight $w(t_{1}-t_{2})$ is given by 
\begin{eqnarray}
w(t) & = & \exp\left\{ -\sum_{q}\left(\frac{\lambda_{q}}{\omega_{q}}\right)^{2}\times\right.\\
 &  & ~\left[(1-\cos\omega_{q}t)\coth(\beta\omega_{q}/2)+i\sin\omega_{q}t\right]\Biggr\}.\nonumber 
\end{eqnarray}
In terms of the bath spectral density $J\left(\omega\right)$, this
can be written as 
\[
w(t)=\exp\left\{ -Q_{2}(t)-iQ_{1}(t)\right\} ,
\]
where
\begin{eqnarray}
Q_{1}(t) & = & \frac{2}{\pi}\int d\omega\frac{J(\omega)}{\omega}\sin\omega t,\\
Q_{2}(t) & = & \frac{2}{\pi}\int d\omega\frac{J(\omega)}{\omega}(1-\cos\omega t)\coth(\beta\omega/2).
\end{eqnarray}

The dressed NCA self energy includes many phonon interactions not
included in the bare NCA. The self energy diagrams composed of the
transformed dot operators $\tilde{d}_{\sigma}$ and $\tilde{d}_{\sigma}^{\dagger}$
can be expanded in terms of the bare dot operators and effectively
contain all the hybridization diagram within the wiggly double lines.
Also, the polaron shift of $U$ and $\epsilon$ is expliciltly included
within the dressed NCA, but not the bare NCA. One might expect it
to be a more appropriate approximation in the polaron limit. On the
other hand, it also misses some contributions that are included in
the bare NCA (see Fig.~\ref{fig:self-energy-dressed}~(b)) and over-emphasizes
others, and at weak coupling to the phonons it might be expected to
be less accurate. The two approximations are therefore somewhat complementary,
if in a non-rigorous sense; it is reasonable to assume that conclusions
supported by both may be robust to the nature of the approximations,
while conclusions supported by only are suspect and should be investigated
further.

\subsection{NCA for asymmetric model\label{sub:NCA-for-asymmetric-model}}

We now briefly discuss the structure of the non--crossing approximation
for the case of an asymmetric Anderson--Holstein model in which the
counter term is not included (i.e. $\delta=0$ in Eq.~(\ref{eq:electron_phonon_coupling_hamiltonian})).
The phonon can then only be created or destroyed in the single electron
state or the doubly occupied state, not in the empty state. Such a
model might be considered a more physically realistic description
of a quantum junction, where one is interested in vibrational states
coupled to electrons.

\begin{figure}[H]
\begin{centering}
\includegraphics{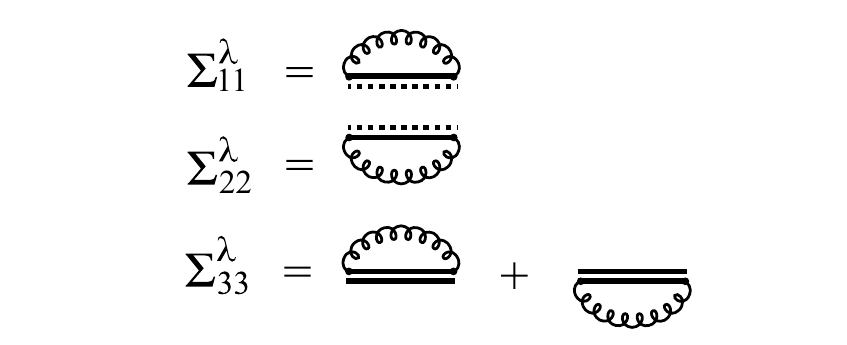}
\par\end{centering}

\caption{The phonon interaction diagrams for asymmetrical model. \label{fig:self-energy-bare-assymetric}}
\end{figure}

In the bare NCA calculation, the phonon coupling lines only connect
points with occupied electron states. The interaction diagrams for
the bare NCA self energy therefore no longer have the symmetric structure
of Fig.~(\ref{fig:self-energy-dressed}), but rather include a different
number of phonon inclusions for each of the matrix elements. This
is illustrated in Fig.~(\ref{fig:self-energy-bare-assymetric}).

For the dressed NCA, the same Lang-Firsov transformation is performed
to eliminate the explicit electron--phonon coupling. The dressed coupling
Hamiltonian then remains the same as in the symmetric case. However,
the renormalized energy becomes 
\begin{equation}
\tilde{\epsilon}=\epsilon-\frac{\lambda^{2}}{\omega_{0}},
\end{equation}
while the renormalized interaction remains the same as Eq.~(\ref{eq:normalized_U}).

With this coupling, $\epsilon=0$ does not correspond to a particle--hole
symmetric point. In the absence of dot--lead coupling, the charge
transfer bands are centered around $\omega_{+}=\frac{U}{2}+\frac{\lambda^{2}}{\omega_{0}}$
and $\omega_{-}=-\frac{U}{2}+3\frac{\lambda^{2}}{\omega_{0}}$.

\section{Results\label{sec:Results}}

We now discuss the application of the two NCA approaches described
above to the Anderson--Holstein impurity model, focusing on a case
where the dot has degenerate spin levels ($\epsilon_{\uparrow}=\epsilon_{\downarrow}=\epsilon_{d}$)
and obeys particle--hole symmetry ($\epsilon_{d}=-\frac{U}{2}$) in
the absence of phonons. The leads are assumed to be flat with a soft
cutoff: $\Gamma_{\ell}(\omega)=\frac{\Gamma_{\ell}}{(1+e^{\nu(\omega-\Omega_{c})})(1+e^{-\nu(\omega+\Omega_{c})})}$,
where $\Omega_{c}=10$ and $\nu=10$. We consider only symmetrical
couplings to the left and right leads, $\Gamma_{L}=\Gamma_{R}=0.5\Gamma$,
and apply the bias voltage $V$ symmetrically such that the chemical
potentials are given by $\mu_{L}=-\mu_{R}=0.5V$.

The methods we have described are suitable for the exploration of
systems containing multiple electron and phonon baths with complicated
densities of states, but we focus on a phonon bath with single mode,
$H_{b}=\omega_{0}b^{\dagger}b$. The electron--phonon coupling Hamiltonian
becomes $V_{b}=\lambda(b^{\dagger}+b)\left(n_{d}-\delta\right)$ and
the strength is characterized by by the parameter $\lambda$. We
assume that all baths are initially at the same inverse temperature
$\beta=10/\Gamma$.

To calculate the spectral function $A(\omega)$ by the double probe
scheme, we attach a pair of auxiliary leads to the system and measure
the corresponding auxiliary currents. The spectral density of the
auxiliary leads is a Gaussian delta function $\Gamma_{a}(\omega,\omega')=\frac{\eta}{\delta_{a}\sqrt{\pi}}e^{-[(\omega-\omega')/\delta_{a}]^{2}}$
where $\eta=10^{-4}\Gamma$ and $\delta_{a}=10^{-2}\Gamma$. The
dot is assumed to be initially empty, and the coupling to the thermally
equilibrated leads and phonon bath is turned on at time $t=0$. The
auxiliary spectral function exhibits some transient behavior, and
approaches the physical steady state spectral function at sufficiently
long time, as discussed in Ref.~\onlinecite{Cohen2014}.

\subsection{Symmetric Model}

We first consider the system which includes the counter term, $\delta=1$.
For this case, the electron--phonon coupling does not break particle--hole
symmetry and the spectral function remains symmetric.

\subsubsection{Transient dynamics}

\begin{figure}
\includegraphics[bb=0bp 14bp 243bp 90bp,clip]{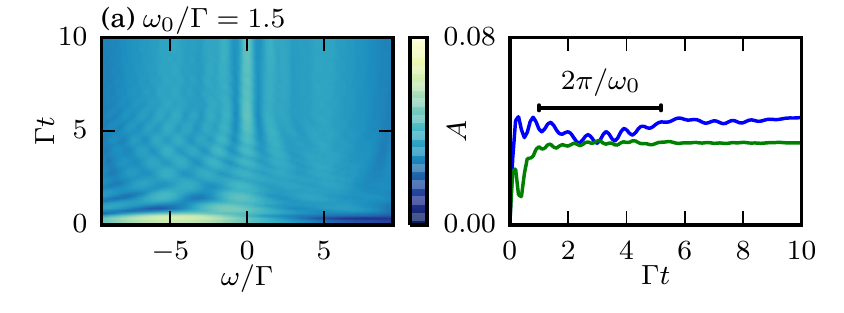}

\begin{spacing}{0.90000000000000002}
\includegraphics[bb=0bp 14bp 243bp 90bp,clip]{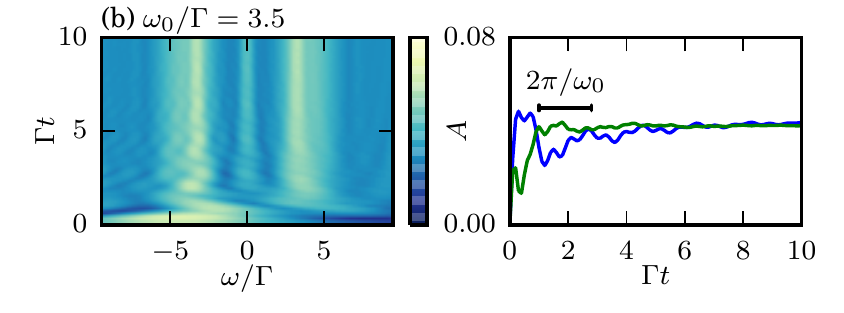}

\includegraphics[bb=0bp 0bp 252bp 88bp,clip]{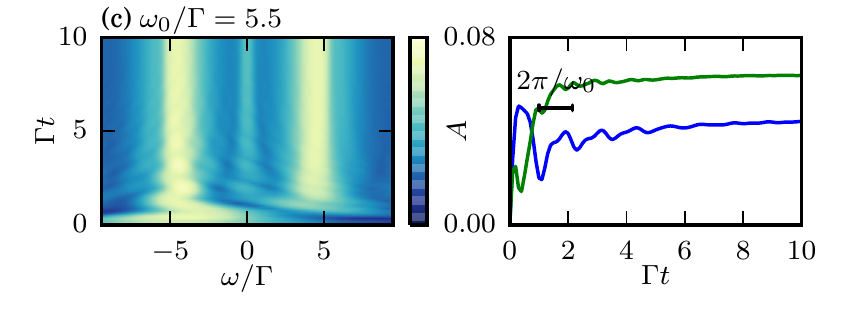}
\end{spacing}

\caption{(left panels) The time evolution of the spectral function $A(\omega;t)$
within the \textbf{bare NCA }is shown for different phonon frequencies.
(right panels) Time dependence of cuts at $\omega=0$ (blue) and $\omega=U/2$
(green). The time scale $2\pi/\omega_{0}$ related to the phonon frequency
is also plotted for comparison. A symmetric dot with $U=-2\epsilon=10\Gamma$
is considered at equilibrium $V=0$. The phonon coupling is set to
$\lambda=1.5\Gamma$ and the counter term is symmetric ($\delta=1$).
The inverse temperature of all baths is $\beta=10/\Gamma$.\label{fig:spectral_bare-time-evolution_symm}}

\includegraphics[bb=0bp 14bp 243bp 90bp,clip]{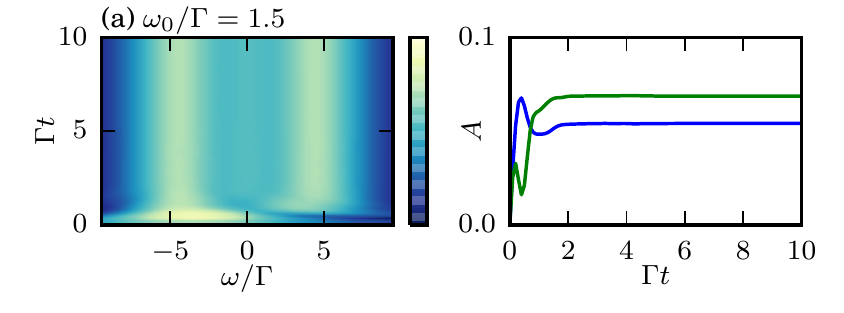}

\includegraphics[bb=0bp 14bp 243bp 90bp,clip]{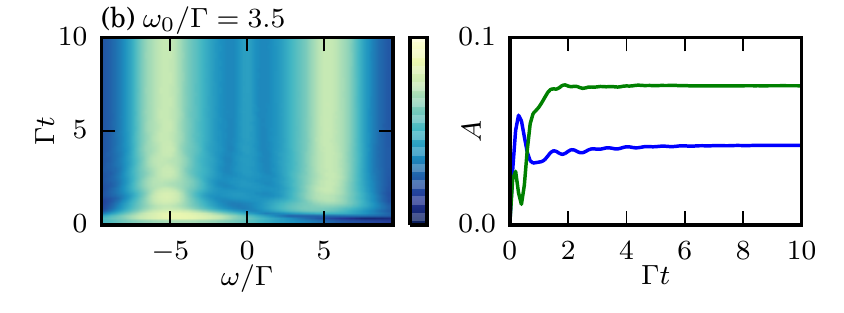}

\includegraphics[bb=0bp 15bp 252bp 94bp]{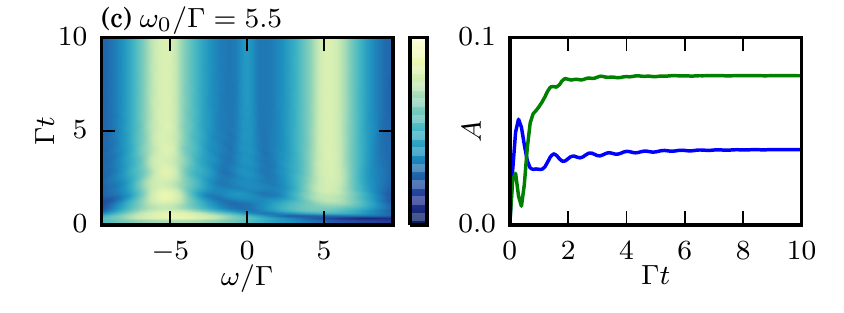}

\caption{The same as Fig.~\ref{fig:spectral_bare-time-evolution_symm} within
the \textbf{dressed NCA}. A symmetric dot with $U=-2\epsilon=10\Gamma$
is considered at equilibrium $V=0$. The phonon coupling is symmetric
with $\lambda=1.5\Gamma$ and the inverse temperature of all baths
is $\beta=10/\Gamma$.\label{fig:spectral_dress-time-evolution_symm}}
\end{figure}

The left panels of Fig.~\ref{fig:spectral_bare-time-evolution_symm}
and Fig.~\ref{fig:spectral_dress-time-evolution_symm} show the transient
evolution of the spectral function $A(\omega;t)$. The corresponding
right panels display single frequency cuts through this data, highlighting
the time evolution of the central peak ($\omega=0$) and the charge
transfer (CT) peak ($\omega/U=0.5$). We observe an overshooting of
the spectral function at short time due to the instantaneous coupling
between the dot and the leads. The bare NCA results (Fig.~\ref{fig:spectral_bare-time-evolution_symm})
exhibit oscillatory behavior in the amplitude of the central peak.
We observe that this is composed of a slower oscillation with a period
of $2\pi/\omega_{0}$, which is associated with the phonon frequency;
and a rapid oscillation with a period of $2\pi/U$, which comes from
the static energetics of the system. However, in the dressed NCA results
(Fig.~\ref{fig:spectral_dress-time-evolution_symm}), oscillatory
behavior consistent with the phonon frequency is not apparent. The
oscillatory behavior predicted by the bare NCA is consistent with
predictions made for the Anderson--Holstein model in the spinless
$U=0$\cite{Klatt2015} and $U=\infty$ cases,\cite{Goker2011} where
the local density of states at $\omega=0$ approaches the steady state
in an oscillating manner with the periodicity of the phonon mode.
Here, the time-evolution of the entire frequency dependent auxiliary
spectral function additionally reveals the transient effect of electron--phonon
coupling on the charge transfer peaks.

At long times, the bare NCA exhibits a strong suppression of the CT
peaks when the phonon frequency is small. However, this suppression
of the CT peaks is not nearly as evident in the dressed NCA results.
Conversely, the dressed NCA shows a strong enhancement of the central
peak at low phonon frequencies, which is not present in the bare NCA
results.

\subsubsection{Equilibrium steady state spectral function}

We next explore the equilibrium spectral function $A\left(\omega\right)$
of the system in the limit of long times, where the system has reached
its steady or equilibrium state. We consider two types of cuts through
the parameter space: the first is the dependence on the phonon frequency
$\omega_{0}$ at constant dot-phonon coupling strength $\lambda$,
and the second is the $\lambda$ dependence at constant $\omega_{0}$.
Here, too, the bare and dressed NCA predict qualitatively different
behaviors.

In Fig.~\ref{fig:spectral_omega_symm}, $A\left(\omega\right)$
is shown for a range of phonon frequencies at intermediate electron--phonon
coupling $\lambda=1.5$. Within bare NCA, shown in panel (a), a set
of features at $\omega=\pm n\omega_{0}$ with $n\in\left\{ 1,2,3\right\} $
is visible at low frequencies. These features, corresponding to Kondo
replicas or sidebands \cite{Hewson2002,Paaske2005b,Jovchev2013,Laakso2014,SeoaneSouto2014,Albrecht2015},
appear as a sequence of positive peaks at $\omega=\pm\left(2n+1\right)\omega_{0}$
and negative peaks at $\omega=\pm2n\omega_{0}$, and are related to
interference effects. In the literature, the Anderson--Holstein impurity
model is mostly assumed to be spinless ($U=0$), and one observes
multiple positive side bands due to a resonance with the phonon. For
a generic Anderson--Holstein model, negative peaks have previously
been predicted in the $T\sim0$ regime by perturbation theory, but
not are exhibited within numerical renormalization group calculation.\cite{Hewson2002,Laakso2014}
However, our calculation shows both positive and negative side peaks
exist at a finite temperature for generic Anderson--Holstein model.
 In the high-frequency regime, the Kondo replicas die out and the
CT peaks appear. The CT peaks are suppressed by coupling to a low
frequency phonon mode, which implies that phonon-induced tunneling
dominates the single particle excitation spectrum in this regime.

Replica-like features can also be observed at $\omega=\pm\omega_{0}$
in the dressed NCA, which is plotted in Fig.~\ref{fig:spectral_omega_symm}~(b).
However, these side peaks are substantially weaker than those observed
in the bare NCA calculation. In the dressed NCA the CT peaks are shifted
by the reorganization energy, such that their central frequencies
are located at $\omega_{\pm}=\pm\left(\epsilon+\frac{\lambda^{2}}{\omega_{0}}\right)$
(as illustrated by the dashed line). A significant enhancement in
$A\left(\omega\right)$ occurs when the two renormalized CT peaks
cross each other. In the low frequency regime $\omega_{0}\leq\frac{\lambda^{2}}{|\epsilon|}$,
the two CT peaks merge and form a wide central peak which is clearly
unrelated to the Kondo effect. The Kondo peak only develops in the
high frequency regime, and in general it is strongly suppressed for
a wide range of parameters.

The $\omega_{0}$ dependence of the central peak $A\left(\omega=0\right)$
exhibits consistent behavior for the two flavors of NCA only at high
frequencies (Fig.~\ref{fig:spectral_omega_symm}~(c)). At low frequencies,
both approximations exhibit enhancement of the central peak, but the
context and perhaps the mechanism of the enhancement is different
between the two cases. In the bare NCA, the amplitude of the Kondo
peak is enhanced as $\omega_{0}$ decreases because the replicas of
the Kondo peak merge when the phonon quasi-states become nearly-degenerate
as $\omega_{0}$ decreases. In the dressed NCA, on the other hand,
the enhancement is maximal where the two CT peaks merge at $\omega_{0}^{*}=\lambda^{2}/\epsilon$.
The contrast with the bare case is even more notable when one considers
that in the bare NCA the CT peaks are almost entirely suppressed at
low frequencies.

\begin{center}
\begin{figure}
\begin{raggedright}
(a)\vspace{-16bp}

\par\end{raggedright}

\begin{raggedleft}
\includegraphics[bb=10bp 10bp 238bp 170bp,clip]{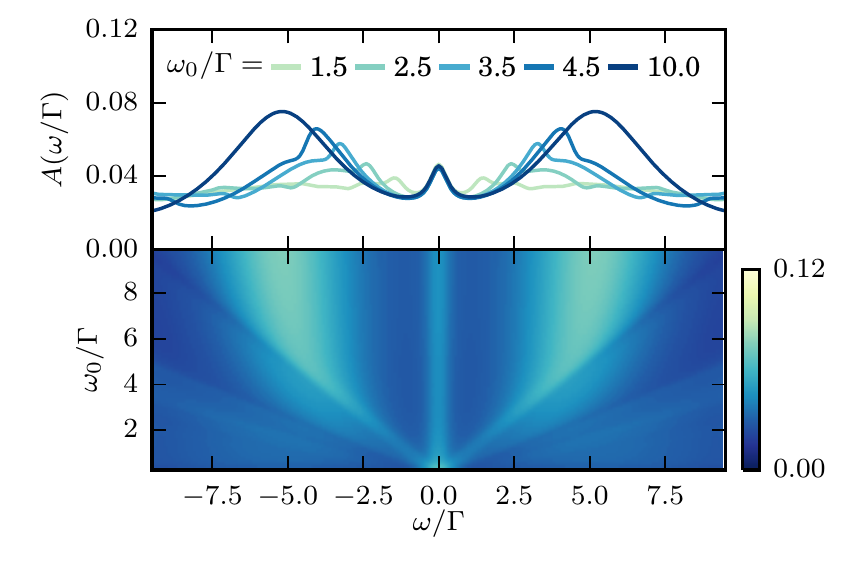}
\par\end{raggedleft}

\begin{raggedright}
(b)\vspace{-16bp}

\par\end{raggedright}

\begin{raggedleft}
\includegraphics[bb=10bp 10bp 238bp 170bp,clip]{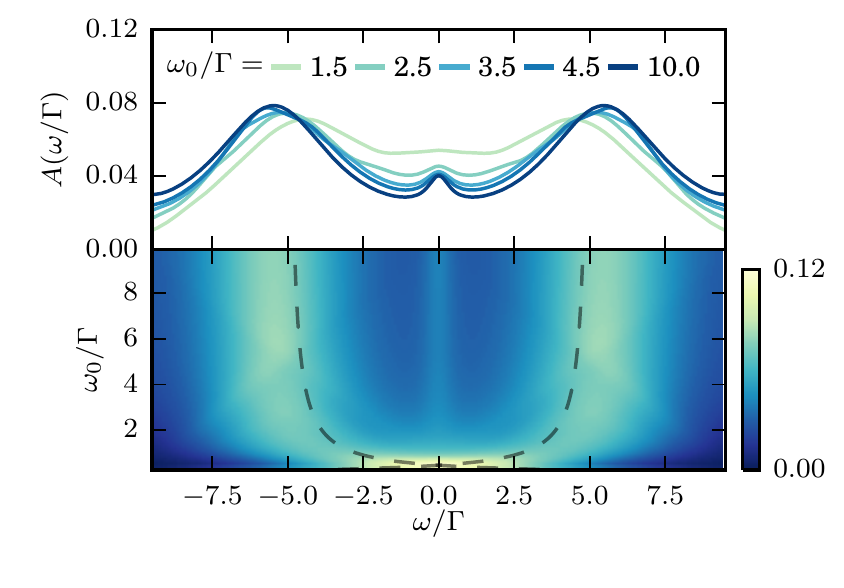}
\par\end{raggedleft}

\begin{raggedright}
(c)\vspace{-24bp}

\par\end{raggedright}

\begin{centering}
\includegraphics[bb=10bp 0bp 220bp 140bp,clip]{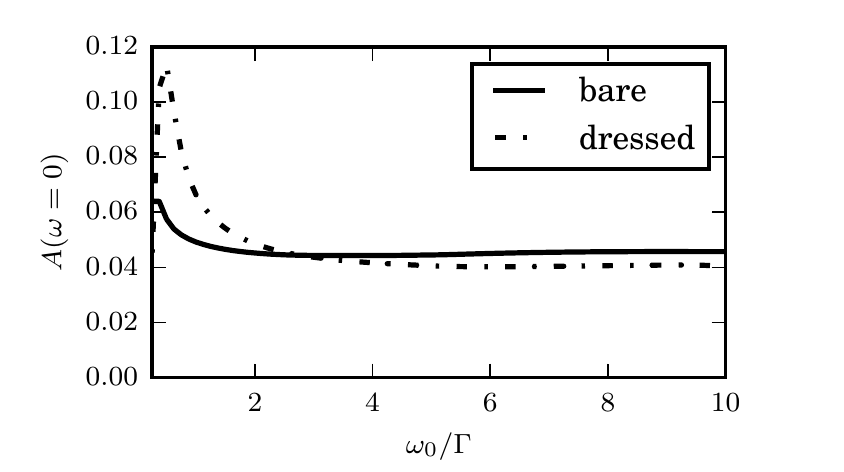}
\par\end{centering}

\caption{The $\omega_{0}$ dependence of the spectral function $A(\omega)$
is calculated by (a)\textbf{ bare NCA} and (b) \textbf{dressed NCA}
for a symmetric dot at equilibrium $V=0$ with $U=-2\epsilon=10\Gamma$.
The phonon coupling is $\lambda=1.5\Gamma$ and the counter term is
symmetric ($\delta=1$). All baths at the same inverse temperature
$\beta=10/\Gamma$. The dashed lines indicate the renormalized charge
transfer peak at $\omega_{\pm}=\pm\left(\epsilon+\frac{\lambda^{2}}{\omega_{0}}\right)$.
The $\omega_{0}$-dependence of the central peak at $\omega=0$ is
plotted in (c).\label{fig:spectral_omega_symm}}
\end{figure}

\par\end{center}

\begin{figure}
\begin{raggedright}
(a)\vspace{-16bp}

\par\end{raggedright}

\begin{raggedleft}
\includegraphics[bb=10bp 10bp 238bp 170bp,clip]{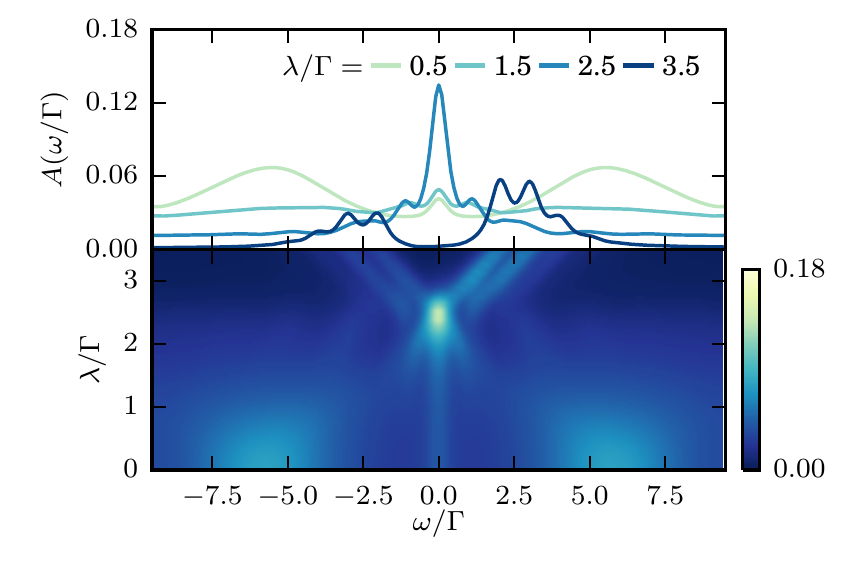}
\par\end{raggedleft}

\begin{raggedright}
(b)\vspace{-16bp}

\par\end{raggedright}

\begin{raggedleft}
\includegraphics[bb=10bp 10bp 238bp 170bp,clip]{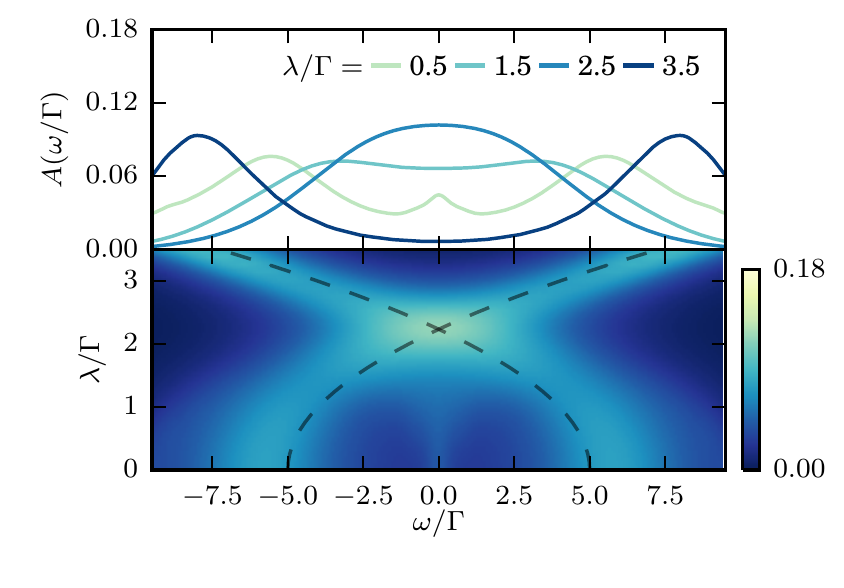}
\par\end{raggedleft}

\begin{raggedright}
(c)\vspace{-24bp}

\par\end{raggedright}

\begin{centering}
\includegraphics[bb=10bp 0bp 220bp 140bp,clip]{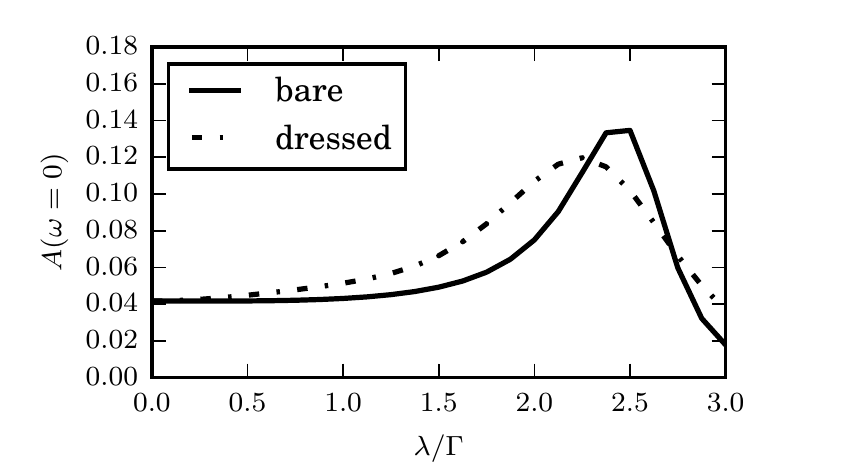}
\par\end{centering}

\caption{The $\lambda$ dependence of the spectral function $A(\omega)$ as
calculated within the (a)\textbf{ bare NCA} and (b) \textbf{dressed
NCA} for a symmetric dot with $U=-2\epsilon=10\Gamma$ at equilibrium
$V=0$. The phonon coupling is $\omega_{0}=1.0\Gamma$ and the counter
term is symmetric ($\delta=1$). All baths at the same inverse temperature
$\beta=10/\Gamma$. The dashed lines indicate the renormalized charge
transfer peak at $\omega_{\pm}=\pm\left(\epsilon+\frac{\lambda^{2}}{\omega_{0}}\right)$.
The $\lambda$-dependence of the central peak at $\omega=0$ is plotted
in (c).\label{fig:spectral_lambda_symm}}
\end{figure}

In Fig.~\ref{fig:spectral_lambda_symm} we repeat the previous analysis
in a different plane of the parameter space, by taking a cut at a
constant (low) phonon frequency $\omega_{0}$ and a range of $\lambda$
values. In the bare NCA (Fig~\ref{fig:spectral_lambda_symm}~(a)),
the CT peaks are suppressed as $\lambda$ increases. One can observe
a set of ridge-like features developing along with a strong enhancement
of the central Kondo peak. In the large $\lambda$ regime, the developed
side peaks shifted linearly with $\lambda$ with a spacing of approximately
$\omega_{0}$ between peaks in frequency. These features resemble
Kondo replicas,\cite{Hewson2002,Paaske2005b,SeoaneSouto2014,Albrecht2015,Laakso2014}
but a closer inspection reveals behavior more complicated than simply
side peaks generated at the phonon frequency $|\omega|=n\omega_{0}$.
A sharp Kondo peak is only apparent before the crossing point of the
ridges. It is significantly enhanced at the crossing point, and is
either completely suppressed or split beyond this point.

No Kondo replicas are observed within the dressed NCA(Fig.~\ref{fig:spectral_lambda_symm}~(b)).
The CT peaks are again renormalized, and appear centered at $\mbox{\ensuremath{\omega}}_{\pm}\approx\pm\left(\epsilon+\frac{\lambda^{2}}{\omega_{0}}\right)$
as illustrated by the dashed lines. The crossing at $\lambda^{*}=\sqrt{\epsilon\omega_{0}}$
leads to a strong enhancement near $\omega=0$. The Kondo peak is
only observable for $\lambda<\lambda^{*}$, and is widened beyond
the point where it can be distinguished from the CT bands before the
crossing point is reached. This widening effect is not observed in
the bare NCA. Past the crossing point, no central feature is visible,
in agreement with the bare NCA.

While the striking non-monotonic enhancement of the $\omega=0$ spectral
function is predicted by both approximations, it occurs at a different
value of $\lambda$ in each case (see Fig.~\ref{fig:spectral_lambda_symm}~(c)).
The peak in the dressed NCA occurs precisely at the value of $\lambda$
for which the effective, dressed $\tilde{U}$ change sign. In this
regard, the result is reminiscent of the NRG prediction of Hewson
and Meyer,\cite{Hewson2002} where the negative--$\tilde{U}$ Anderson--Holstein
model flows to the $U=0$ behavior. Within the bare NCA, the peak
value of $A(\omega=0)$ occurs for a slightly larger value of $\lambda$.
Here, the self--consistency of the perturbation theory presumably
captures, in an approximate manner, the terms leading to negative--$\tilde{U}$
behavior as well. Lastly, it should be mentioned that this non-monotonic
behavior is consistent with the prediction of Ref.~\onlinecite{Cornaglia2004}.
We return to this point later in the manuscript.

\subsubsection{Nonequilibrium steady state spectral function}

We now consider a nonequilibrium system driven by a bias voltage
$V=2\Gamma$. The $\omega_{0}$ dependence of $A\left(\omega\right)$
is plotted in Fig.~\ref{fig:spectral_omega_symm_noeq}. The voltage
splitting of Kondo peak\cite{Meir1993,Cohen2014a} can be observed
in both approximations. The central Kondo peak splits into two peaks
at $\omega=\pm V/2$ independently of the phonon frequency. Kondo
replicas are not clearly distinguishable, since the splitting smears
out the associated features. However, a set of linearly dependent
signatures remains visible.

\begin{center}
\begin{figure}
\begin{raggedright}
(a)\vspace{-16bp}

\par\end{raggedright}

\begin{raggedleft}
\includegraphics[bb=10bp 10bp 238bp 170bp,clip]{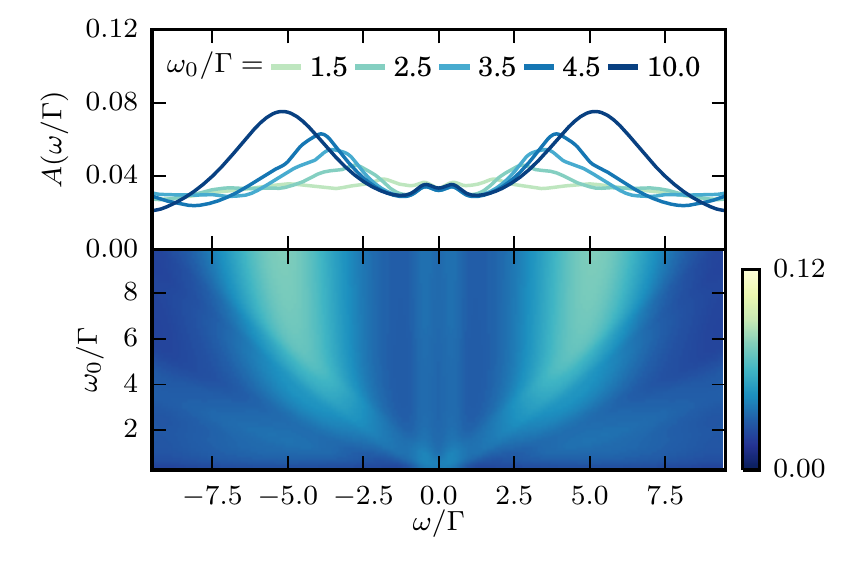}
\par\end{raggedleft}

\begin{raggedright}
(b)\vspace{-16bp}

\par\end{raggedright}

\begin{raggedleft}
\includegraphics[bb=10bp 10bp 238bp 170bp,clip]{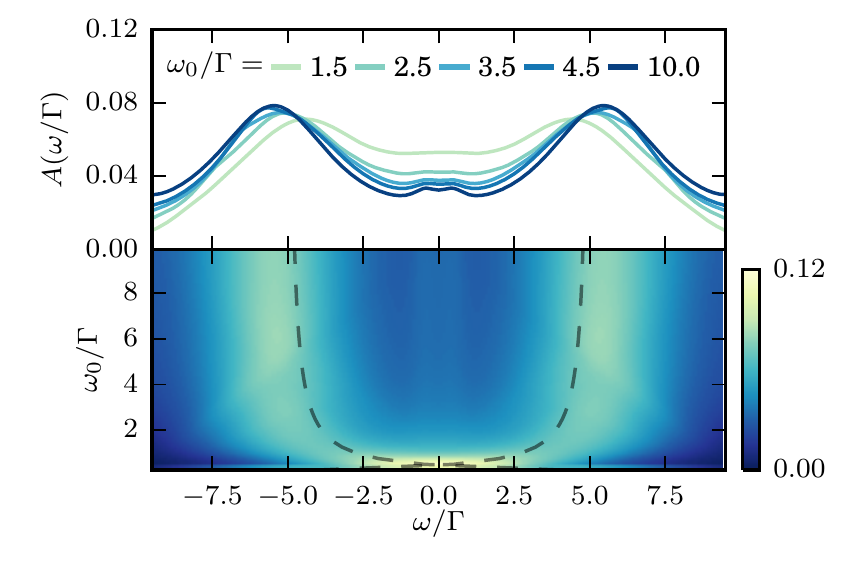}
\par\end{raggedleft}

\caption{The $\omega_{0}$-dependence of the spectral function $A(\omega)$
for a symmetric dot with $U=-2\epsilon=10\Gamma$ under a nonequilibrium
symmetrically applied bias voltage $V=2\Gamma$ within the (a)\textbf{
bare NCA} and (b) \textbf{dressed NCA}. The phonon coupling is $\lambda=1.5\Gamma$
and the counter term is symmetric ($\delta=1$). All baths at the
same inverse temperature $\beta=10/\Gamma$.\label{fig:spectral_omega_symm_noeq}}
\end{figure}

\par\end{center}

\subsection{Asymmetric Model}

In the following subsection, we consider an Anderson--Holstein model
without a counter term, i.e. $\delta=0$ in Eq.~(\ref{eq:electron_phonon_coupling_hamiltonian}).
While the isolated dot Hamiltonian is still assumed to remain particle-hole
symmetric, the electron--phonon coupling breaks the particle--hole
symmetry of the system and results in an asymmetric spectral function.
The two NCA formulations we employ take this asymmetry into account
in different ways, as pointed out in sec.~\ref{sub:NCA-for-asymmetric-model}.
In addition to the spectral function, we study the effects of the
symmetry breaking on transport properties. This is of particular interest,
because under a symmetrically applied bias the differential conductance
is a symmetric function of frequency even without particle--hole symmetry.
Additionally, one may not be able to observe the replicas directly
in a transport experiment, due to the nonequilibrium shifting or suppression
of the Kondo peak, which would also affect the replicas. We show that
an indirect experimental signal of the replica effect may remain.

\subsubsection{Transient dynamics}

\begin{figure}
\begin{raggedright}
\includegraphics[bb=0bp 14bp 243bp 90bp,clip]{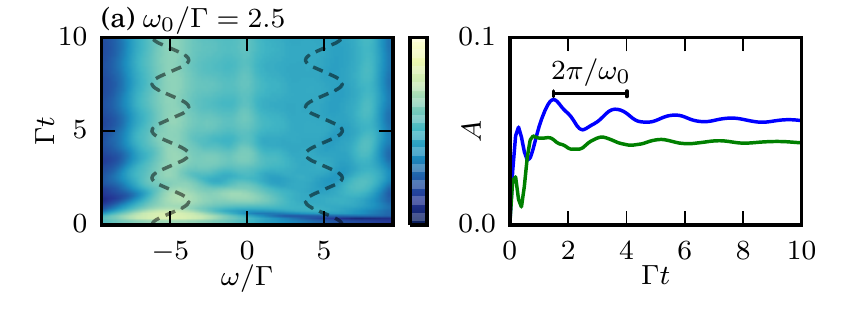}
\par\end{raggedright}

\begin{raggedright}
\includegraphics[bb=0bp 14bp 243bp 90bp,clip]{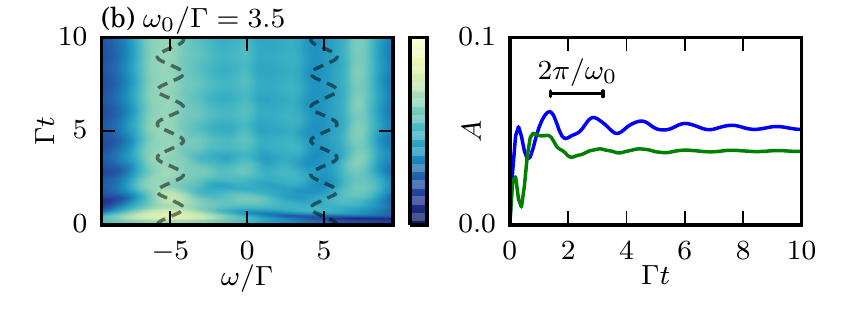}
\par\end{raggedright}

\begin{raggedright}
\includegraphics[bb=0bp 5bp 252bp 88bp,clip]{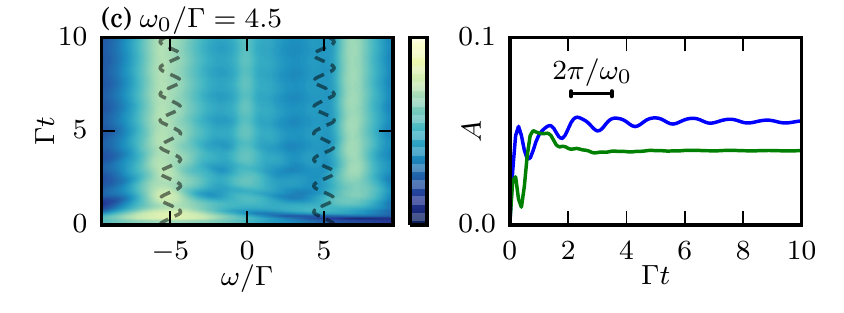}
\par\end{raggedright}

\caption{(left panels) The time evolution of the spectral function $A(\omega;t)$
within the \textbf{bare NCA }is shown for different phonon frequencies.
The frequency oscillations of the CT peaks along with an illustration
of the expected energy oscillations in the adiabatic limit (dash lines)
are also exhibited. (right panels) Time dependence of cuts at $\omega=0$
(blue) and $\omega=U/2$ (green). The time scale $2\pi/\omega_{0}$
related to the phonon frequency is also plotted for comparison. The
dot is symmetric with $U=-2\epsilon=10\Gamma$ at equilibrium $V=0$.
The phonon coupling is $\lambda=1.5\Gamma$ and the counter term is
asymmetric ($\delta=0$). The inverse temperature of all baths is
$\beta=10/\Gamma$.\label{fig:spectral_bare-time-evolution_asym}}

\begin{raggedright}
\includegraphics[bb=0bp 14bp 243bp 90bp,clip]{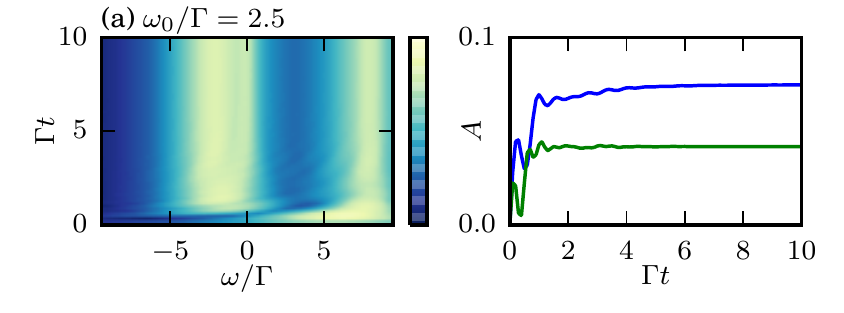}
\par\end{raggedright}

\begin{raggedright}
\includegraphics[bb=0bp 14bp 243bp 90bp,clip]{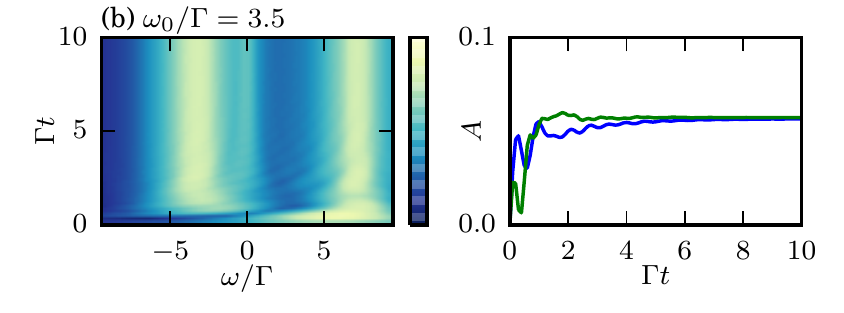}
\par\end{raggedright}

\begin{raggedright}
\includegraphics[bb=0bp 5bp 243bp 90bp,clip]{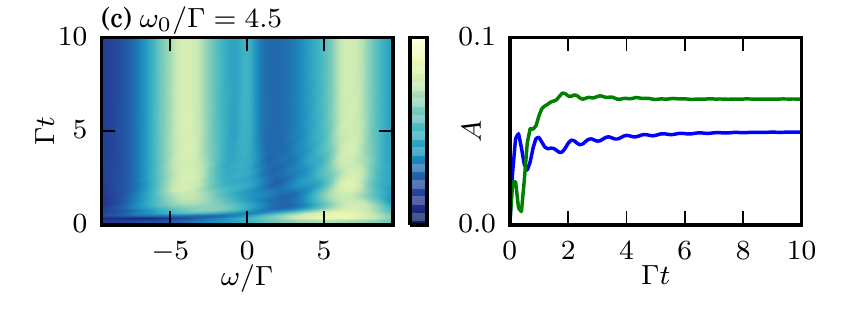}
\par\end{raggedright}

\caption{ The same as Fig.~\ref{fig:spectral_bare-time-evolution_asym} within
the \textbf{dressed NCA}. The dot is symmetric with $U=-2\epsilon=10\Gamma$
at equilibrium $V=0$. The phonon coupling is asymmetric ($\delta=0$)
with $\lambda=1.5\Gamma$ and the inverse temperature is $\beta=10/\Gamma$.\label{fig:spectral_dress-time-evolution_asym}}
\end{figure}

Within the bare NCA, the CT peaks and Kondo peak oscillate at the
phonon frequency $\omega_{0}$, but the oscillations are manifested
in different ways (Fig.~\ref{fig:spectral_bare-time-evolution_asym},
left panels). In particular, the CT peaks oscillate in \emph{frequency},
while the Kondo peak oscillates in \emph{amplitude}. At short times
and in the adiabatic limit, the CT peak oscillations can be explained
by oscillating energy levels ($\tilde{\epsilon}_{\sigma}=\epsilon_{\sigma}+\frac{2\lambda}{\omega_{0}}\sin(\omega_{0}t+\phi_{0})$)
with some unknown initial phase. This is illustrated by the black
dashed lines in the left panels of Fig.~\ref{fig:spectral_bare-time-evolution_asym}.
All these features are washed out in the dressed NCA.

\subsubsection{Steady state spectral function}

To explore the effects of phonons on the equilibrium spectral function,
we once again plot first the $\omega_{0}$ dependence at constant
$\lambda$, and then the $\lambda$ dependence at constant $\omega_{0}$.
Within the bare NCA, the Kondo replica features can clearly be seen
in Fig.~\ref{fig:spectral_omega_asym}~(a), but harder to distinguish
in the cuts. They are mixed with a variety of other effect including
the low-frequency smearing of the Kondo resonance and the suppression
of the positive CT peak. The replica effect and the above-mentioned
CT suppression are both stronger at positive frequencies. At small
phonon frequencies, the Kondo resonance merges with the negative CT
peak.

At the intermediate phonon frequency $\omega_{0}=|\epsilon_{\sigma}-U|$
where the replicas are aligned with the CT peaks, a non-monotonic
enhancement of the central peak is evident, and is especially strong
at large $\lambda.$ This can be seen more clearly in the cut shown
in Fig.~\ref{fig:conductance_omega_asym}~(c). We believe this is
due to a phonon-assisted process which is similar to the Kondo spin-flip
process, and which becomes possible for electrons with energies closed
to the chemical potential.\cite{Hewson1993,Cornaglia2004} The effects
described here are largely washed out in the dressed NCA. 

\begin{center}
\begin{figure}
\begin{raggedright}

\par\end{raggedright}

\begin{raggedright}
(a)\vspace{-16bp}

\par\end{raggedright}

\begin{raggedleft}
\includegraphics[bb=10bp 10bp 238bp 170bp,clip]{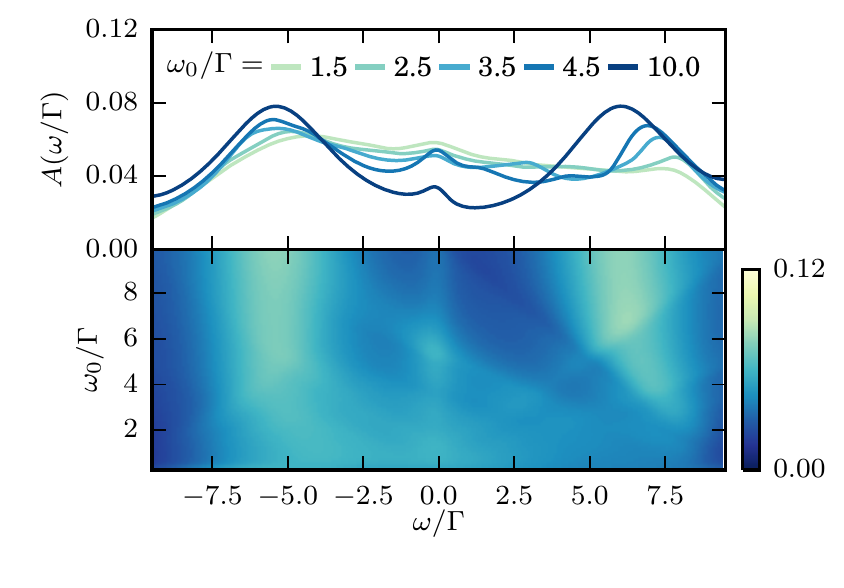}
\par\end{raggedleft}

\begin{raggedright}
(b)\vspace{-16bp}

\par\end{raggedright}

\begin{raggedleft}
\includegraphics[bb=10bp 10bp 238bp 170bp,clip]{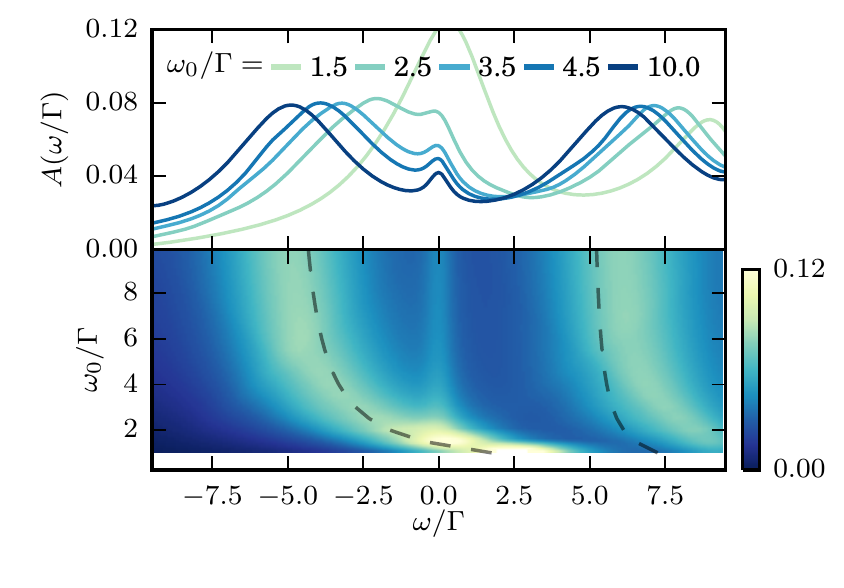}
\par\end{raggedleft}

\caption{The $\omega_{0}$-dependence of the spectral function $A(\omega)$
for a dot in equilibrium as calculated within the (a)\textbf{ bare
NCA} and (b) \textbf{dressed NCA}. The electron--phonon coupling is
asymmetric ($\delta=0$) and the coupling strength is $\lambda=1.5\Gamma$.
The dot is symmetric with $U=-2\epsilon=10\Gamma$. All baths at the
same inverse temperature $\beta=10/\Gamma$.\label{fig:spectral_omega_asym}}
\end{figure}

\par\end{center}

We continue to investigate the $\lambda$ dependence at constant $\omega_{0}$.
Here, we plot the results for both approximations at a relatively
large $\omega_{0}$ (Fig.~\ref{fig:spectral_lambda_asym}). The bare
NCA (panel (a)) shows a suppression of the charge transfer bands and
a widening of the Kondo peak. The dressed NCA (panel (b)) shows an
asymmetric shift of the CT peaks to approximately $\omega_{+}=\frac{U}{2}+\frac{\lambda^{2}}{\omega_{0}}$
and $\omega_{-}=-\frac{U}{2}+3\frac{\lambda^{2}}{\omega_{0}}$, as
might be expected in the anti-adiabatic limit. Some deviation from
this occurs, especially for the positive CT band. More interestingly,
as the CT peak merges with the Kondo peak at $\lambda=\sqrt{\frac{U\omega_{0}}{6}}$,
a strong enhancement occurs. This enhancement is not observed in the
bare NCA.

\begin{center}
\begin{figure}
\begin{raggedright}
(a)\vspace{-16bp}

\par\end{raggedright}

\begin{raggedleft}
\includegraphics[bb=10bp 10bp 238bp 170bp,clip]{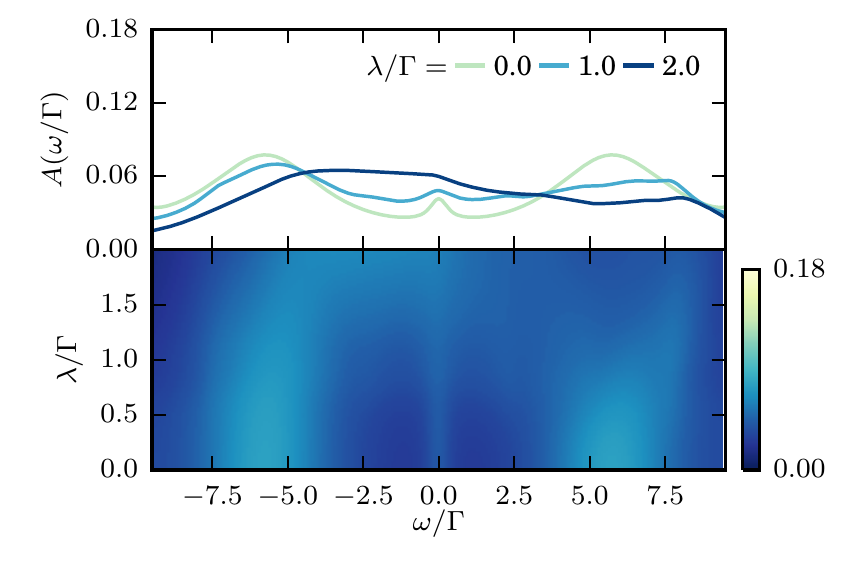}
\par\end{raggedleft}

\begin{raggedright}
(b)\vspace{-16bp}

\par\end{raggedright}

\begin{raggedleft}
\includegraphics[bb=10bp 10bp 238bp 170bp,clip]{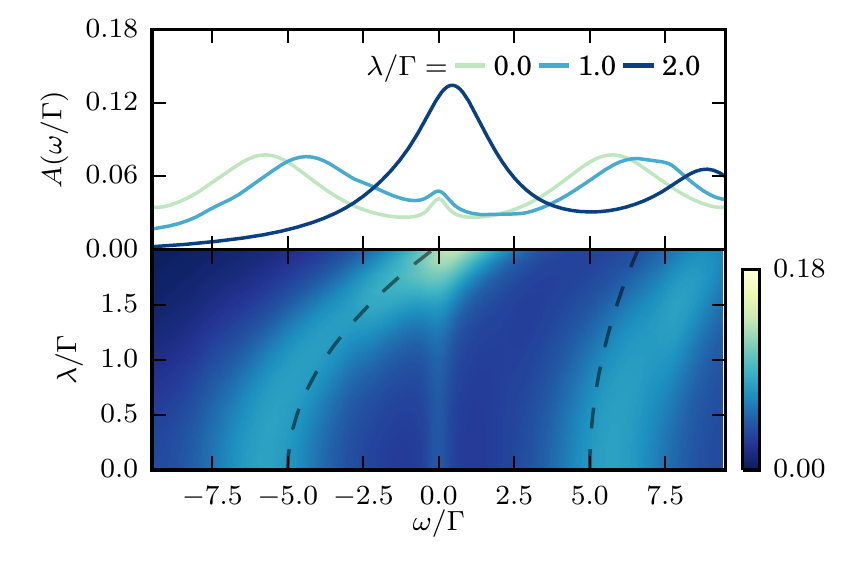}
\par\end{raggedleft}

\caption{The $\lambda$-dependence of the spectral function $A(\omega)$ as
calculated within the (a)\textbf{ bare NCA} and the (b) \textbf{dressed
NCA} for an equilibrium symmetric dot with $U=-2\epsilon=10\Gamma$.
The phonon frequency is $\omega_{0}/\Gamma=2.5$. The dashed lines
indicate the center of the CT peaks as estimated by the energy renormalization
at the anti-adiabatic limit $\omega_{CT}^{+}/\Gamma=-\epsilon+\frac{\lambda^{2}}{\omega_{0}}$
and $\omega_{CT}^{-}/\Gamma=\epsilon+3\frac{\lambda^{2}}{\omega_{0}}$.
All baths at the same inverse temperature $\beta=10/\Gamma$.\label{fig:spectral_lambda_asym}}
\end{figure}

\par\end{center}

\subsubsection{Steady state conductance }

\begin{figure}
\begin{raggedright}
(a) $\lambda/\Gamma=1$
\par\end{raggedright}

\begin{raggedleft}
\textbf{\includegraphics[bb=10bp 10bp 238bp 170bp,clip]{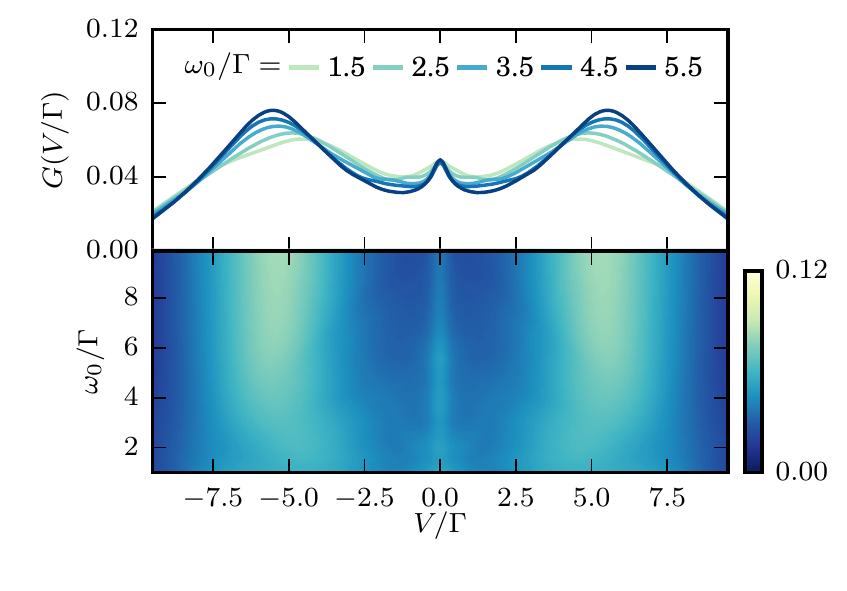}}
\par\end{raggedleft}

\begin{raggedright}
(b) $\lambda/\Gamma=2$
\par\end{raggedright}

\begin{raggedleft}
\includegraphics[bb=10bp 10bp 238bp 170bp,clip]{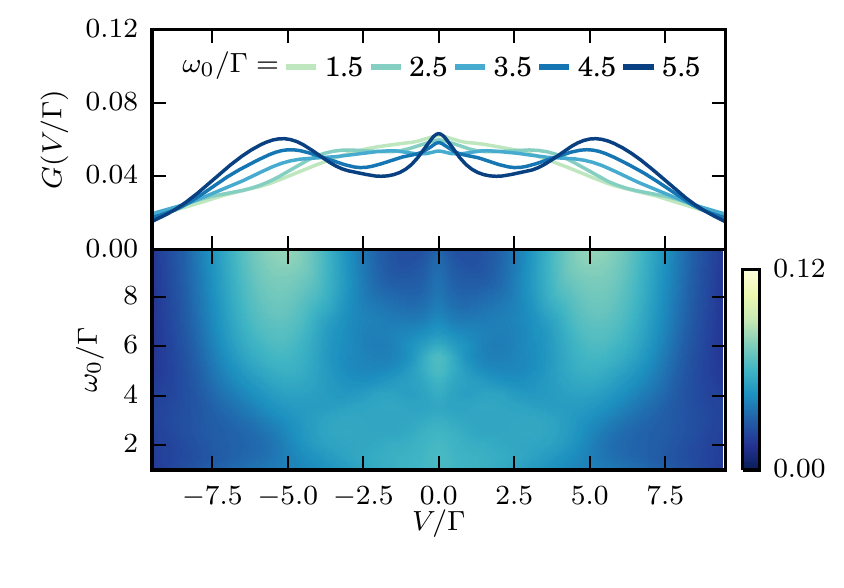}
\par\end{raggedleft}

\begin{raggedright}
(c)\vspace{-24bp}

\par\end{raggedright}

\begin{centering}
\includegraphics[bb=0bp 0bp 252bp 145bp,clip]{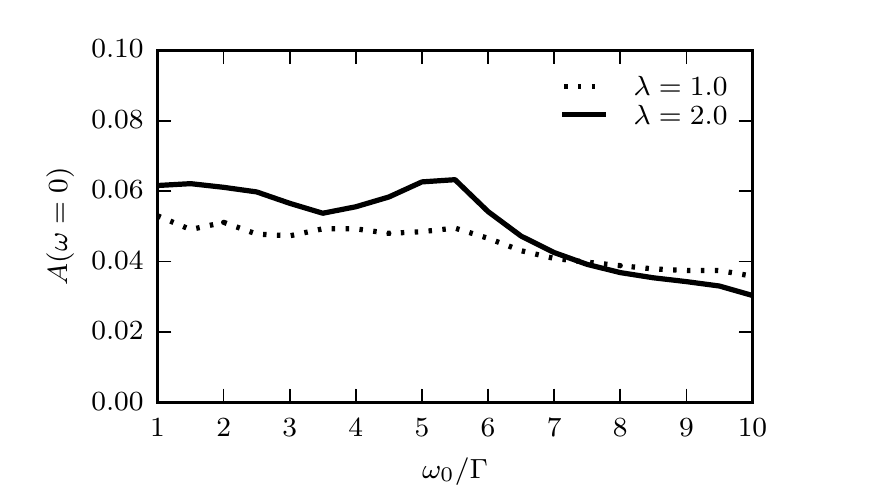}
\par\end{centering}

\caption{The conductance $G(V)$ as calculated within the\textbf{ bare NCA}
for different electron--phonon coupling (a)$\lambda/\Gamma=1$ and
(b)$\lambda/\Gamma=2$ with a symmetrically applied bias $\mu_{L}=\mu_{R}=V$.
The dot is also symmetric with $U=-2\epsilon=10\Gamma$. Panel (c)
shows the $\omega_{0}$-dependence of the central peak at $\omega=0$.
All baths at the same inverse temperature $\beta=10/\Gamma$. \label{fig:conductance_omega_asym}}
\end{figure}

Despite the symmetry breaking of the spectral function, the differential
conductance $G(V)\equiv\frac{dI}{dV}(V)$ under a symmetrically applied
bias ($\mu_{L}=-\mu_{R}=V/2$) remains a symmetric function of frequency
even without the counter term. The replica effect and the non-monotonic
enhancement, as visible in, \emph{e.g.,} Fig.~\ref{fig:spectral_omega_asym},
appears in the spectral function, which could in principle be accessible
in spectroscopic experiments. However, spectroscopic studies of single
molecules in junctions and mesoscopic quantum dots are difficult to
perform, and transport experiments are far more common. It is interesting
to consider whether these effects are observable in the differential
conductance as well as the spectral function; outside of linear response
these quantities may differ qualitatively.\cite{Cohen2014a} Fig.~\ref{fig:conductance_omega_asym}
shows the differential conductance as it varies under the effect of
the phonon frequency $\omega_{0}$ at two different phonon coupling
strengths $\lambda$. The non-monotonic enhancement remains clearly
visible, while the side peaks are substantially weaker than their
counterparts in the spectral function. The bare NCA therefore predicts
that the non-monotonicity could be observed in transport experiments.
Since it is related to the side bands merging with the charge transfer
bands, an experimental observation of it could also be considered
an indirect confirmation of the replica effect. We note that the dressed
NCA also predicts a non-monotonicity, but one which does not appear
related to the replica effect. It will take a more sophisticated theoretical
treatment to determine whether this effect is real or an artifact
of the two NCA approaches, and to understand more deeply the mechanism
that lies behind it. 

In Ref.~\onlinecite{Cornaglia2004}, a non-monotonic effective Kondo
temperature and zero-bias conductance has been predicted in the Anderson--Holstein
model via the consideration of two limiting cases. In particular,
for weak electron--phonon coupling $2\lambda^{2}/\omega_{0}\ll U$,
the low--energy excitations of the Anderson--Holstein model can be
approximated by an isotropic Kondo Hamiltonian with the coupling to
phonons leading to an increase in the effective Kondo temperature.
On the other hand, for strong electron--phonon coupling $2\lambda^{2}/\omega_{0}\gg U$,
the low-energy excitations can be approximated by an anisotropic Kondo
Hamiltonian in which the effective Kondo temperature decreases with
increasing $\lambda$. This crossover behavior is observed in both
NCAs, though the implied maximum in the spectral function occurs at
a different $\lambda$ (see also Fig.~\ref{fig:spectral_lambda_symm}c).
Interestingly, when examining the spectral function at all energies
simultaneously, a set of higher energy features which appear to be
shifted replicas of the maximum is also revealed.

\section{Conclusions\label{sec:Conclusions}}

In this paper we formulate and compare two distinct non-crossing approximations
for the study of the Anderson--Holstein model. The first approximation,
which we call the bare NCA, is a self-consistent resummation based
on a self energy which contains the electron--phonon coupling and
hybridization with the leads to lowest order. Within the second approximation,
which we term the dressed NCA, a Lang--Firsov transformation is first
applied, and the resulting transformed set of interactions are then
included in a self-consistent, lowest order self energy. We focus
on the predictions of both approximations with regard to transient
dynamics as well as the non-equilibrium steady state behavior of the
spectral function. In general, it should be expected that any flavor
of NCA will be inaccurate for low--frequency properties. For example,
NCA predicts a broadened and suppressed Kondo resonance when compared
with exact numerics.\cite{Cohen2014} Due to the paucity of exact
and global information related to the dynamical properties of the
model, a detailed assessment of the success and failure of the respective
methods is not possible even for higher frequency features. On the
other hand, we believe it is plausible to favor the bare NCA when
the electron--phonon coupling is weak, the dressed NCA when it is
strong, and both approaches when they produce consistent results in
the intermediate coupling regime. Since the two approximation are
based on disparate limits of the electron--phonon portion of the problem,
we focus on the intermediate coupling regime in an attempt to assess
the validity of the two approximations.

We find that several features appear to be robust within both flavors
of NCA. First, the Kondo peak is enhanced in particular regimes, but
is universally suppressed in the large electron--phonon coupling regime.
Second, low energy tunneling occurs and charge transfer peaks are
suppressed when phonon frequency is small compared to other relevant
energy scales. Lastly, the voltage splitting of the Kondo peak robustly
occurs in the non-equilibrium regime. We expect these features to
be real and experimentally reproducible behaviors in the Anderson--Holstein
model.

Conversely, several striking dynamical properties appear only within
one type of NCA approximation. In particular, the oscillatory transient
behavior exhibited in Fig.~\ref{fig:spectral_bare-time-evolution_asym}
and the replication of the Kondo peak is only observed within the
bare NCA, while polaronic shifts of the charge transfer peaks occur
only in the dressed NCA approximation. It is important to note that
these observations do not necessarily imply that such behaviors are
artifacts. In particular, since the bare NCA is expected to capture
accurately the weak electron--phonon situation, it is plausible that
the features revealed in Fig.~\ref{fig:spectral_lambda_symm}~and~\ref{fig:spectral_bare-time-evolution_asym}
are real properties of the model in this regime. The dressed NCA may
not predict this behavior due to the fact that several low order diagrams
associated with the interplay between hybridization and electron--phonon
coupling are absent. On the other hand, polaronic effects may only
be captured within the dressed NCA, and thus strong coupling shifts
of the charge transfer peaks should be expected once the coupling
to phonons is sizable.

Perhaps the most important aspect of the work presented here is that
it lays the foundation for exact real-time QMC approaches based on
expansion around the NCA approximation. These ``bold-line'' approaches
have been successful in the treatment of the simpler Anderson model,
and have enabled the simulation of relatively long real time information
before the dynamical sign problem becomes problematic. Convergence
of these approaches depends crucially on having a reasonably accurate
partial summation of diagrams from the outset. With respect to the
work presented here, we expect that the bare and dressed NCA approximations
should provide a good starting point in the weak and strong electron--phonon
coupling regimes, respectively. In addition to validating or falsifying
the predictions made by the individual NCA approximations of this
paper, real-time QMC approaches that make use of the bare and dressed
NCA techniques should allow for the exact simulation of the Anderson--Holstein
model in regimes that are currently inaccessible.

\begin{acknowledgments}
We would like to thank Philipp Werner and Andrés Montoya-Castillo
for helpful comments and discussions. DRR acknowledges support from
NSF CHE--146802. AJM acknowledges support from NSF DMR--1308236.
\end{acknowledgments}

\appendix

\section{Comparison with DMFT-based Monte Carlo results}

The top panel of Fig.~3 of Ref.~\onlinecite{Werner2010a} illustrates
the behavior of the spectral function of an Anderson--Holstein problem
computed via analytical continuation of \textit{exact} imaginary-time
quantum Monte Carlo as a function of increasing electron--phonon coupling,
and is analogous to our Fig.~\ref{fig:spectral_lambda_symm}. While
it is difficult to make a direct comparison between these results
and the results presented in our work due to the fact that the previous
results were obtained self-consistently in the context of dynamical
mean field theory, we have computed the spectral function for the
same model and parameters within the NCA approaches outlined in this
paper. In this sense, the results of Fig.~16 represent a type of
non-iterated NCA impurity solution in the DMFT context. The electron--phonon
coupling parameters used in Fig.~3 of Ref.~\onlinecite{Werner2010a}
are sufficiently large to render the bare NCA unstable. On the other
hand, the dressed NCA is in qualitative agreement with the analytically
continued results.

Quantitatively, the dressed NCA produces peaks in positions similar
to those obtained by Monte Carlo for large $\lambda$, but the $\omega=0$
and low frequency peaks are broadened and suppressed when compared
to those of the analytically continued exact data. This broadening
and suppression appears to be a general feature of NCA.\cite{Cohen2014}
While the behavior of the gap closing feature can be observed in both
the NCA and the analytically continued Monte Carlo data, it is still
unclear to what degree the differences in the spectral functions are
due to the effects of analytical continuation and the self-consistency
of the DMFT calculation.

\begin{figure}
\begin{centering}
\includegraphics[bb=600bp 135bp 1200bp 540bp,clip,scale=0.21]{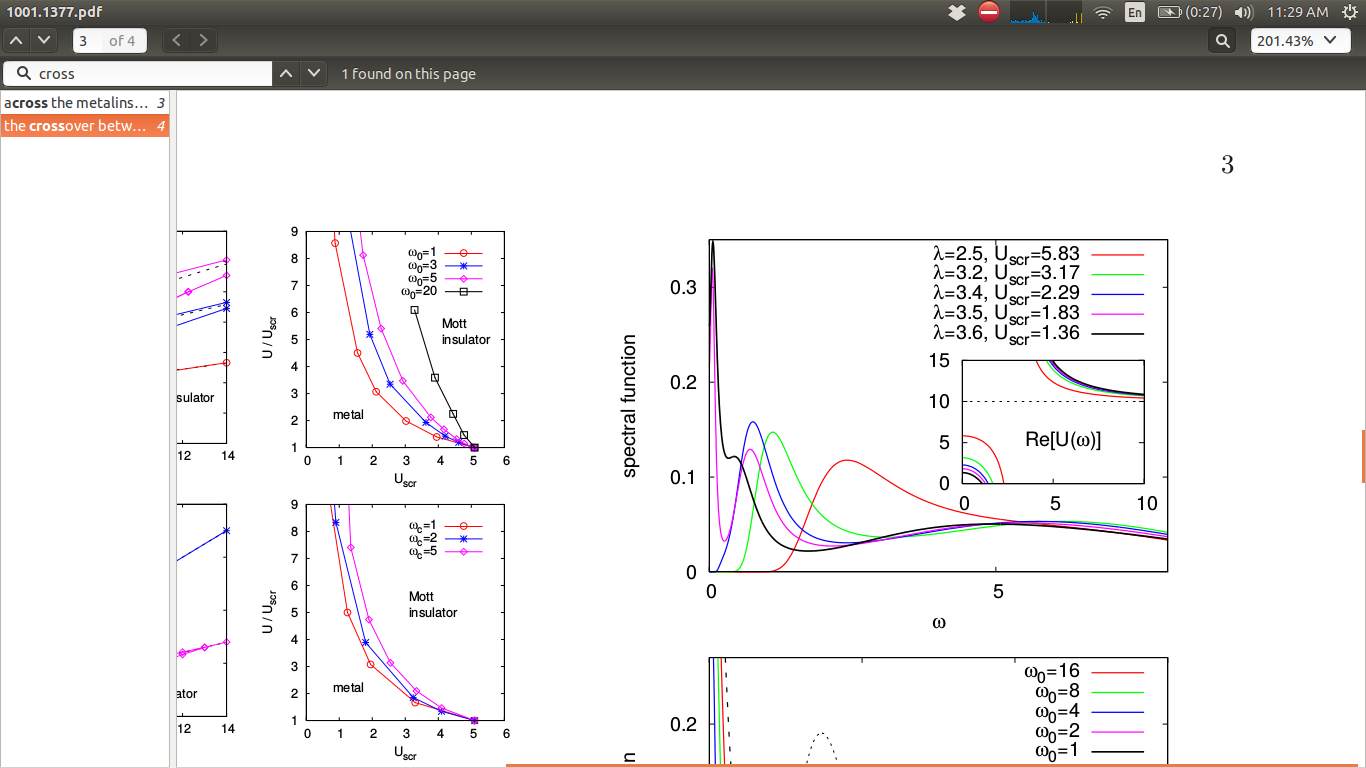}\includegraphics[scale=0.8]{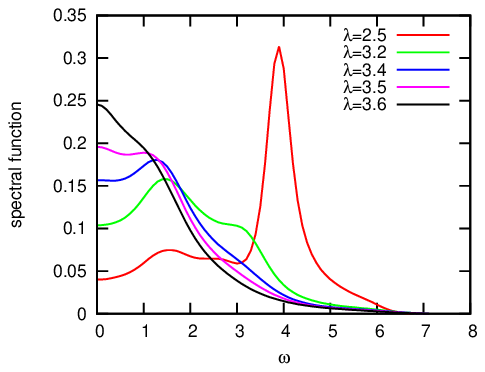}
\par\end{centering}

\caption{Left: Evolution of the spectral function across the metal--insulator
transition (gap closing) by increasing the phonon coupling. Right:
The spectral function $A(\omega)$ in the \textbf{strong} coupling
regime is calculated within the dressed NCA for a symmetric dot with
$U=-2\epsilon=10\Gamma$ at equilibrium $V=0$. The density of state
is of the semi-circular form $\Gamma\left(\omega\right)=\sqrt{4t^{2}-\omega^{2}}$
with $t=1$. The phonon coupling is $\omega_{0}=3.0\Gamma$ and the
counter term is symmetric ($\delta=1$). The baths are maintained
at a temperature $\beta\Gamma=50$. }
\end{figure}

\bibliographystyle{apsrev4-1}
\bibliography{NCA-Anderson-Holstein}

\begin{thebibliography}{100}%
\makeatletter
\providecommand \@ifxundefined [1]{%
 \@ifx{#1\undefined}
}%
\providecommand \@ifnum [1]{%
 \ifnum #1\expandafter \@firstoftwo
 \else \expandafter \@secondoftwo
 \fi
}%
\providecommand \@ifx [1]{%
 \ifx #1\expandafter \@firstoftwo
 \else \expandafter \@secondoftwo
 \fi
}%
\providecommand \natexlab [1]{#1}%
\providecommand \enquote  [1]{``#1''}%
\providecommand \bibnamefont  [1]{#1}%
\providecommand \bibfnamefont [1]{#1}%
\providecommand \citenamefont [1]{#1}%
\providecommand \href@noop [0]{\@secondoftwo}%
\providecommand \href [0]{\begingroup \@sanitize@url \@href}%
\providecommand \@href[1]{\@@startlink{#1}\@@href}%
\providecommand \@@href[1]{\endgroup#1\@@endlink}%
\providecommand \@sanitize@url [0]{\catcode `\\12\catcode `\$12\catcode
  `\&12\catcode `\#12\catcode `\^12\catcode `\_12\catcode `\%12\relax}%
\providecommand \@@startlink[1]{}%
\providecommand \@@endlink[0]{}%
\providecommand \url  [0]{\begingroup\@sanitize@url \@url }%
\providecommand \@url [1]{\endgroup\@href {#1}{\urlprefix }}%
\providecommand \urlprefix  [0]{URL }%
\providecommand \Eprint [0]{\href }%
\providecommand \doibase [0]{http://dx.doi.org/}%
\providecommand \selectlanguage [0]{\@gobble}%
\providecommand \bibinfo  [0]{\@secondoftwo}%
\providecommand \bibfield  [0]{\@secondoftwo}%
\providecommand \translation [1]{[#1]}%
\providecommand \BibitemOpen [0]{}%
\providecommand \bibitemStop [0]{}%
\providecommand \bibitemNoStop [0]{.\EOS\space}%
\providecommand \EOS [0]{\spacefactor3000\relax}%
\providecommand \BibitemShut  [1]{\csname bibitem#1\endcsname}%
\let\auto@bib@innerbib\@empty
\bibitem [{\citenamefont {Ashcroft}\ and\ \citenamefont
  {Mermin}(1976)}]{AshcroftandMermin1976}%
  \BibitemOpen
  \bibfield  {author} {\bibinfo {author} {\bibfnamefont {N.~W.}\ \bibnamefont
  {Ashcroft}}\ and\ \bibinfo {author} {\bibfnamefont {N.~D.}\ \bibnamefont
  {Mermin}},\ }\href@noop {} {\emph {\bibinfo {title} {{Solid State
  Physics}}}}\ (\bibinfo  {publisher} {Saunders College},\ \bibinfo {year}
  {1976})\BibitemShut {NoStop}%
\bibitem [{\citenamefont {Aradhya}\ and\ \citenamefont
  {Venkataraman}(2013)}]{Aradhya2013}%
  \BibitemOpen
  \bibfield  {author} {\bibinfo {author} {\bibfnamefont {S.~V.}\ \bibnamefont
  {Aradhya}}\ and\ \bibinfo {author} {\bibfnamefont {L.}~\bibnamefont
  {Venkataraman}},\ }\href {\doibase 10.1038/nnano.2013.91} {\bibfield
  {journal} {\bibinfo  {journal} {Nat. Nanotechnol.}\ }\textbf {\bibinfo
  {volume} {8}},\ \bibinfo {pages} {399} (\bibinfo {year} {2013})}\BibitemShut
  {NoStop}%
\bibitem [{\citenamefont {Nitzan}\ and\ \citenamefont
  {Ratner}(2003)}]{Nitzan2003}%
  \BibitemOpen
  \bibfield  {author} {\bibinfo {author} {\bibfnamefont {A.}~\bibnamefont
  {Nitzan}}\ and\ \bibinfo {author} {\bibfnamefont {M.~A.}\ \bibnamefont
  {Ratner}},\ }\href {\doibase 10.1126/science.1081572} {\bibfield  {journal}
  {\bibinfo  {journal} {Science}\ }\textbf {\bibinfo {volume} {300}},\ \bibinfo
  {pages} {1384} (\bibinfo {year} {2003})}\BibitemShut {NoStop}%
\bibitem [{\citenamefont {Qiu}\ \emph {et~al.}(2003)\citenamefont {Qiu},
  \citenamefont {Nazin},\ and\ \citenamefont {Ho}}]{Qiu2003}%
  \BibitemOpen
  \bibfield  {author} {\bibinfo {author} {\bibfnamefont {X.~H.}\ \bibnamefont
  {Qiu}}, \bibinfo {author} {\bibfnamefont {G.~V.}\ \bibnamefont {Nazin}}, \
  and\ \bibinfo {author} {\bibfnamefont {W.}~\bibnamefont {Ho}},\ }\href
  {\doibase 10.1126/science.1078675} {\bibfield  {journal} {\bibinfo  {journal}
  {Science}\ }\textbf {\bibinfo {volume} {299}},\ \bibinfo {pages} {542}
  (\bibinfo {year} {2003})}\BibitemShut {NoStop}%
\bibitem [{\citenamefont {Joachim}\ and\ \citenamefont
  {Ratner}(2005)}]{Joachim2005}%
  \BibitemOpen
  \bibfield  {author} {\bibinfo {author} {\bibfnamefont {C.}~\bibnamefont
  {Joachim}}\ and\ \bibinfo {author} {\bibfnamefont {M.~A.}\ \bibnamefont
  {Ratner}},\ }\href {\doibase 10.1073/pnas.0500075102} {\bibfield  {journal}
  {\bibinfo  {journal} {Proc. Natl. Acad. Sci. U. S. A.}\ }\textbf {\bibinfo
  {volume} {102}},\ \bibinfo {pages} {8801} (\bibinfo {year}
  {2005})}\BibitemShut {NoStop}%
\bibitem [{\citenamefont {Chen}\ \emph {et~al.}(2003)\citenamefont {Chen},
  \citenamefont {Zwolak},\ and\ \citenamefont {{Di Ventra}}}]{Chen2003}%
  \BibitemOpen
  \bibfield  {author} {\bibinfo {author} {\bibfnamefont {Y.~C.}\ \bibnamefont
  {Chen}}, \bibinfo {author} {\bibfnamefont {M.}~\bibnamefont {Zwolak}}, \ and\
  \bibinfo {author} {\bibfnamefont {M.}~\bibnamefont {{Di Ventra}}},\ }\href
  {\doibase 10.1021/nl0348544} {\bibfield  {journal} {\bibinfo  {journal} {Nano
  Lett.}\ }\textbf {\bibinfo {volume} {3}},\ \bibinfo {pages} {1691} (\bibinfo
  {year} {2003})}\BibitemShut {NoStop}%
\bibitem [{\citenamefont {Li}\ \emph {et~al.}(2012)\citenamefont {Li},
  \citenamefont {Ren}, \citenamefont {Wang}, \citenamefont {Zhang},
  \citenamefont {H{\"{a}}nggi},\ and\ \citenamefont {Li}}]{Li2012}%
  \BibitemOpen
  \bibfield  {author} {\bibinfo {author} {\bibfnamefont {N.}~\bibnamefont
  {Li}}, \bibinfo {author} {\bibfnamefont {J.}~\bibnamefont {Ren}}, \bibinfo
  {author} {\bibfnamefont {L.}~\bibnamefont {Wang}}, \bibinfo {author}
  {\bibfnamefont {G.}~\bibnamefont {Zhang}}, \bibinfo {author} {\bibfnamefont
  {P.}~\bibnamefont {H{\"{a}}nggi}}, \ and\ \bibinfo {author} {\bibfnamefont
  {B.}~\bibnamefont {Li}},\ }\href {\doibase 10.1103/RevModPhys.84.1045}
  {\bibfield  {journal} {\bibinfo  {journal} {Rev. Mod. Phys.}\ }\textbf
  {\bibinfo {volume} {84}},\ \bibinfo {pages} {1045} (\bibinfo {year}
  {2012})}\BibitemShut {NoStop}%
\bibitem [{\citenamefont {Dubi}\ and\ \citenamefont {{Di
  Ventra}}(2011)}]{Dubi2011}%
  \BibitemOpen
  \bibfield  {author} {\bibinfo {author} {\bibfnamefont {Y.}~\bibnamefont
  {Dubi}}\ and\ \bibinfo {author} {\bibfnamefont {M.}~\bibnamefont {{Di
  Ventra}}},\ }\href {\doibase 10.1103/RevModPhys.83.131} {\bibfield  {journal}
  {\bibinfo  {journal} {Rev. Mod. Phys.}\ }\textbf {\bibinfo {volume} {83}},\
  \bibinfo {pages} {131} (\bibinfo {year} {2011})}\BibitemShut {NoStop}%
\bibitem [{\citenamefont {Scott}\ and\ \citenamefont
  {Natelson}(2010)}]{Scott2010}%
  \BibitemOpen
  \bibfield  {author} {\bibinfo {author} {\bibfnamefont {G.~D.}\ \bibnamefont
  {Scott}}\ and\ \bibinfo {author} {\bibfnamefont {D.}~\bibnamefont
  {Natelson}},\ }\href {\doibase 10.1021/nn100793s} {\bibfield  {journal}
  {\bibinfo  {journal} {ACS Nano}\ }\textbf {\bibinfo {volume} {4}},\ \bibinfo
  {pages} {3560} (\bibinfo {year} {2010})}\BibitemShut {NoStop}%
\bibitem [{\citenamefont {Zimbovskaya}\ and\ \citenamefont
  {Pederson}(2011)}]{Zimbovskaya2011}%
  \BibitemOpen
  \bibfield  {author} {\bibinfo {author} {\bibfnamefont {N.~A.}\ \bibnamefont
  {Zimbovskaya}}\ and\ \bibinfo {author} {\bibfnamefont {M.~R.}\ \bibnamefont
  {Pederson}},\ }\href {\doibase 10.1016/j.physrep.2011.08.002} {\bibfield
  {journal} {\bibinfo  {journal} {Phys. Rep.}\ }\textbf {\bibinfo {volume}
  {509}},\ \bibinfo {pages} {1} (\bibinfo {year} {2011})}\BibitemShut {NoStop}%
\bibitem [{\citenamefont {Yu}\ \emph {et~al.}(2004)\citenamefont {Yu},
  \citenamefont {Keane}, \citenamefont {Ciszek}, \citenamefont {Cheng},
  \citenamefont {Stewart}, \citenamefont {Tour},\ and\ \citenamefont
  {Natelson}}]{Yu2004}%
  \BibitemOpen
  \bibfield  {author} {\bibinfo {author} {\bibfnamefont {L.~H.}\ \bibnamefont
  {Yu}}, \bibinfo {author} {\bibfnamefont {Z.~K.}\ \bibnamefont {Keane}},
  \bibinfo {author} {\bibfnamefont {J.~W.}\ \bibnamefont {Ciszek}}, \bibinfo
  {author} {\bibfnamefont {L.}~\bibnamefont {Cheng}}, \bibinfo {author}
  {\bibfnamefont {M.~P.}\ \bibnamefont {Stewart}}, \bibinfo {author}
  {\bibfnamefont {J.~M.}\ \bibnamefont {Tour}}, \ and\ \bibinfo {author}
  {\bibfnamefont {D.}~\bibnamefont {Natelson}},\ }\href {\doibase
  10.1103/PhysRevLett.93.266802} {\bibfield  {journal} {\bibinfo  {journal}
  {Phys. Rev. Lett.}\ }\textbf {\bibinfo {volume} {93}},\ \bibinfo {pages}
  {266802} (\bibinfo {year} {2004})}\BibitemShut {NoStop}%
\bibitem [{\citenamefont {Rakhmilevitch}\ \emph {et~al.}(2014)\citenamefont
  {Rakhmilevitch}, \citenamefont {Koryt{\'{a}}r}, \citenamefont {Bagrets},
  \citenamefont {Evers},\ and\ \citenamefont {Tal}}]{Rakhmilevitch2014}%
  \BibitemOpen
  \bibfield  {author} {\bibinfo {author} {\bibfnamefont {D.}~\bibnamefont
  {Rakhmilevitch}}, \bibinfo {author} {\bibfnamefont {R.}~\bibnamefont
  {Koryt{\'{a}}r}}, \bibinfo {author} {\bibfnamefont {A.}~\bibnamefont
  {Bagrets}}, \bibinfo {author} {\bibfnamefont {F.}~\bibnamefont {Evers}}, \
  and\ \bibinfo {author} {\bibfnamefont {O.}~\bibnamefont {Tal}},\ }\href
  {\doibase 10.1103/PhysRevLett.113.236603} {\bibfield  {journal} {\bibinfo
  {journal} {Phys. Rev. Lett.}\ }\textbf {\bibinfo {volume} {113}},\ \bibinfo
  {pages} {236603} (\bibinfo {year} {2014})}\BibitemShut {NoStop}%
\bibitem [{\citenamefont {Gaudioso}\ \emph {et~al.}(2000)\citenamefont
  {Gaudioso}, \citenamefont {Lauhon},\ and\ \citenamefont {Ho}}]{Gaudioso2000}%
  \BibitemOpen
  \bibfield  {author} {\bibinfo {author} {\bibfnamefont {J.}~\bibnamefont
  {Gaudioso}}, \bibinfo {author} {\bibfnamefont {L.~J.}\ \bibnamefont
  {Lauhon}}, \ and\ \bibinfo {author} {\bibfnamefont {W.}~\bibnamefont {Ho}},\
  }\href {\doibase 10.1103/PhysRevLett.85.1918} {\bibfield  {journal} {\bibinfo
   {journal} {Phys. Rev. Lett.}\ }\textbf {\bibinfo {volume} {85}},\ \bibinfo
  {pages} {1918} (\bibinfo {year} {2000})}\BibitemShut {NoStop}%
\bibitem [{\citenamefont {Capone}\ \emph {et~al.}(2009)\citenamefont {Capone},
  \citenamefont {Fabrizio}, \citenamefont {Castellani},\ and\ \citenamefont
  {Tosatti}}]{Capone2009}%
  \BibitemOpen
  \bibfield  {author} {\bibinfo {author} {\bibfnamefont {M.}~\bibnamefont
  {Capone}}, \bibinfo {author} {\bibfnamefont {M.}~\bibnamefont {Fabrizio}},
  \bibinfo {author} {\bibfnamefont {C.}~\bibnamefont {Castellani}}, \ and\
  \bibinfo {author} {\bibfnamefont {E.}~\bibnamefont {Tosatti}},\ }\href
  {\doibase 10.1103/RevModPhys.81.943} {\bibfield  {journal} {\bibinfo
  {journal} {Rev. Mod. Phys.}\ }\textbf {\bibinfo {volume} {81}},\ \bibinfo
  {pages} {943} (\bibinfo {year} {2009})}\BibitemShut {NoStop}%
\bibitem [{\citenamefont {{Dal Conte}}\ \emph {et~al.}(2012)\citenamefont {{Dal
  Conte}}, \citenamefont {Giannetti}, \citenamefont {Coslovich}, \citenamefont
  {Cilento}, \citenamefont {Bossini}, \citenamefont {Abebaw}, \citenamefont
  {Banfi}, \citenamefont {Ferrini}, \citenamefont {Eisaki}, \citenamefont
  {Greven}, \citenamefont {Damascelli}, \citenamefont {van~der Marel},\ and\
  \citenamefont {Parmigiani}}]{DalConte2012}%
  \BibitemOpen
  \bibfield  {author} {\bibinfo {author} {\bibfnamefont {S.}~\bibnamefont {{Dal
  Conte}}}, \bibinfo {author} {\bibfnamefont {C.}~\bibnamefont {Giannetti}},
  \bibinfo {author} {\bibfnamefont {G.}~\bibnamefont {Coslovich}}, \bibinfo
  {author} {\bibfnamefont {F.}~\bibnamefont {Cilento}}, \bibinfo {author}
  {\bibfnamefont {D.}~\bibnamefont {Bossini}}, \bibinfo {author} {\bibfnamefont
  {T.}~\bibnamefont {Abebaw}}, \bibinfo {author} {\bibfnamefont
  {F.}~\bibnamefont {Banfi}}, \bibinfo {author} {\bibfnamefont
  {G.}~\bibnamefont {Ferrini}}, \bibinfo {author} {\bibfnamefont
  {H.}~\bibnamefont {Eisaki}}, \bibinfo {author} {\bibfnamefont
  {M.}~\bibnamefont {Greven}}, \bibinfo {author} {\bibfnamefont
  {A.}~\bibnamefont {Damascelli}}, \bibinfo {author} {\bibfnamefont
  {D.}~\bibnamefont {van~der Marel}}, \ and\ \bibinfo {author} {\bibfnamefont
  {F.}~\bibnamefont {Parmigiani}},\ }\href {\doibase 10.1126/science.1216765}
  {\bibfield  {journal} {\bibinfo  {journal} {Science}\ }\textbf {\bibinfo
  {volume} {335}},\ \bibinfo {pages} {1600} (\bibinfo {year}
  {2012})}\BibitemShut {NoStop}%
\bibitem [{\citenamefont {Gadermaier}\ \emph {et~al.}(2010)\citenamefont
  {Gadermaier}, \citenamefont {Alexandrov}, \citenamefont {Kabanov},
  \citenamefont {Kusar}, \citenamefont {Mertelj}, \citenamefont {Yao},
  \citenamefont {Manzoni}, \citenamefont {Brida}, \citenamefont {Cerullo},\
  and\ \citenamefont {Mihailovic}}]{Gadermaier2010}%
  \BibitemOpen
  \bibfield  {author} {\bibinfo {author} {\bibfnamefont {C.}~\bibnamefont
  {Gadermaier}}, \bibinfo {author} {\bibfnamefont {A.~S.}\ \bibnamefont
  {Alexandrov}}, \bibinfo {author} {\bibfnamefont {V.~V.}\ \bibnamefont
  {Kabanov}}, \bibinfo {author} {\bibfnamefont {P.}~\bibnamefont {Kusar}},
  \bibinfo {author} {\bibfnamefont {T.}~\bibnamefont {Mertelj}}, \bibinfo
  {author} {\bibfnamefont {X.}~\bibnamefont {Yao}}, \bibinfo {author}
  {\bibfnamefont {C.}~\bibnamefont {Manzoni}}, \bibinfo {author} {\bibfnamefont
  {D.}~\bibnamefont {Brida}}, \bibinfo {author} {\bibfnamefont
  {G.}~\bibnamefont {Cerullo}}, \ and\ \bibinfo {author} {\bibfnamefont
  {D.}~\bibnamefont {Mihailovic}},\ }\href {\doibase
  10.1103/PhysRevLett.105.257001} {\bibfield  {journal} {\bibinfo  {journal}
  {Phys. Rev. Lett.}\ }\textbf {\bibinfo {volume} {105}},\ \bibinfo {pages}
  {257001} (\bibinfo {year} {2010})}\BibitemShut {NoStop}%
\bibitem [{\citenamefont {Gadermaier}\ \emph {et~al.}(2014)\citenamefont
  {Gadermaier}, \citenamefont {Kabanov}, \citenamefont {Alexandrov},
  \citenamefont {Stojchevska}, \citenamefont {Mertelj}, \citenamefont
  {Manzoni}, \citenamefont {Cerullo}, \citenamefont {Zhigadlo}, \citenamefont
  {Karpinski}, \citenamefont {Cai}, \citenamefont {Yao}, \citenamefont {Toda},
  \citenamefont {Oda}, \citenamefont {Sugai},\ and\ \citenamefont
  {Mihailovic}}]{Gadermaier2014}%
  \BibitemOpen
  \bibfield  {author} {\bibinfo {author} {\bibfnamefont {C.}~\bibnamefont
  {Gadermaier}}, \bibinfo {author} {\bibfnamefont {V.~V.}\ \bibnamefont
  {Kabanov}}, \bibinfo {author} {\bibfnamefont {A.~S.}\ \bibnamefont
  {Alexandrov}}, \bibinfo {author} {\bibfnamefont {L.}~\bibnamefont
  {Stojchevska}}, \bibinfo {author} {\bibfnamefont {T.}~\bibnamefont
  {Mertelj}}, \bibinfo {author} {\bibfnamefont {C.}~\bibnamefont {Manzoni}},
  \bibinfo {author} {\bibfnamefont {G.}~\bibnamefont {Cerullo}}, \bibinfo
  {author} {\bibfnamefont {N.~D.}\ \bibnamefont {Zhigadlo}}, \bibinfo {author}
  {\bibfnamefont {J.}~\bibnamefont {Karpinski}}, \bibinfo {author}
  {\bibfnamefont {Y.~Q.}\ \bibnamefont {Cai}}, \bibinfo {author} {\bibfnamefont
  {X.}~\bibnamefont {Yao}}, \bibinfo {author} {\bibfnamefont {Y.}~\bibnamefont
  {Toda}}, \bibinfo {author} {\bibfnamefont {M.}~\bibnamefont {Oda}}, \bibinfo
  {author} {\bibfnamefont {S.}~\bibnamefont {Sugai}}, \ and\ \bibinfo {author}
  {\bibfnamefont {D.}~\bibnamefont {Mihailovic}},\ }\href {\doibase
  10.1103/PhysRevX.4.011056} {\bibfield  {journal} {\bibinfo  {journal} {Phys.
  Rev. X}\ }\textbf {\bibinfo {volume} {4}},\ \bibinfo {pages} {011056}
  (\bibinfo {year} {2014})}\BibitemShut {NoStop}%
\bibitem [{\citenamefont {Perfetti}\ \emph {et~al.}(2006)\citenamefont
  {Perfetti}, \citenamefont {Loukakos}, \citenamefont {Lisowski}, \citenamefont
  {Bovensiepen}, \citenamefont {Berger}, \citenamefont {Biermann},
  \citenamefont {Cornaglia}, \citenamefont {Georges},\ and\ \citenamefont
  {Wolf}}]{Perfetti2006}%
  \BibitemOpen
  \bibfield  {author} {\bibinfo {author} {\bibfnamefont {L.}~\bibnamefont
  {Perfetti}}, \bibinfo {author} {\bibfnamefont {P.~A.}\ \bibnamefont
  {Loukakos}}, \bibinfo {author} {\bibfnamefont {M.}~\bibnamefont {Lisowski}},
  \bibinfo {author} {\bibfnamefont {U.}~\bibnamefont {Bovensiepen}}, \bibinfo
  {author} {\bibfnamefont {H.}~\bibnamefont {Berger}}, \bibinfo {author}
  {\bibfnamefont {S.}~\bibnamefont {Biermann}}, \bibinfo {author}
  {\bibfnamefont {P.~S.}\ \bibnamefont {Cornaglia}}, \bibinfo {author}
  {\bibfnamefont {A.}~\bibnamefont {Georges}}, \ and\ \bibinfo {author}
  {\bibfnamefont {M.}~\bibnamefont {Wolf}},\ }\href {\doibase
  10.1103/PhysRevLett.97.067402} {\bibfield  {journal} {\bibinfo  {journal}
  {Phys. Rev. Lett.}\ }\textbf {\bibinfo {volume} {97}},\ \bibinfo {pages}
  {067402} (\bibinfo {year} {2006})}\BibitemShut {NoStop}%
\bibitem [{\citenamefont {Perfetti}\ \emph {et~al.}(2008)\citenamefont
  {Perfetti}, \citenamefont {Loukakos}, \citenamefont {Lisowski}, \citenamefont
  {Bovensiepen}, \citenamefont {Wolf}, \citenamefont {Berger}, \citenamefont
  {Biermann},\ and\ \citenamefont {Georges}}]{Perfetti2008}%
  \BibitemOpen
  \bibfield  {author} {\bibinfo {author} {\bibfnamefont {L.}~\bibnamefont
  {Perfetti}}, \bibinfo {author} {\bibfnamefont {P.~A.}\ \bibnamefont
  {Loukakos}}, \bibinfo {author} {\bibfnamefont {M.}~\bibnamefont {Lisowski}},
  \bibinfo {author} {\bibfnamefont {U.}~\bibnamefont {Bovensiepen}}, \bibinfo
  {author} {\bibfnamefont {M.}~\bibnamefont {Wolf}}, \bibinfo {author}
  {\bibfnamefont {H.}~\bibnamefont {Berger}}, \bibinfo {author} {\bibfnamefont
  {S.}~\bibnamefont {Biermann}}, \ and\ \bibinfo {author} {\bibfnamefont
  {A.}~\bibnamefont {Georges}},\ }\href {\doibase
  10.1088/1367-2630/10/5/053019} {\bibfield  {journal} {\bibinfo  {journal}
  {New J. Phys.}\ }\textbf {\bibinfo {volume} {10}},\ \bibinfo {pages} {053019}
  (\bibinfo {year} {2008})}\BibitemShut {NoStop}%
\bibitem [{\citenamefont {Capone}\ \emph {et~al.}(2010)\citenamefont {Capone},
  \citenamefont {Castellani},\ and\ \citenamefont {Grilli}}]{Capone2010}%
  \BibitemOpen
  \bibfield  {author} {\bibinfo {author} {\bibfnamefont {M.}~\bibnamefont
  {Capone}}, \bibinfo {author} {\bibfnamefont {C.}~\bibnamefont {Castellani}},
  \ and\ \bibinfo {author} {\bibfnamefont {M.}~\bibnamefont {Grilli}},\
  }\href@noop {} {\bibfield  {journal} {\bibinfo  {journal} {Adv. Condens.
  Matter Phys.}\ }\textbf {\bibinfo {volume} {2010}},\ \bibinfo {pages}
  {920860} (\bibinfo {year} {2010})}\BibitemShut {NoStop}%
\bibitem [{\citenamefont {Fausti}\ \emph {et~al.}(2011)\citenamefont {Fausti},
  \citenamefont {Tobey}, \citenamefont {Dean}, \citenamefont {Kaiser},
  \citenamefont {Dienst}, \citenamefont {Hoffmann}, \citenamefont {Pyon},
  \citenamefont {Takayama}, \citenamefont {Takagi},\ and\ \citenamefont
  {Cavalleri}}]{Fausti2011}%
  \BibitemOpen
  \bibfield  {author} {\bibinfo {author} {\bibfnamefont {D.}~\bibnamefont
  {Fausti}}, \bibinfo {author} {\bibfnamefont {R.~I.}\ \bibnamefont {Tobey}},
  \bibinfo {author} {\bibfnamefont {N.}~\bibnamefont {Dean}}, \bibinfo {author}
  {\bibfnamefont {S.}~\bibnamefont {Kaiser}}, \bibinfo {author} {\bibfnamefont
  {A.}~\bibnamefont {Dienst}}, \bibinfo {author} {\bibfnamefont {M.~C.}\
  \bibnamefont {Hoffmann}}, \bibinfo {author} {\bibfnamefont {S.}~\bibnamefont
  {Pyon}}, \bibinfo {author} {\bibfnamefont {T.}~\bibnamefont {Takayama}},
  \bibinfo {author} {\bibfnamefont {H.}~\bibnamefont {Takagi}}, \ and\ \bibinfo
  {author} {\bibfnamefont {A.}~\bibnamefont {Cavalleri}},\ }\href {\doibase
  10.1126/science.1197294} {\bibfield  {journal} {\bibinfo  {journal}
  {Science}\ }\textbf {\bibinfo {volume} {331}},\ \bibinfo {pages} {189}
  (\bibinfo {year} {2011})}\BibitemShut {NoStop}%
\bibitem [{\citenamefont {Kaiser}\ \emph {et~al.}(2014)\citenamefont {Kaiser},
  \citenamefont {Clark}, \citenamefont {Nicoletti}, \citenamefont {Cotugno},
  \citenamefont {Tobey}, \citenamefont {Dean}, \citenamefont {Lupi},
  \citenamefont {Okamoto}, \citenamefont {Hasegawa}, \citenamefont {Jaksch},\
  and\ \citenamefont {Cavalleri}}]{Kaiser2014}%
  \BibitemOpen
  \bibfield  {author} {\bibinfo {author} {\bibfnamefont {S.}~\bibnamefont
  {Kaiser}}, \bibinfo {author} {\bibfnamefont {S.~R.}\ \bibnamefont {Clark}},
  \bibinfo {author} {\bibfnamefont {D.}~\bibnamefont {Nicoletti}}, \bibinfo
  {author} {\bibfnamefont {G.}~\bibnamefont {Cotugno}}, \bibinfo {author}
  {\bibfnamefont {R.~I.}\ \bibnamefont {Tobey}}, \bibinfo {author}
  {\bibfnamefont {N.}~\bibnamefont {Dean}}, \bibinfo {author} {\bibfnamefont
  {S.}~\bibnamefont {Lupi}}, \bibinfo {author} {\bibfnamefont {H.}~\bibnamefont
  {Okamoto}}, \bibinfo {author} {\bibfnamefont {T.}~\bibnamefont {Hasegawa}},
  \bibinfo {author} {\bibfnamefont {D.}~\bibnamefont {Jaksch}}, \ and\ \bibinfo
  {author} {\bibfnamefont {A.}~\bibnamefont {Cavalleri}},\ }\href {\doibase
  10.1038/srep03823} {\bibfield  {journal} {\bibinfo  {journal} {Sci. Rep.}\
  }\textbf {\bibinfo {volume} {4}},\ \bibinfo {pages} {3823} (\bibinfo {year}
  {2014})}\BibitemShut {NoStop}%
\bibitem [{\citenamefont {Anderson}(1961)}]{Anderson1961}%
  \BibitemOpen
  \bibfield  {author} {\bibinfo {author} {\bibfnamefont {P.~W.}\ \bibnamefont
  {Anderson}},\ }\href {\doibase 10.1103/PhysRev.124.41} {\bibfield  {journal}
  {\bibinfo  {journal} {Phys. Rev.}\ }\textbf {\bibinfo {volume} {124}},\
  \bibinfo {pages} {41} (\bibinfo {year} {1961})}\BibitemShut {NoStop}%
\bibitem [{\citenamefont {Holstein}(1959)}]{Holstein1959}%
  \BibitemOpen
  \bibfield  {author} {\bibinfo {author} {\bibfnamefont {T.}~\bibnamefont
  {Holstein}},\ }\href {\doibase 10.1016/0003-4916(59)90002-8} {\bibfield
  {journal} {\bibinfo  {journal} {Ann. Phys. (N. Y).}\ }\textbf {\bibinfo
  {volume} {8}},\ \bibinfo {pages} {325} (\bibinfo {year} {1959})}\BibitemShut
  {NoStop}%
\bibitem [{\citenamefont {Hewson}\ and\ \citenamefont
  {Meyer}(2001)}]{Hewson2002}%
  \BibitemOpen
  \bibfield  {author} {\bibinfo {author} {\bibfnamefont {A.}~\bibnamefont
  {Hewson}}\ and\ \bibinfo {author} {\bibfnamefont {D.}~\bibnamefont {Meyer}},\
  }\href {\doibase 10.1088/0953-8984/14/3/312} {\bibfield  {journal} {\bibinfo
  {journal} {J. Phys. Condens. Matter}\ }\textbf {\bibinfo {volume} {14}},\
  \bibinfo {pages} {23} (\bibinfo {year} {2001})}\BibitemShut {NoStop}%
\bibitem [{\citenamefont {Cornaglia}\ \emph {et~al.}(2004)\citenamefont
  {Cornaglia}, \citenamefont {Ness},\ and\ \citenamefont
  {Grempel}}]{Cornaglia2004}%
  \BibitemOpen
  \bibfield  {author} {\bibinfo {author} {\bibfnamefont {P.~S.}\ \bibnamefont
  {Cornaglia}}, \bibinfo {author} {\bibfnamefont {H.}~\bibnamefont {Ness}}, \
  and\ \bibinfo {author} {\bibfnamefont {D.~R.}\ \bibnamefont {Grempel}},\
  }\href {\doibase 10.1103/PhysRevLett.93.147201} {\bibfield  {journal}
  {\bibinfo  {journal} {Phys. Rev. Lett.}\ }\textbf {\bibinfo {volume} {93}},\
  \bibinfo {pages} {147201} (\bibinfo {year} {2004})}\BibitemShut {NoStop}%
\bibitem [{\citenamefont {Werner}\ and\ \citenamefont
  {Millis}(2007)}]{Werner2007}%
  \BibitemOpen
  \bibfield  {author} {\bibinfo {author} {\bibfnamefont {P.}~\bibnamefont
  {Werner}}\ and\ \bibinfo {author} {\bibfnamefont {A.~J.}\ \bibnamefont
  {Millis}},\ }\href {\doibase 10.1103/PhysRevLett.99.146404} {\bibfield
  {journal} {\bibinfo  {journal} {Phys. Rev. Lett.}\ }\textbf {\bibinfo
  {volume} {99}},\ \bibinfo {pages} {146404} (\bibinfo {year}
  {2007})}\BibitemShut {NoStop}%
\bibitem [{\citenamefont {Han}(2010)}]{Han2010}%
  \BibitemOpen
  \bibfield  {author} {\bibinfo {author} {\bibfnamefont {J.~E.}\ \bibnamefont
  {Han}},\ }\href {\doibase 10.1103/PhysRevB.81.113106} {\bibfield  {journal}
  {\bibinfo  {journal} {Phys. Rev. B}\ }\textbf {\bibinfo {volume} {81}},\
  \bibinfo {pages} {113106} (\bibinfo {year} {2010})}\BibitemShut {NoStop}%
\bibitem [{\citenamefont {Georges}\ \emph {et~al.}(1996)\citenamefont
  {Georges}, \citenamefont {Kotliar}, \citenamefont {Krauth},\ and\
  \citenamefont {Rozenberg}}]{Georges1996}%
  \BibitemOpen
  \bibfield  {author} {\bibinfo {author} {\bibfnamefont {A.}~\bibnamefont
  {Georges}}, \bibinfo {author} {\bibfnamefont {G.}~\bibnamefont {Kotliar}},
  \bibinfo {author} {\bibfnamefont {W.}~\bibnamefont {Krauth}}, \ and\ \bibinfo
  {author} {\bibfnamefont {M.~J.}\ \bibnamefont {Rozenberg}},\ }\href {\doibase
  10.1103/RevModPhys.68.13} {\bibfield  {journal} {\bibinfo  {journal} {Rev.
  Mod. Phys.}\ }\textbf {\bibinfo {volume} {68}},\ \bibinfo {pages} {13}
  (\bibinfo {year} {1996})}\BibitemShut {NoStop}%
\bibitem [{\citenamefont {Werner}\ and\ \citenamefont
  {Eckstein}(2013)}]{Werner2013}%
  \BibitemOpen
  \bibfield  {author} {\bibinfo {author} {\bibfnamefont {P.}~\bibnamefont
  {Werner}}\ and\ \bibinfo {author} {\bibfnamefont {M.}~\bibnamefont
  {Eckstein}},\ }\href {\doibase 10.1103/PhysRevB.88.165108} {\bibfield
  {journal} {\bibinfo  {journal} {Phys. Rev. B}\ }\textbf {\bibinfo {volume}
  {88}},\ \bibinfo {pages} {165108} (\bibinfo {year} {2013})}\BibitemShut
  {NoStop}%
\bibitem [{\citenamefont {Gole{\v{z}}}\ \emph {et~al.}(2015)\citenamefont
  {Gole{\v{z}}}, \citenamefont {Eckstein},\ and\ \citenamefont
  {Werner}}]{Golez2015}%
  \BibitemOpen
  \bibfield  {author} {\bibinfo {author} {\bibfnamefont {D.}~\bibnamefont
  {Gole{\v{z}}}}, \bibinfo {author} {\bibfnamefont {M.}~\bibnamefont
  {Eckstein}}, \ and\ \bibinfo {author} {\bibfnamefont {P.}~\bibnamefont
  {Werner}},\ }\href {\doibase 10.1103/PhysRevB.92.195123} {\bibfield
  {journal} {\bibinfo  {journal} {Phys. Rev. B}\ }\textbf {\bibinfo {volume}
  {92}},\ \bibinfo {pages} {195123} (\bibinfo {year} {2015})}\BibitemShut
  {NoStop}%
\bibitem [{\citenamefont {Paaske}\ and\ \citenamefont
  {Flensberg}(2005)}]{Paaske2005b}%
  \BibitemOpen
  \bibfield  {author} {\bibinfo {author} {\bibfnamefont {J.}~\bibnamefont
  {Paaske}}\ and\ \bibinfo {author} {\bibfnamefont {K.}~\bibnamefont
  {Flensberg}},\ }\href {\doibase 10.1103/PhysRevLett.94.176801} {\bibfield
  {journal} {\bibinfo  {journal} {Phys. Rev. Lett.}\ }\textbf {\bibinfo
  {volume} {94}},\ \bibinfo {pages} {176801} (\bibinfo {year}
  {2005})}\BibitemShut {NoStop}%
\bibitem [{\citenamefont {Jovchev}\ and\ \citenamefont
  {Anders}(2013)}]{Jovchev2013}%
  \BibitemOpen
  \bibfield  {author} {\bibinfo {author} {\bibfnamefont {A.}~\bibnamefont
  {Jovchev}}\ and\ \bibinfo {author} {\bibfnamefont {F.~B.}\ \bibnamefont
  {Anders}},\ }\href {\doibase 10.1103/PhysRevB.87.195112} {\bibfield
  {journal} {\bibinfo  {journal} {Phys. Rev. B}\ }\textbf {\bibinfo {volume}
  {87}},\ \bibinfo {pages} {195112} (\bibinfo {year} {2013})}\BibitemShut
  {NoStop}%
\bibitem [{\citenamefont {{Seoane Souto}}\ \emph {et~al.}(2014)\citenamefont
  {{Seoane Souto}}, \citenamefont {Yeyati}, \citenamefont
  {Mart{\'{\i}}n-Rodero},\ and\ \citenamefont {Monreal}}]{SeoaneSouto2014}%
  \BibitemOpen
  \bibfield  {author} {\bibinfo {author} {\bibfnamefont {R.}~\bibnamefont
  {{Seoane Souto}}}, \bibinfo {author} {\bibfnamefont {A.~L.}\ \bibnamefont
  {Yeyati}}, \bibinfo {author} {\bibfnamefont {A.}~\bibnamefont
  {Mart{\'{\i}}n-Rodero}}, \ and\ \bibinfo {author} {\bibfnamefont {R.~C.}\
  \bibnamefont {Monreal}},\ }\href {\doibase 10.1103/PhysRevB.89.085412}
  {\bibfield  {journal} {\bibinfo  {journal} {Phys. Rev. B}\ }\textbf {\bibinfo
  {volume} {89}},\ \bibinfo {pages} {085412} (\bibinfo {year}
  {2014})}\BibitemShut {NoStop}%
\bibitem [{\citenamefont {Albrecht}\ \emph
  {et~al.}(2013{\natexlab{a}})\citenamefont {Albrecht}, \citenamefont
  {Martin-Rodero}, \citenamefont {Monreal}, \citenamefont {M{\"{u}}hlbacher},\
  and\ \citenamefont {{Levy Yeyati}}}]{Albrecht2013a}%
  \BibitemOpen
  \bibfield  {author} {\bibinfo {author} {\bibfnamefont {K.~F.}\ \bibnamefont
  {Albrecht}}, \bibinfo {author} {\bibfnamefont {A.}~\bibnamefont
  {Martin-Rodero}}, \bibinfo {author} {\bibfnamefont {R.~C.}\ \bibnamefont
  {Monreal}}, \bibinfo {author} {\bibfnamefont {L.}~\bibnamefont
  {M{\"{u}}hlbacher}}, \ and\ \bibinfo {author} {\bibfnamefont
  {A.}~\bibnamefont {{Levy Yeyati}}},\ }\href {\doibase
  10.1103/PhysRevB.87.085127} {\bibfield  {journal} {\bibinfo  {journal} {Phys.
  Rev. B}\ }\textbf {\bibinfo {volume} {87}},\ \bibinfo {pages} {085127}
  (\bibinfo {year} {2013}{\natexlab{a}})}\BibitemShut {NoStop}%
\bibitem [{\citenamefont {Albrecht}\ \emph {et~al.}(2015)\citenamefont
  {Albrecht}, \citenamefont {Martin-Rodero}, \citenamefont {Schachenmayer},\
  and\ \citenamefont {M{\"{u}}hlbacher}}]{Albrecht2015}%
  \BibitemOpen
  \bibfield  {author} {\bibinfo {author} {\bibfnamefont {K.~F.}\ \bibnamefont
  {Albrecht}}, \bibinfo {author} {\bibfnamefont {A.}~\bibnamefont
  {Martin-Rodero}}, \bibinfo {author} {\bibfnamefont {J.}~\bibnamefont
  {Schachenmayer}}, \ and\ \bibinfo {author} {\bibfnamefont {L.}~\bibnamefont
  {M{\"{u}}hlbacher}},\ }\href {\doibase 10.1103/PhysRevB.91.064305} {\bibfield
   {journal} {\bibinfo  {journal} {Phys. Rev. B}\ }\textbf {\bibinfo {volume}
  {91}},\ \bibinfo {pages} {064305} (\bibinfo {year} {2015})}\BibitemShut
  {NoStop}%
\bibitem [{\citenamefont {Mitra}\ \emph {et~al.}(2004)\citenamefont {Mitra},
  \citenamefont {Aleiner},\ and\ \citenamefont {Millis}}]{Mitra2004}%
  \BibitemOpen
  \bibfield  {author} {\bibinfo {author} {\bibfnamefont {A.}~\bibnamefont
  {Mitra}}, \bibinfo {author} {\bibfnamefont {I.}~\bibnamefont {Aleiner}}, \
  and\ \bibinfo {author} {\bibfnamefont {A.~J.}\ \bibnamefont {Millis}},\
  }\href {\doibase 10.1103/PhysRevB.69.245302} {\bibfield  {journal} {\bibinfo
  {journal} {Phys. Rev. B}\ }\textbf {\bibinfo {volume} {69}},\ \bibinfo
  {pages} {245302} (\bibinfo {year} {2004})}\BibitemShut {NoStop}%
\bibitem [{\citenamefont {Mitra}\ \emph {et~al.}(2005)\citenamefont {Mitra},
  \citenamefont {Aleiner},\ and\ \citenamefont {Millis}}]{Mitra2005}%
  \BibitemOpen
  \bibfield  {author} {\bibinfo {author} {\bibfnamefont {A.}~\bibnamefont
  {Mitra}}, \bibinfo {author} {\bibfnamefont {I.}~\bibnamefont {Aleiner}}, \
  and\ \bibinfo {author} {\bibfnamefont {A.~J.}\ \bibnamefont {Millis}},\
  }\href {\doibase 10.1103/PhysRevLett.94.076404} {\bibfield  {journal}
  {\bibinfo  {journal} {Phys. Rev. Lett.}\ }\textbf {\bibinfo {volume} {94}},\
  \bibinfo {pages} {076404} (\bibinfo {year} {2005})}\BibitemShut {NoStop}%
\bibitem [{\citenamefont {H{\"{a}}rtle}\ \emph {et~al.}(2009)\citenamefont
  {H{\"{a}}rtle}, \citenamefont {Benesch},\ and\ \citenamefont
  {Thoss}}]{Hartle2009}%
  \BibitemOpen
  \bibfield  {author} {\bibinfo {author} {\bibfnamefont {R.}~\bibnamefont
  {H{\"{a}}rtle}}, \bibinfo {author} {\bibfnamefont {C.}~\bibnamefont
  {Benesch}}, \ and\ \bibinfo {author} {\bibfnamefont {M.}~\bibnamefont
  {Thoss}},\ }\href {\doibase 10.1103/PhysRevLett.102.146801} {\bibfield
  {journal} {\bibinfo  {journal} {Phys. Rev. Lett.}\ }\textbf {\bibinfo
  {volume} {102}},\ \bibinfo {pages} {146801} (\bibinfo {year}
  {2009})}\BibitemShut {NoStop}%
\bibitem [{\citenamefont {Schultz}\ and\ \citenamefont {von
  Oppen}(2009)}]{Schultz2009}%
  \BibitemOpen
  \bibfield  {author} {\bibinfo {author} {\bibfnamefont {M.~G.}\ \bibnamefont
  {Schultz}}\ and\ \bibinfo {author} {\bibfnamefont {F.}~\bibnamefont {von
  Oppen}},\ }\href {\doibase 10.1103/PhysRevB.80.033302} {\bibfield  {journal}
  {\bibinfo  {journal} {Phys. Rev. B}\ }\textbf {\bibinfo {volume} {80}},\
  \bibinfo {pages} {033302} (\bibinfo {year} {2009})}\BibitemShut {NoStop}%
\bibitem [{\citenamefont {Esposito}\ and\ \citenamefont
  {Galperin}(2009)}]{Esposito2009}%
  \BibitemOpen
  \bibfield  {author} {\bibinfo {author} {\bibfnamefont {M.}~\bibnamefont
  {Esposito}}\ and\ \bibinfo {author} {\bibfnamefont {M.}~\bibnamefont
  {Galperin}},\ }\href {\doibase 10.1103/PhysRevB.79.205303} {\bibfield
  {journal} {\bibinfo  {journal} {Phys. Rev. B}\ }\textbf {\bibinfo {volume}
  {79}},\ \bibinfo {pages} {205303} (\bibinfo {year} {2009})}\BibitemShut
  {NoStop}%
\bibitem [{\citenamefont {Esposito}\ and\ \citenamefont
  {Galperin}(2010)}]{Esposito2010}%
  \BibitemOpen
  \bibfield  {author} {\bibinfo {author} {\bibfnamefont {M.}~\bibnamefont
  {Esposito}}\ and\ \bibinfo {author} {\bibfnamefont {M.}~\bibnamefont
  {Galperin}},\ }\href {\doibase 10.1021/jp103369s} {\bibfield  {journal}
  {\bibinfo  {journal} {J. Phys. Chem. C}\ }\textbf {\bibinfo {volume} {114}},\
  \bibinfo {pages} {20362} (\bibinfo {year} {2010})}\BibitemShut {NoStop}%
\bibitem [{\citenamefont {Dou}\ \emph {et~al.}(2015)\citenamefont {Dou},
  \citenamefont {Nitzan},\ and\ \citenamefont {Subotnik}}]{Dou2015}%
  \BibitemOpen
  \bibfield  {author} {\bibinfo {author} {\bibfnamefont {W.}~\bibnamefont
  {Dou}}, \bibinfo {author} {\bibfnamefont {A.}~\bibnamefont {Nitzan}}, \ and\
  \bibinfo {author} {\bibfnamefont {J.~E.}\ \bibnamefont {Subotnik}},\ }\href
  {\doibase 10.1063/1.4908034} {\bibfield  {journal} {\bibinfo  {journal} {J.
  Chem. Phys.}\ }\textbf {\bibinfo {volume} {142}},\ \bibinfo {pages} {084110}
  (\bibinfo {year} {2015})}\BibitemShut {NoStop}%
\bibitem [{\citenamefont {Tikhodeev}\ \emph {et~al.}(2001)\citenamefont
  {Tikhodeev}, \citenamefont {Natario}, \citenamefont {Makoshi}, \citenamefont
  {Mii},\ and\ \citenamefont {Ueba}}]{Tikhodeev2001}%
  \BibitemOpen
  \bibfield  {author} {\bibinfo {author} {\bibfnamefont {S.}~\bibnamefont
  {Tikhodeev}}, \bibinfo {author} {\bibfnamefont {M.}~\bibnamefont {Natario}},
  \bibinfo {author} {\bibfnamefont {K.}~\bibnamefont {Makoshi}}, \bibinfo
  {author} {\bibfnamefont {T.}~\bibnamefont {Mii}}, \ and\ \bibinfo {author}
  {\bibfnamefont {H.}~\bibnamefont {Ueba}},\ }\href {\doibase
  10.1016/S0039-6028(01)01190-6} {\bibfield  {journal} {\bibinfo  {journal}
  {Surf. Sci.}\ }\textbf {\bibinfo {volume} {493}},\ \bibinfo {pages} {63}
  (\bibinfo {year} {2001})}\BibitemShut {NoStop}%
\bibitem [{\citenamefont {Mii}\ \emph {et~al.}(2002)\citenamefont {Mii},
  \citenamefont {Tikhodeev},\ and\ \citenamefont {Ueba}}]{Mii2002}%
  \BibitemOpen
  \bibfield  {author} {\bibinfo {author} {\bibfnamefont {T.}~\bibnamefont
  {Mii}}, \bibinfo {author} {\bibfnamefont {S.}~\bibnamefont {Tikhodeev}}, \
  and\ \bibinfo {author} {\bibfnamefont {H.}~\bibnamefont {Ueba}},\ }\href
  {\doibase 10.1016/S0039-6028(01)01894-5} {\bibfield  {journal} {\bibinfo
  {journal} {Surf. Sci.}\ }\textbf {\bibinfo {volume} {502-503}},\ \bibinfo
  {pages} {26} (\bibinfo {year} {2002})}\BibitemShut {NoStop}%
\bibitem [{\citenamefont {Galperin}\ \emph {et~al.}(2004)\citenamefont
  {Galperin}, \citenamefont {Ratner},\ and\ \citenamefont
  {Nitzan}}]{Galperin2004}%
  \BibitemOpen
  \bibfield  {author} {\bibinfo {author} {\bibfnamefont {M.}~\bibnamefont
  {Galperin}}, \bibinfo {author} {\bibfnamefont {M.~A.}\ \bibnamefont
  {Ratner}}, \ and\ \bibinfo {author} {\bibfnamefont {A.}~\bibnamefont
  {Nitzan}},\ }\href {\doibase 10.1063/1.1814076} {\bibfield  {journal}
  {\bibinfo  {journal} {J. Chem. Phys.}\ }\textbf {\bibinfo {volume} {121}},\
  \bibinfo {pages} {11965} (\bibinfo {year} {2004})}\BibitemShut {NoStop}%
\bibitem [{\citenamefont {Ueda}\ and\ \citenamefont {Eto}(2006)}]{Ueda2006}%
  \BibitemOpen
  \bibfield  {author} {\bibinfo {author} {\bibfnamefont {A.}~\bibnamefont
  {Ueda}}\ and\ \bibinfo {author} {\bibfnamefont {M.}~\bibnamefont {Eto}},\
  }\href {\doibase 10.1103/PhysRevB.73.235353} {\bibfield  {journal} {\bibinfo
  {journal} {Phys. Rev. B}\ }\textbf {\bibinfo {volume} {73}},\ \bibinfo
  {pages} {235353} (\bibinfo {year} {2006})}\BibitemShut {NoStop}%
\bibitem [{\citenamefont {Dash}\ \emph {et~al.}(2010)\citenamefont {Dash},
  \citenamefont {Ness},\ and\ \citenamefont {Godby}}]{Dash2010}%
  \BibitemOpen
  \bibfield  {author} {\bibinfo {author} {\bibfnamefont {L.~K.}\ \bibnamefont
  {Dash}}, \bibinfo {author} {\bibfnamefont {H.}~\bibnamefont {Ness}}, \ and\
  \bibinfo {author} {\bibfnamefont {R.~W.}\ \bibnamefont {Godby}},\ }\href
  {\doibase 10.1063/1.3339390} {\bibfield  {journal} {\bibinfo  {journal} {J.
  Chem. Phys.}\ }\textbf {\bibinfo {volume} {132}},\ \bibinfo {pages} {104113}
  (\bibinfo {year} {2010})}\BibitemShut {NoStop}%
\bibitem [{\citenamefont {Dash}\ \emph {et~al.}(2011)\citenamefont {Dash},
  \citenamefont {Ness},\ and\ \citenamefont {Godby}}]{Dash2011}%
  \BibitemOpen
  \bibfield  {author} {\bibinfo {author} {\bibfnamefont {L.~K.}\ \bibnamefont
  {Dash}}, \bibinfo {author} {\bibfnamefont {H.}~\bibnamefont {Ness}}, \ and\
  \bibinfo {author} {\bibfnamefont {R.~W.}\ \bibnamefont {Godby}},\ }\href
  {\doibase 10.1103/PhysRevB.84.085433} {\bibfield  {journal} {\bibinfo
  {journal} {Phys. Rev. B}\ }\textbf {\bibinfo {volume} {84}},\ \bibinfo
  {pages} {085433} (\bibinfo {year} {2011})}\BibitemShut {NoStop}%
\bibitem [{\citenamefont {Dong}\ \emph {et~al.}(2013)\citenamefont {Dong},
  \citenamefont {Ding},\ and\ \citenamefont {Lei}}]{Dong2013}%
  \BibitemOpen
  \bibfield  {author} {\bibinfo {author} {\bibfnamefont {B.}~\bibnamefont
  {Dong}}, \bibinfo {author} {\bibfnamefont {G.~H.}\ \bibnamefont {Ding}}, \
  and\ \bibinfo {author} {\bibfnamefont {X.~L.}\ \bibnamefont {Lei}},\ }\href
  {\doibase 10.1103/PhysRevB.88.075414} {\bibfield  {journal} {\bibinfo
  {journal} {Phys. Rev. B}\ }\textbf {\bibinfo {volume} {88}},\ \bibinfo
  {pages} {075414} (\bibinfo {year} {2013})}\BibitemShut {NoStop}%
\bibitem [{\citenamefont {Galperin}\ \emph {et~al.}(2006)\citenamefont
  {Galperin}, \citenamefont {Nitzan},\ and\ \citenamefont
  {Ratner}}]{Galperin2006a}%
  \BibitemOpen
  \bibfield  {author} {\bibinfo {author} {\bibfnamefont {M.}~\bibnamefont
  {Galperin}}, \bibinfo {author} {\bibfnamefont {A.}~\bibnamefont {Nitzan}}, \
  and\ \bibinfo {author} {\bibfnamefont {M.~A.}\ \bibnamefont {Ratner}},\
  }\href {\doibase 10.1103/PhysRevB.73.045314} {\bibfield  {journal} {\bibinfo
  {journal} {Phys. Rev. B}\ }\textbf {\bibinfo {volume} {73}},\ \bibinfo
  {pages} {45314} (\bibinfo {year} {2006})}\BibitemShut {NoStop}%
\bibitem [{\citenamefont {Galperin}\ \emph {et~al.}(2007)\citenamefont
  {Galperin}, \citenamefont {Nitzan},\ and\ \citenamefont
  {Ratner}}]{Galperin2007}%
  \BibitemOpen
  \bibfield  {author} {\bibinfo {author} {\bibfnamefont {M.}~\bibnamefont
  {Galperin}}, \bibinfo {author} {\bibfnamefont {A.}~\bibnamefont {Nitzan}}, \
  and\ \bibinfo {author} {\bibfnamefont {M.~A.}\ \bibnamefont {Ratner}},\
  }\href {\doibase 10.1103/PhysRevB.76.035301} {\bibfield  {journal} {\bibinfo
  {journal} {Phys. Rev. B}\ }\textbf {\bibinfo {volume} {76}},\ \bibinfo
  {pages} {035301} (\bibinfo {year} {2007})}\BibitemShut {NoStop}%
\bibitem [{\citenamefont {Monreal}\ \emph {et~al.}(2010)\citenamefont
  {Monreal}, \citenamefont {Flores},\ and\ \citenamefont
  {Martin-Rodero}}]{Monreal2010a}%
  \BibitemOpen
  \bibfield  {author} {\bibinfo {author} {\bibfnamefont {R.~C.}\ \bibnamefont
  {Monreal}}, \bibinfo {author} {\bibfnamefont {F.}~\bibnamefont {Flores}}, \
  and\ \bibinfo {author} {\bibfnamefont {A.}~\bibnamefont {Martin-Rodero}},\
  }\href {\doibase 10.1103/PhysRevB.82.235412} {\bibfield  {journal} {\bibinfo
  {journal} {Phys. Rev. B}\ }\textbf {\bibinfo {volume} {82}},\ \bibinfo
  {pages} {235412} (\bibinfo {year} {2010})}\BibitemShut {NoStop}%
\bibitem [{\citenamefont {Sayyad}\ and\ \citenamefont
  {Eckstein}(2015)}]{Sayyad2015}%
  \BibitemOpen
  \bibfield  {author} {\bibinfo {author} {\bibfnamefont {S.}~\bibnamefont
  {Sayyad}}\ and\ \bibinfo {author} {\bibfnamefont {M.}~\bibnamefont
  {Eckstein}},\ }\href {\doibase 10.1103/PhysRevB.91.104301} {\bibfield
  {journal} {\bibinfo  {journal} {Phys. Rev. B}\ }\textbf {\bibinfo {volume}
  {91}},\ \bibinfo {pages} {104301} (\bibinfo {year} {2015})}\BibitemShut
  {NoStop}%
\bibitem [{\citenamefont {Martin-Rodero}\ \emph {et~al.}(2008)\citenamefont
  {Martin-Rodero}, \citenamefont {{Levy Yeyati}}, \citenamefont {Flores},\ and\
  \citenamefont {Monreal}}]{Martin-Rodero2008}%
  \BibitemOpen
  \bibfield  {author} {\bibinfo {author} {\bibfnamefont {A.}~\bibnamefont
  {Martin-Rodero}}, \bibinfo {author} {\bibfnamefont {A.}~\bibnamefont {{Levy
  Yeyati}}}, \bibinfo {author} {\bibfnamefont {F.}~\bibnamefont {Flores}}, \
  and\ \bibinfo {author} {\bibfnamefont {R.~C.}\ \bibnamefont {Monreal}},\
  }\href {\doibase 10.1103/PhysRevB.78.235112} {\bibfield  {journal} {\bibinfo
  {journal} {Phys. Rev. B}\ }\textbf {\bibinfo {volume} {78}},\ \bibinfo
  {pages} {235112} (\bibinfo {year} {2008})}\BibitemShut {NoStop}%
\bibitem [{\citenamefont {M{\"{u}}hlbacher}\ and\ \citenamefont
  {Rabani}(2008)}]{Muhlbacher2008}%
  \BibitemOpen
  \bibfield  {author} {\bibinfo {author} {\bibfnamefont {L.}~\bibnamefont
  {M{\"{u}}hlbacher}}\ and\ \bibinfo {author} {\bibfnamefont {E.}~\bibnamefont
  {Rabani}},\ }\href {\doibase 10.1103/PhysRevLett.100.176403} {\bibfield
  {journal} {\bibinfo  {journal} {Phys. Rev. Lett.}\ }\textbf {\bibinfo
  {volume} {100}},\ \bibinfo {pages} {176403} (\bibinfo {year}
  {2008})}\BibitemShut {NoStop}%
\bibitem [{\citenamefont {Klatt}\ \emph {et~al.}(2015)\citenamefont {Klatt},
  \citenamefont {M{\"{u}}hlbacher},\ and\ \citenamefont {Komnik}}]{Klatt2015}%
  \BibitemOpen
  \bibfield  {author} {\bibinfo {author} {\bibfnamefont {J.}~\bibnamefont
  {Klatt}}, \bibinfo {author} {\bibfnamefont {L.}~\bibnamefont
  {M{\"{u}}hlbacher}}, \ and\ \bibinfo {author} {\bibfnamefont
  {A.}~\bibnamefont {Komnik}},\ }\href {\doibase 10.1103/PhysRevB.91.155306}
  {\bibfield  {journal} {\bibinfo  {journal} {Phys. Rev. B}\ }\textbf {\bibinfo
  {volume} {91}},\ \bibinfo {pages} {155306} (\bibinfo {year}
  {2015})}\BibitemShut {NoStop}%
\bibitem [{\citenamefont {H{\"{u}}tzen}\ \emph {et~al.}(2012)\citenamefont
  {H{\"{u}}tzen}, \citenamefont {Weiss}, \citenamefont {Thorwart},\ and\
  \citenamefont {Egger}}]{Hutzen2012}%
  \BibitemOpen
  \bibfield  {author} {\bibinfo {author} {\bibfnamefont {R.}~\bibnamefont
  {H{\"{u}}tzen}}, \bibinfo {author} {\bibfnamefont {S.}~\bibnamefont {Weiss}},
  \bibinfo {author} {\bibfnamefont {M.}~\bibnamefont {Thorwart}}, \ and\
  \bibinfo {author} {\bibfnamefont {R.}~\bibnamefont {Egger}},\ }\href
  {\doibase 10.1103/PhysRevB.85.121408} {\bibfield  {journal} {\bibinfo
  {journal} {Phys. Rev. B}\ }\textbf {\bibinfo {volume} {85}},\ \bibinfo
  {pages} {121408} (\bibinfo {year} {2012})}\BibitemShut {NoStop}%
\bibitem [{\citenamefont {Simine}\ and\ \citenamefont
  {Segal}(2013)}]{Simine2013a}%
  \BibitemOpen
  \bibfield  {author} {\bibinfo {author} {\bibfnamefont {L.}~\bibnamefont
  {Simine}}\ and\ \bibinfo {author} {\bibfnamefont {D.}~\bibnamefont {Segal}},\
  }\href {\doibase 10.1063/1.4808108} {\bibfield  {journal} {\bibinfo
  {journal} {J. Chem. Phys.}\ }\textbf {\bibinfo {volume} {138}},\ \bibinfo
  {pages} {214111} (\bibinfo {year} {2013})}\BibitemShut {NoStop}%
\bibitem [{\citenamefont {Simine}\ and\ \citenamefont
  {Segal}(2014)}]{Simine2014}%
  \BibitemOpen
  \bibfield  {author} {\bibinfo {author} {\bibfnamefont {L.}~\bibnamefont
  {Simine}}\ and\ \bibinfo {author} {\bibfnamefont {D.}~\bibnamefont {Segal}},\
  }\href {\doibase 10.1063/1.4885051} {\bibfield  {journal} {\bibinfo
  {journal} {J. Chem. Phys.}\ }\textbf {\bibinfo {volume} {141}},\ \bibinfo
  {pages} {014704} (\bibinfo {year} {2014})}\BibitemShut {NoStop}%
\bibitem [{\citenamefont {Albrecht}\ \emph {et~al.}(2012)\citenamefont
  {Albrecht}, \citenamefont {Wang}, \citenamefont {M{\"{u}}hlbacher},
  \citenamefont {Thoss},\ and\ \citenamefont {Komnik}}]{Albrecht2012}%
  \BibitemOpen
  \bibfield  {author} {\bibinfo {author} {\bibfnamefont {K.~F.}\ \bibnamefont
  {Albrecht}}, \bibinfo {author} {\bibfnamefont {H.}~\bibnamefont {Wang}},
  \bibinfo {author} {\bibfnamefont {L.}~\bibnamefont {M{\"{u}}hlbacher}},
  \bibinfo {author} {\bibfnamefont {M.}~\bibnamefont {Thoss}}, \ and\ \bibinfo
  {author} {\bibfnamefont {A.}~\bibnamefont {Komnik}},\ }\href {\doibase
  10.1103/PhysRevB.86.081412} {\bibfield  {journal} {\bibinfo  {journal} {Phys.
  Rev. B}\ }\textbf {\bibinfo {volume} {86}},\ \bibinfo {pages} {081412}
  (\bibinfo {year} {2012})}\BibitemShut {NoStop}%
\bibitem [{\citenamefont {Wang}\ and\ \citenamefont {Thoss}(2013)}]{Wang2013}%
  \BibitemOpen
  \bibfield  {author} {\bibinfo {author} {\bibfnamefont {H.}~\bibnamefont
  {Wang}}\ and\ \bibinfo {author} {\bibfnamefont {M.}~\bibnamefont {Thoss}},\
  }\href {\doibase 10.1021/jp401464b} {\bibfield  {journal} {\bibinfo
  {journal} {J. Phys. Chem. A}\ }\textbf {\bibinfo {volume} {117}},\ \bibinfo
  {pages} {7431} (\bibinfo {year} {2013})}\BibitemShut {NoStop}%
\bibitem [{\citenamefont {Albrecht}\ \emph
  {et~al.}(2013{\natexlab{b}})\citenamefont {Albrecht}, \citenamefont {Soller},
  \citenamefont {M{\"{u}}hlbacher},\ and\ \citenamefont
  {Komnik}}]{Albrecht2013}%
  \BibitemOpen
  \bibfield  {author} {\bibinfo {author} {\bibfnamefont {K.~F.}\ \bibnamefont
  {Albrecht}}, \bibinfo {author} {\bibfnamefont {H.}~\bibnamefont {Soller}},
  \bibinfo {author} {\bibfnamefont {L.}~\bibnamefont {M{\"{u}}hlbacher}}, \
  and\ \bibinfo {author} {\bibfnamefont {A.}~\bibnamefont {Komnik}},\ }\href
  {\doibase 10.1016/j.physe.2013.05.019} {\bibfield  {journal} {\bibinfo
  {journal} {Phys. E}\ }\textbf {\bibinfo {volume} {54}},\ \bibinfo {pages}
  {15} (\bibinfo {year} {2013}{\natexlab{b}})}\BibitemShut {NoStop}%
\bibitem [{\citenamefont {Entin-Wohlman}\ \emph {et~al.}(2005)\citenamefont
  {Entin-Wohlman}, \citenamefont {Aharony},\ and\ \citenamefont
  {Meir}}]{Entin-Wohlman2005}%
  \BibitemOpen
  \bibfield  {author} {\bibinfo {author} {\bibfnamefont {O.}~\bibnamefont
  {Entin-Wohlman}}, \bibinfo {author} {\bibfnamefont {A.}~\bibnamefont
  {Aharony}}, \ and\ \bibinfo {author} {\bibfnamefont {Y.}~\bibnamefont
  {Meir}},\ }\href {\doibase 10.1103/PhysRevB.71.035333} {\bibfield  {journal}
  {\bibinfo  {journal} {Phys. Rev. B}\ }\textbf {\bibinfo {volume} {71}},\
  \bibinfo {pages} {035333} (\bibinfo {year} {2005})}\BibitemShut {NoStop}%
\bibitem [{\citenamefont {Koch}\ and\ \citenamefont {{Von
  Oppen}}(2005)}]{Koch2005}%
  \BibitemOpen
  \bibfield  {author} {\bibinfo {author} {\bibfnamefont {J.}~\bibnamefont
  {Koch}}\ and\ \bibinfo {author} {\bibfnamefont {F.}~\bibnamefont {{Von
  Oppen}}},\ }\href {\doibase 10.1103/PhysRevLett.94.206804} {\bibfield
  {journal} {\bibinfo  {journal} {Phys. Rev. Lett.}\ }\textbf {\bibinfo
  {volume} {94}},\ \bibinfo {pages} {206804} (\bibinfo {year}
  {2005})}\BibitemShut {NoStop}%
\bibitem [{\citenamefont {Koch}\ \emph {et~al.}(2006)\citenamefont {Koch},
  \citenamefont {Semmelhack}, \citenamefont {{Von Oppen}},\ and\ \citenamefont
  {Nitzan}}]{Koch2006}%
  \BibitemOpen
  \bibfield  {author} {\bibinfo {author} {\bibfnamefont {J.}~\bibnamefont
  {Koch}}, \bibinfo {author} {\bibfnamefont {M.}~\bibnamefont {Semmelhack}},
  \bibinfo {author} {\bibfnamefont {F.}~\bibnamefont {{Von Oppen}}}, \ and\
  \bibinfo {author} {\bibfnamefont {A.}~\bibnamefont {Nitzan}},\ }\href
  {\doibase 10.1103/PhysRevB.73.155306} {\bibfield  {journal} {\bibinfo
  {journal} {Phys. Rev. B}\ }\textbf {\bibinfo {volume} {73}},\ \bibinfo
  {pages} {155306} (\bibinfo {year} {2006})}\BibitemShut {NoStop}%
\bibitem [{\citenamefont {Monreal}\ and\ \citenamefont
  {Martin-Rodero}(2009)}]{Monreal2009}%
  \BibitemOpen
  \bibfield  {author} {\bibinfo {author} {\bibfnamefont {R.~C.}\ \bibnamefont
  {Monreal}}\ and\ \bibinfo {author} {\bibfnamefont {A.}~\bibnamefont
  {Martin-Rodero}},\ }\href {\doibase 10.1103/PhysRevB.79.115140} {\bibfield
  {journal} {\bibinfo  {journal} {Phys. Rev. B}\ }\textbf {\bibinfo {volume}
  {79}},\ \bibinfo {pages} {115140} (\bibinfo {year} {2009})}\BibitemShut
  {NoStop}%
\bibitem [{\citenamefont {Maekawa}\ \emph {et~al.}(2003)\citenamefont
  {Maekawa}, \citenamefont {Ko}, \citenamefont {Scho}, \citenamefont
  {Martinek}, \citenamefont {Utsumi}, \citenamefont {Imamura},\ and\
  \citenamefont {Barnas}}]{Swirkowicz2008}%
  \BibitemOpen
  \bibfield  {author} {\bibinfo {author} {\bibfnamefont {S.}~\bibnamefont
  {Maekawa}}, \bibinfo {author} {\bibfnamefont {J.}~\bibnamefont {Ko}},
  \bibinfo {author} {\bibfnamefont {G.}~\bibnamefont {Scho}}, \bibinfo {author}
  {\bibfnamefont {J.}~\bibnamefont {Martinek}}, \bibinfo {author}
  {\bibfnamefont {Y.}~\bibnamefont {Utsumi}}, \bibinfo {author} {\bibfnamefont
  {H.}~\bibnamefont {Imamura}}, \ and\ \bibinfo {author} {\bibfnamefont
  {J.}~\bibnamefont {Barnas}},\ }\href {\doibase 10.1103/PhysRevLett.91.127203}
  {\bibfield  {journal} {\bibinfo  {journal} {Phys. B}\ }\textbf {\bibinfo
  {volume} {91}},\ \bibinfo {pages} {20} (\bibinfo {year} {2003})}\BibitemShut
  {NoStop}%
\bibitem [{\citenamefont {Goker}\ and\ \citenamefont
  {Uyanik}(2012)}]{Goker2011}%
  \BibitemOpen
  \bibfield  {author} {\bibinfo {author} {\bibfnamefont {A.}~\bibnamefont
  {Goker}}\ and\ \bibinfo {author} {\bibfnamefont {B.}~\bibnamefont {Uyanik}},\
  }\href {\doibase 10.1016/j.physleta.2012.07.025} {\bibfield  {journal}
  {\bibinfo  {journal} {Phys. Lett. A}\ }\textbf {\bibinfo {volume} {376}},\
  \bibinfo {pages} {2735} (\bibinfo {year} {2012})}\BibitemShut {NoStop}%
\bibitem [{\citenamefont {Roura-Bas}\ \emph {et~al.}(2013)\citenamefont
  {Roura-Bas}, \citenamefont {Tosi},\ and\ \citenamefont
  {Aligia}}]{Roura-Bas2013}%
  \BibitemOpen
  \bibfield  {author} {\bibinfo {author} {\bibfnamefont {P.}~\bibnamefont
  {Roura-Bas}}, \bibinfo {author} {\bibfnamefont {L.}~\bibnamefont {Tosi}}, \
  and\ \bibinfo {author} {\bibfnamefont {A.~A.}\ \bibnamefont {Aligia}},\
  }\href {\doibase 10.1103/PhysRevB.87.195136} {\bibfield  {journal} {\bibinfo
  {journal} {Phys. Rev. B}\ }\textbf {\bibinfo {volume} {87}},\ \bibinfo
  {pages} {195136} (\bibinfo {year} {2013})}\BibitemShut {NoStop}%
\bibitem [{\citenamefont {H{\"{a}}rtle}\ and\ \citenamefont
  {Thoss}(2011)}]{Hartle2011a}%
  \BibitemOpen
  \bibfield  {author} {\bibinfo {author} {\bibfnamefont {R.}~\bibnamefont
  {H{\"{a}}rtle}}\ and\ \bibinfo {author} {\bibfnamefont {M.}~\bibnamefont
  {Thoss}},\ }\href {\doibase 10.1103/PhysRevB.83.115414} {\bibfield  {journal}
  {\bibinfo  {journal} {Phys. Rev. B}\ }\textbf {\bibinfo {volume} {83}},\
  \bibinfo {pages} {115414} (\bibinfo {year} {2011})}\BibitemShut {NoStop}%
\bibitem [{\citenamefont {Wilner}\ \emph {et~al.}(2013)\citenamefont {Wilner},
  \citenamefont {Wang}, \citenamefont {Cohen}, \citenamefont {Thoss},\ and\
  \citenamefont {Rabani}}]{Wilner2013}%
  \BibitemOpen
  \bibfield  {author} {\bibinfo {author} {\bibfnamefont {E.~Y.}\ \bibnamefont
  {Wilner}}, \bibinfo {author} {\bibfnamefont {H.}~\bibnamefont {Wang}},
  \bibinfo {author} {\bibfnamefont {G.}~\bibnamefont {Cohen}}, \bibinfo
  {author} {\bibfnamefont {M.}~\bibnamefont {Thoss}}, \ and\ \bibinfo {author}
  {\bibfnamefont {E.}~\bibnamefont {Rabani}},\ }\href {\doibase
  10.1103/PhysRevB.88.045137} {\bibfield  {journal} {\bibinfo  {journal} {Phys.
  Rev. B}\ }\textbf {\bibinfo {volume} {88}},\ \bibinfo {pages} {045137}
  (\bibinfo {year} {2013})}\BibitemShut {NoStop}%
\bibitem [{\citenamefont {Wilner}\ \emph {et~al.}(2014)\citenamefont {Wilner},
  \citenamefont {Wang}, \citenamefont {Thoss},\ and\ \citenamefont
  {Rabani}}]{Wilner2014}%
  \BibitemOpen
  \bibfield  {author} {\bibinfo {author} {\bibfnamefont {E.~Y.}\ \bibnamefont
  {Wilner}}, \bibinfo {author} {\bibfnamefont {H.}~\bibnamefont {Wang}},
  \bibinfo {author} {\bibfnamefont {M.}~\bibnamefont {Thoss}}, \ and\ \bibinfo
  {author} {\bibfnamefont {E.}~\bibnamefont {Rabani}},\ }\href {\doibase
  10.1103/PhysRevB.89.205129} {\bibfield  {journal} {\bibinfo  {journal} {Phys.
  Rev. B}\ }\textbf {\bibinfo {volume} {89}},\ \bibinfo {pages} {205129}
  (\bibinfo {year} {2014})}\BibitemShut {NoStop}%
\bibitem [{\citenamefont {Cornaglia}\ \emph {et~al.}(2005)\citenamefont
  {Cornaglia}, \citenamefont {Grempel},\ and\ \citenamefont
  {Ness}}]{Cornaglia2005}%
  \BibitemOpen
  \bibfield  {author} {\bibinfo {author} {\bibfnamefont {P.~S.}\ \bibnamefont
  {Cornaglia}}, \bibinfo {author} {\bibfnamefont {D.~R.}\ \bibnamefont
  {Grempel}}, \ and\ \bibinfo {author} {\bibfnamefont {H.}~\bibnamefont
  {Ness}},\ }\href {\doibase 10.1103/PhysRevB.71.075320} {\bibfield  {journal}
  {\bibinfo  {journal} {Phys. Rev. B}\ }\textbf {\bibinfo {volume} {71}},\
  \bibinfo {pages} {075320} (\bibinfo {year} {2005})}\BibitemShut {NoStop}%
\bibitem [{\citenamefont {Cornaglia}\ \emph {et~al.}(2007)\citenamefont
  {Cornaglia}, \citenamefont {Usaj},\ and\ \citenamefont
  {Balseiro}}]{Cornaglia2007}%
  \BibitemOpen
  \bibfield  {author} {\bibinfo {author} {\bibfnamefont {P.~S.}\ \bibnamefont
  {Cornaglia}}, \bibinfo {author} {\bibfnamefont {G.}~\bibnamefont {Usaj}}, \
  and\ \bibinfo {author} {\bibfnamefont {C.~A.}\ \bibnamefont {Balseiro}},\
  }\href {\doibase 10.1103/PhysRevB.76.241403} {\bibfield  {journal} {\bibinfo
  {journal} {Phys. Rev. B}\ }\textbf {\bibinfo {volume} {76}},\ \bibinfo
  {pages} {241403} (\bibinfo {year} {2007})}\BibitemShut {NoStop}%
\bibitem [{\citenamefont {Eidelstein}\ \emph {et~al.}(2013)\citenamefont
  {Eidelstein}, \citenamefont {Goberman},\ and\ \citenamefont
  {Schiller}}]{Eidelstein2013}%
  \BibitemOpen
  \bibfield  {author} {\bibinfo {author} {\bibfnamefont {E.}~\bibnamefont
  {Eidelstein}}, \bibinfo {author} {\bibfnamefont {D.}~\bibnamefont
  {Goberman}}, \ and\ \bibinfo {author} {\bibfnamefont {A.}~\bibnamefont
  {Schiller}},\ }\href {\doibase 10.1103/PhysRevB.87.075319} {\bibfield
  {journal} {\bibinfo  {journal} {Phys. Rev. B}\ }\textbf {\bibinfo {volume}
  {87}},\ \bibinfo {pages} {075319} (\bibinfo {year} {2013})}\BibitemShut
  {NoStop}%
\bibitem [{\citenamefont {Laakso}\ \emph {et~al.}(2014)\citenamefont {Laakso},
  \citenamefont {Kennes}, \citenamefont {Jakobs},\ and\ \citenamefont
  {Meden}}]{Laakso2014}%
  \BibitemOpen
  \bibfield  {author} {\bibinfo {author} {\bibfnamefont {M.~A.}\ \bibnamefont
  {Laakso}}, \bibinfo {author} {\bibfnamefont {D.~M.}\ \bibnamefont {Kennes}},
  \bibinfo {author} {\bibfnamefont {S.~G.}\ \bibnamefont {Jakobs}}, \ and\
  \bibinfo {author} {\bibfnamefont {V.}~\bibnamefont {Meden}},\ }\href
  {\doibase 10.1088/1367-2630/16/2/023007} {\bibfield  {journal} {\bibinfo
  {journal} {New J. Phys.}\ }\textbf {\bibinfo {volume} {16}},\ \bibinfo
  {pages} {023007} (\bibinfo {year} {2014})}\BibitemShut {NoStop}%
\bibitem [{\citenamefont {Arrachea}\ and\ \citenamefont
  {Rozenberg}(2005)}]{Arrachea2005}%
  \BibitemOpen
  \bibfield  {author} {\bibinfo {author} {\bibfnamefont {L.}~\bibnamefont
  {Arrachea}}\ and\ \bibinfo {author} {\bibfnamefont {M.~J.}\ \bibnamefont
  {Rozenberg}},\ }\href {\doibase 10.1103/PhysRevB.72.041301} {\bibfield
  {journal} {\bibinfo  {journal} {Phys. Rev. B}\ }\textbf {\bibinfo {volume}
  {72}},\ \bibinfo {pages} {41301} (\bibinfo {year} {2005})}\BibitemShut
  {NoStop}%
\bibitem [{\citenamefont {Cohen}\ \emph
  {et~al.}(2014{\natexlab{a}})\citenamefont {Cohen}, \citenamefont {Reichman},
  \citenamefont {Millis},\ and\ \citenamefont {Gull}}]{Cohen2014}%
  \BibitemOpen
  \bibfield  {author} {\bibinfo {author} {\bibfnamefont {G.}~\bibnamefont
  {Cohen}}, \bibinfo {author} {\bibfnamefont {D.~R.}\ \bibnamefont {Reichman}},
  \bibinfo {author} {\bibfnamefont {A.~J.}\ \bibnamefont {Millis}}, \ and\
  \bibinfo {author} {\bibfnamefont {E.}~\bibnamefont {Gull}},\ }\href {\doibase
  10.1103/PhysRevB.89.115139} {\bibfield  {journal} {\bibinfo  {journal} {Phys.
  Rev. B}\ }\textbf {\bibinfo {volume} {89}},\ \bibinfo {pages} {115139}
  (\bibinfo {year} {2014}{\natexlab{a}})}\BibitemShut {NoStop}%
\bibitem [{\citenamefont {K{\"{o}}nig}\ \emph {et~al.}(1996)\citenamefont
  {K{\"{o}}nig}, \citenamefont {Schoeller},\ and\ \citenamefont
  {Sch{\"{o}}n}}]{Konig1996}%
  \BibitemOpen
  \bibfield  {author} {\bibinfo {author} {\bibfnamefont {J.}~\bibnamefont
  {K{\"{o}}nig}}, \bibinfo {author} {\bibfnamefont {H.}~\bibnamefont
  {Schoeller}}, \ and\ \bibinfo {author} {\bibfnamefont {G.}~\bibnamefont
  {Sch{\"{o}}n}},\ }\href {\doibase 10.1103/PhysRevLett.76.1715} {\bibfield
  {journal} {\bibinfo  {journal} {Phys. Rev. Lett.}\ }\textbf {\bibinfo
  {volume} {76}},\ \bibinfo {pages} {1715} (\bibinfo {year}
  {1996})}\BibitemShut {NoStop}%
\bibitem [{\citenamefont {Schir{\'{o}}}\ and\ \citenamefont
  {Fabrizio}(2009)}]{Schiro2009}%
  \BibitemOpen
  \bibfield  {author} {\bibinfo {author} {\bibfnamefont {M.}~\bibnamefont
  {Schir{\'{o}}}}\ and\ \bibinfo {author} {\bibfnamefont {M.}~\bibnamefont
  {Fabrizio}},\ }\href {\doibase 10.1103/PhysRevB.79.153302} {\bibfield
  {journal} {\bibinfo  {journal} {Phys. Rev. B}\ }\textbf {\bibinfo {volume}
  {79}},\ \bibinfo {pages} {153302} (\bibinfo {year} {2009})}\BibitemShut
  {NoStop}%
\bibitem [{\citenamefont {Gull}\ \emph
  {et~al.}(2011{\natexlab{a}})\citenamefont {Gull}, \citenamefont {Millis},
  \citenamefont {Lichtenstein}, \citenamefont {Rubtsov}, \citenamefont
  {Troyer},\ and\ \citenamefont {Werner}}]{Gull2011}%
  \BibitemOpen
  \bibfield  {author} {\bibinfo {author} {\bibfnamefont {E.}~\bibnamefont
  {Gull}}, \bibinfo {author} {\bibfnamefont {A.~J.}\ \bibnamefont {Millis}},
  \bibinfo {author} {\bibfnamefont {A.~I.}\ \bibnamefont {Lichtenstein}},
  \bibinfo {author} {\bibfnamefont {A.~N.}\ \bibnamefont {Rubtsov}}, \bibinfo
  {author} {\bibfnamefont {M.}~\bibnamefont {Troyer}}, \ and\ \bibinfo {author}
  {\bibfnamefont {P.}~\bibnamefont {Werner}},\ }\href {\doibase
  10.1103/RevModPhys.83.349} {\bibfield  {journal} {\bibinfo  {journal} {Rev.
  Mod. Phys.}\ }\textbf {\bibinfo {volume} {83}},\ \bibinfo {pages} {349}
  (\bibinfo {year} {2011}{\natexlab{a}})}\BibitemShut {NoStop}%
\bibitem [{\citenamefont {Cohen}\ and\ \citenamefont
  {Rabani}(2011)}]{Cohen2011}%
  \BibitemOpen
  \bibfield  {author} {\bibinfo {author} {\bibfnamefont {G.}~\bibnamefont
  {Cohen}}\ and\ \bibinfo {author} {\bibfnamefont {E.}~\bibnamefont {Rabani}},\
  }\href {\doibase 10.1103/PhysRevB.84.075150} {\bibfield  {journal} {\bibinfo
  {journal} {Phys. Rev. B}\ }\textbf {\bibinfo {volume} {84}},\ \bibinfo
  {pages} {075150} (\bibinfo {year} {2011})}\BibitemShut {NoStop}%
\bibitem [{\citenamefont {Cohen}\ \emph
  {et~al.}(2013{\natexlab{a}})\citenamefont {Cohen}, \citenamefont {Wilner},\
  and\ \citenamefont {Rabani}}]{Cohen2013}%
  \BibitemOpen
  \bibfield  {author} {\bibinfo {author} {\bibfnamefont {G.}~\bibnamefont
  {Cohen}}, \bibinfo {author} {\bibfnamefont {E.~Y.}\ \bibnamefont {Wilner}}, \
  and\ \bibinfo {author} {\bibfnamefont {E.}~\bibnamefont {Rabani}},\ }\href
  {\doibase 10.1088/1367-2630/15/7/073018} {\bibfield  {journal} {\bibinfo
  {journal} {New J. Phys.}\ }\textbf {\bibinfo {volume} {15}},\ \bibinfo
  {pages} {073018} (\bibinfo {year} {2013}{\natexlab{a}})}\BibitemShut
  {NoStop}%
\bibitem [{\citenamefont {Cohen}\ \emph
  {et~al.}(2013{\natexlab{b}})\citenamefont {Cohen}, \citenamefont {Gull},
  \citenamefont {Reichman}, \citenamefont {Millis},\ and\ \citenamefont
  {Rabani}}]{Cohen2013a}%
  \BibitemOpen
  \bibfield  {author} {\bibinfo {author} {\bibfnamefont {G.}~\bibnamefont
  {Cohen}}, \bibinfo {author} {\bibfnamefont {E.}~\bibnamefont {Gull}},
  \bibinfo {author} {\bibfnamefont {D.~R.}\ \bibnamefont {Reichman}}, \bibinfo
  {author} {\bibfnamefont {A.~J.}\ \bibnamefont {Millis}}, \ and\ \bibinfo
  {author} {\bibfnamefont {E.}~\bibnamefont {Rabani}},\ }\href {\doibase
  10.1103/PhysRevB.87.195108} {\bibfield  {journal} {\bibinfo  {journal} {Phys.
  Rev. B}\ }\textbf {\bibinfo {volume} {87}},\ \bibinfo {pages} {195108}
  (\bibinfo {year} {2013}{\natexlab{b}})}\BibitemShut {NoStop}%
\bibitem [{\citenamefont {Gull}\ \emph {et~al.}(2010)\citenamefont {Gull},
  \citenamefont {Reichman},\ and\ \citenamefont {Millis}}]{Gull2010}%
  \BibitemOpen
  \bibfield  {author} {\bibinfo {author} {\bibfnamefont {E.}~\bibnamefont
  {Gull}}, \bibinfo {author} {\bibfnamefont {D.~R.}\ \bibnamefont {Reichman}},
  \ and\ \bibinfo {author} {\bibfnamefont {A.~J.}\ \bibnamefont {Millis}},\
  }\href {\doibase 10.1103/PhysRevB.82.075109} {\bibfield  {journal} {\bibinfo
  {journal} {Phys. Rev. B}\ }\textbf {\bibinfo {volume} {82}},\ \bibinfo
  {pages} {075109} (\bibinfo {year} {2010})}\BibitemShut {NoStop}%
\bibitem [{\citenamefont {Bickers}(1987)}]{Bickers1987b}%
  \BibitemOpen
  \bibfield  {author} {\bibinfo {author} {\bibfnamefont {N.~E.}\ \bibnamefont
  {Bickers}},\ }\href {\doibase 10.1103/RevModPhys.59.845} {\bibfield
  {journal} {\bibinfo  {journal} {Rev. Mod. Phys.}\ }\textbf {\bibinfo {volume}
  {59}},\ \bibinfo {pages} {845} (\bibinfo {year} {1987})}\BibitemShut
  {NoStop}%
\bibitem [{\citenamefont {Pruschke}\ and\ \citenamefont
  {Grewe}(1989)}]{Pruschke1989}%
  \BibitemOpen
  \bibfield  {author} {\bibinfo {author} {\bibfnamefont {T.}~\bibnamefont
  {Pruschke}}\ and\ \bibinfo {author} {\bibfnamefont {N.}~\bibnamefont
  {Grewe}},\ }\href {\doibase 10.1007/BF01311391} {\bibfield  {journal}
  {\bibinfo  {journal} {Zeitschrift f{\"{u}}r Phys. B Condens. Matter}\
  }\textbf {\bibinfo {volume} {74}},\ \bibinfo {pages} {439} (\bibinfo {year}
  {1989})}\BibitemShut {NoStop}%
\bibitem [{\citenamefont {Wingreen}\ and\ \citenamefont
  {Meir}(1994)}]{Wingreen1994}%
  \BibitemOpen
  \bibfield  {author} {\bibinfo {author} {\bibfnamefont {N.~S.}\ \bibnamefont
  {Wingreen}}\ and\ \bibinfo {author} {\bibfnamefont {Y.}~\bibnamefont
  {Meir}},\ }\href {\doibase 10.1103/PhysRevB.49.11040} {\bibfield  {journal}
  {\bibinfo  {journal} {Phys. Rev. B}\ }\textbf {\bibinfo {volume} {49}},\
  \bibinfo {pages} {40} (\bibinfo {year} {1994})}\BibitemShut {NoStop}%
\bibitem [{\citenamefont {Haule}\ \emph {et~al.}(2001)\citenamefont {Haule},
  \citenamefont {Kirchner}, \citenamefont {Kroha},\ and\ \citenamefont
  {W{\"{o}}lfle}}]{Haule2001}%
  \BibitemOpen
  \bibfield  {author} {\bibinfo {author} {\bibfnamefont {K.}~\bibnamefont
  {Haule}}, \bibinfo {author} {\bibfnamefont {S.}~\bibnamefont {Kirchner}},
  \bibinfo {author} {\bibfnamefont {J.}~\bibnamefont {Kroha}}, \ and\ \bibinfo
  {author} {\bibfnamefont {P.}~\bibnamefont {W{\"{o}}lfle}},\ }\href {\doibase
  10.1103/PhysRevB.64.155111} {\bibfield  {journal} {\bibinfo  {journal} {Phys.
  Rev. B}\ }\textbf {\bibinfo {volume} {64}},\ \bibinfo {pages} {155111}
  (\bibinfo {year} {2001})}\BibitemShut {NoStop}%
\bibitem [{\citenamefont {Eckstein}\ and\ \citenamefont
  {Werner}(2010)}]{Eckstein2010}%
  \BibitemOpen
  \bibfield  {author} {\bibinfo {author} {\bibfnamefont {M.}~\bibnamefont
  {Eckstein}}\ and\ \bibinfo {author} {\bibfnamefont {P.}~\bibnamefont
  {Werner}},\ }\href {\doibase 10.1103/PhysRevB.82.115115} {\bibfield
  {journal} {\bibinfo  {journal} {Phys. Rev. B}\ }\textbf {\bibinfo {volume}
  {82}},\ \bibinfo {pages} {115115} (\bibinfo {year} {2010})}\BibitemShut
  {NoStop}%
\bibitem [{\citenamefont {Gull}\ \emph
  {et~al.}(2011{\natexlab{b}})\citenamefont {Gull}, \citenamefont {Reichman},\
  and\ \citenamefont {Millis}}]{Gull2011a}%
  \BibitemOpen
  \bibfield  {author} {\bibinfo {author} {\bibfnamefont {E.}~\bibnamefont
  {Gull}}, \bibinfo {author} {\bibfnamefont {D.~R.}\ \bibnamefont {Reichman}},
  \ and\ \bibinfo {author} {\bibfnamefont {A.~J.}\ \bibnamefont {Millis}},\
  }\href {\doibase 10.1103/PhysRevB.84.085134} {\bibfield  {journal} {\bibinfo
  {journal} {Phys. Rev. B}\ }\textbf {\bibinfo {volume} {84}},\ \bibinfo
  {pages} {085134} (\bibinfo {year} {2011}{\natexlab{b}})}\BibitemShut
  {NoStop}%
\bibitem [{\citenamefont {Cohen}\ \emph
  {et~al.}(2014{\natexlab{b}})\citenamefont {Cohen}, \citenamefont {Gull},
  \citenamefont {Reichman},\ and\ \citenamefont {Millis}}]{Cohen2014a}%
  \BibitemOpen
  \bibfield  {author} {\bibinfo {author} {\bibfnamefont {G.}~\bibnamefont
  {Cohen}}, \bibinfo {author} {\bibfnamefont {E.}~\bibnamefont {Gull}},
  \bibinfo {author} {\bibfnamefont {D.~R.}\ \bibnamefont {Reichman}}, \ and\
  \bibinfo {author} {\bibfnamefont {A.~J.}\ \bibnamefont {Millis}},\ }\href
  {\doibase 10.1103/PhysRevLett.112.146802} {\bibfield  {journal} {\bibinfo
  {journal} {Phys. Rev. Lett.}\ }\textbf {\bibinfo {volume} {112}},\ \bibinfo
  {pages} {146802} (\bibinfo {year} {2014}{\natexlab{b}})}\BibitemShut
  {NoStop}%
\bibitem [{\citenamefont {Cohen}\ \emph {et~al.}(2015)\citenamefont {Cohen},
  \citenamefont {Gull}, \citenamefont {Reichman},\ and\ \citenamefont
  {Millis}}]{Cohen2015}%
  \BibitemOpen
  \bibfield  {author} {\bibinfo {author} {\bibfnamefont {G.}~\bibnamefont
  {Cohen}}, \bibinfo {author} {\bibfnamefont {E.}~\bibnamefont {Gull}},
  \bibinfo {author} {\bibfnamefont {D.~R.}\ \bibnamefont {Reichman}}, \ and\
  \bibinfo {author} {\bibfnamefont {A.~J.}\ \bibnamefont {Millis}},\ }\href
  {\doibase 10.1103/PhysRevLett.115.266802} {\bibfield  {journal} {\bibinfo
  {journal} {Phys. Rev. Lett.}\ }\textbf {\bibinfo {volume} {115}},\ \bibinfo
  {pages} {266802} (\bibinfo {year} {2015})}\BibitemShut {NoStop}%
\bibitem [{\citenamefont {White}\ and\ \citenamefont
  {Galperin}(2012)}]{White2012}%
  \BibitemOpen
  \bibfield  {author} {\bibinfo {author} {\bibfnamefont {A.~J.}\ \bibnamefont
  {White}}\ and\ \bibinfo {author} {\bibfnamefont {M.}~\bibnamefont
  {Galperin}},\ }\href {\doibase 10.1039/c2cp41017f} {\bibfield  {journal}
  {\bibinfo  {journal} {Phys. Chem. Chem. Phys.}\ }\textbf {\bibinfo {volume}
  {14}},\ \bibinfo {pages} {13809} (\bibinfo {year} {2012})}\BibitemShut
  {NoStop}%
\bibitem [{\citenamefont {Lebanon}\ and\ \citenamefont
  {Schiller}(2001)}]{Lebanon2001}%
  \BibitemOpen
  \bibfield  {author} {\bibinfo {author} {\bibfnamefont {E.}~\bibnamefont
  {Lebanon}}\ and\ \bibinfo {author} {\bibfnamefont {A.}~\bibnamefont
  {Schiller}},\ }\href {\doibase 10.1103/PhysRevB.65.035308} {\bibfield
  {journal} {\bibinfo  {journal} {Phys. Rev. B}\ }\textbf {\bibinfo {volume}
  {65}},\ \bibinfo {pages} {035308} (\bibinfo {year} {2001})}\BibitemShut
  {NoStop}%
\bibitem [{\citenamefont {Sun}\ and\ \citenamefont {Guo}(2001)}]{Sun2001}%
  \BibitemOpen
  \bibfield  {author} {\bibinfo {author} {\bibfnamefont {Q.~F.}\ \bibnamefont
  {Sun}}\ and\ \bibinfo {author} {\bibfnamefont {H.}~\bibnamefont {Guo}},\
  }\href {\doibase 10.1103/PhysRevB.64.153306} {\bibfield  {journal} {\bibinfo
  {journal} {Phys. Rev. B}\ }\textbf {\bibinfo {volume} {64}},\ \bibinfo
  {pages} {153306} (\bibinfo {year} {2001})}\BibitemShut {NoStop}%
\bibitem [{\citenamefont {Werner}\ and\ \citenamefont
  {Millis}(2010)}]{Werner2010a}%
  \BibitemOpen
  \bibfield  {author} {\bibinfo {author} {\bibfnamefont {P.}~\bibnamefont
  {Werner}}\ and\ \bibinfo {author} {\bibfnamefont {A.~J.}\ \bibnamefont
  {Millis}},\ }\href {\doibase 10.1103/PhysRevLett.104.146401} {\bibfield
  {journal} {\bibinfo  {journal} {Phys. Rev. Lett.}\ }\textbf {\bibinfo
  {volume} {104}},\ \bibinfo {pages} {146401} (\bibinfo {year}
  {2010})}\BibitemShut {NoStop}%
\bibitem [{\citenamefont {Meir}\ \emph {et~al.}(1993)\citenamefont {Meir},
  \citenamefont {Wingreen},\ and\ \citenamefont {Lee}}]{Meir1993}%
  \BibitemOpen
  \bibfield  {author} {\bibinfo {author} {\bibfnamefont {Y.}~\bibnamefont
  {Meir}}, \bibinfo {author} {\bibfnamefont {N.~S.}\ \bibnamefont {Wingreen}},
  \ and\ \bibinfo {author} {\bibfnamefont {P.~A.}\ \bibnamefont {Lee}},\ }\href
  {\doibase 10.1103/PhysRevLett.70.2601} {\bibfield  {journal} {\bibinfo
  {journal} {Phys. Rev. Lett.}\ }\textbf {\bibinfo {volume} {70}},\ \bibinfo
  {pages} {2601} (\bibinfo {year} {1993})}\BibitemShut {NoStop}%
\bibitem [{\citenamefont {Hewson}(1993)}]{Hewson1993}%
  \BibitemOpen
  \bibfield  {author} {\bibinfo {author} {\bibfnamefont {A.}~\bibnamefont
  {Hewson}},\ }\href
  {http://www.cambridge.org/gb/knowledge/isbn/item1156358/?site{\_}locale=en{\_}GB}
  {\emph {\bibinfo {title} {{The Kondo Problem to Heavy Fermions}}}}\ (\bibinfo
   {publisher} {Cambridge University Press},\ \bibinfo {year}
  {1993})\BibitemShut {NoStop}%
\end{thebibliography}%

\end{document}